\documentclass{aa}
\usepackage[utf8]{inputenc}
\usepackage{natbib}
\usepackage{amsmath}
\usepackage{gensymb}
\usepackage{placeins}
\usepackage{afterpage}
\usepackage{tabularx}
\usepackage{booktabs}
\usepackage{multirow}
\usepackage{kantlipsum}
\usepackage{soul}
\usepackage{dblfloatfix}
\setstcolor{red}
\usepackage{hyperref}
\hypersetup{
    colorlinks=true,
    linkcolor=blue,
    citecolor=blue}
    
\usepackage{color}
\usepackage{graphicx}
\usepackage[export]{adjustbox}

\usepackage{txfonts}

\begin{document} 

\title{Distribution of solids in the rings of the HD 163296 disk: a multiwavelength study}
\author{G. Guidi \inst{1}
    \and
    A. Isella \inst{2}
    \and L. Testi \inst{3}
  \and  C. J. Chandler\inst{4}
\and H. B. Liu \inst{5}
\and H. M. Schmid \inst{1}
\and G. Rosotti \inst{6,7}
\and C. Meng \inst{2}
\and J. Jennings \inst{8}
\and J. P. Williams \inst{9}
\and J. M. Carpenter \inst{10}
\and I. de Gregorio-Monsalvo \inst{11}
\and H. Li \inst{12}
\and S.F. Liu \inst{13}
\and S. Ortolani \inst{14}
\and S. P. Quanz \inst{1}
\and L. Ricci \inst{15}
\and M. Tazzari \inst{8}
}

  \institute{ETH Zurich, Institute for Particle Physics and Astrophysics, 
Wolfgang-Pauli-Str. 27,CH-8093 Zurich, Switzerland\\ \email{gguidi@ethz.ch}
  \and
  Department of Physics and Astronomy, Rice University 6100 Main Street, MS-108, Houston, TX 77005, USA
  \and
  ESO, Karl Schwarzschild str. 2, 85748 Garching bei M\"unchen, Germany
  \and
  National Radio Astronomy Observatory, P.O. Box O, Socorro, NM 87801, USA
\and Institute of Astronomy and Astrophysics, Academia Sinica, Roosevelt Rd, Taipei 10617, Taiwan, ROC
\and Leiden Observatory, Leiden University, P.O. Box 9513, NL-2300 RA Leiden, the Netherlands 
\and School of Physics and Astronomy, University of Leicester, Leicester LE1 7RH, UK
\and Institute of Astronomy, University of Cambridge, Madingley Road, CB3 0HA Cambridge, UK
\and Institute for Astronomy, University of Hawaii at Manoa, Honolulu, HI 96822, USA
\and Joint ALMA Observatory, Avenida Alonso de C\'ordova 3107, Vitacura, Santiago, Chile 
\and European Southern Observatory, Alonso de Córdova 3107, Casilla 19001, Santiago de Chile, Chile
\and Theoretical Division, Los Alamos National Laboratory, Los Alamo, NM 87545, USA
\and School of Physics and Astronomy, Sun Yat-sen University, 2 Daxue Road, Zhuhai 519082, Guangdong Province, People’s Republic of China
\and Universit\'a di Padova, Dipartimento di Astronomia, Vicolo dell’Osservatorio 2, I--35122 Padova, Italy
\and Department of Physics and Astronomy, California State University Northridge, 18111 Nordhoff Street, Northridge, CA 91330, USA
}



\titlerunning{Distribution of solids in the rings of the HD 163296 disk}
\authorrunning {Guidi et al.}
\abstract
{Observations at millimeter wavelengths of bright protoplanetary disks have shown the ubiquitous presence of structures such as rings and spirals in the continuum emission. The derivation of the underlying properties of the emitting material is nontrivial because of the complex radiative processes involved.}
{In this paper we analyze new observations from the Atacama Large Millimeter/submillimeter Array (ALMA) and the \textit{Karl G. Jansky Very Large Array} (VLA) at high angular resolutions corresponding to 5 -- 8\,au to determine the dust spatial distribution and grain properties in the ringed disk of HD~163296.} 
{We fit the spectral energy distribution as a function of the radius at five wavelengths from 0.9 to 9\,mm, using a simple power law and a physical model based on an analytic description of radiative transfer that includes isothermal scattering. We considered eight dust populations and compared the models' performance using Bayesian evidence.}
{Our analysis shows that the moderately high optical depth ($\tau$>1) at $\lambda \leq$1.3\,mm in the dust rings  artificially lower the millimeter spectral index, which should therefore not be considered as a reliable direct proxy of the dust properties and especially the grain size. 
We find that the outer disk is composed of small grains on the order of 200\,$\mu$m with no significant difference between rings at 66 and 100\,au and the adjacent gaps, while in the innermost $\sim$30\,au, larger grains ($\geq$mm) could be present. We show that the assumptions on the dust composition have a strong impact on the derived surface densities and grain size. In particular, increasing the porosity of the grains to 80\% results in a total dust mass  about five times higher with respect to grains with 25\% porosity. Finally, we find that the derived opacities as a function of frequency deviate from a simple power law and that grains with a lower porosity seem to better reproduce the observations of HD~163296.} 
{While we do not find evidence of differential trapping in the rings of HD~163296, our overall results are consistent with the postulated presence of giant planets affecting the dust temperature structure and surface density, and possibly originating a second-generation dust population of small grains.}

\maketitle

\section{introduction}
The evolution of protoplanetary disks during their lifetime of a few million years \citep{2007ApJ...662.1067H} is intrinsically connected with the birth of planets from their solid and gaseous material. The multiplicity of processes at play make a global description of disks extremely 
challenging, although the increasing number of disk observations taken in the recent years -- from radio to optical and infrared frequencies -- are providing new and precious information that is helping to  unravel the puzzle. 

The emerging picture is that all the disks observed at high resolution present, even at early stages \citep[e.g., 1 Myr for HL~Tau,][]{ALMAHLTau}, a variety of small-scale structures (cavities, rings, spirals), indicating that local processes play a major role in the disk evolution \citep[e.g.][]{2018ApJ...869L..41A,2018ApJ...869...17L}. 
This opposes the previous view of disks with ``smooth'' and monotonically decreasing surface density profiles, which rapidly disperse under the action of global mechanisms such as viscous evolution \citep{1974MNRAS.168..603L}, consistent with low to medium resolution ($\geq$0\farcs5) observations in the submillimeter/millimeter available until $\sim$10 years ago. 
Substantial effort in the theoretical interpretation of this plenitude of new observations is being carried out, but the origin of the observed rings and spiral structures is still under debate. The most popular scenarios invoke planets perturbing the disk dust and gas distribution \citep[e.g.,][]{2018ApJ...869L..47Z}, photoevaporation \citep{2017MNRAS.464L..95E}, and gravitational instabilities \citep{2016Sci...353.1519P}, but also several magnetized, non self-gravitating disk instabilities, such as vortices via
Rossby waves \citep{2018ApJ...867....3H}, vertical shear instabilities \citep{2020arXiv200811195P,2021arXiv210800907B}, Magnetohydrodynamics (MHD) winds \citep{2020A&A...639A..95R}, warps and their induced instabilities \citep{2020arXiv201000862D}. Chemical effects, such as rapid dust growth around condensation fronts of the most abundant molecular species \citep{RosJohansen2013} or a pileup resulting from the sintering effect \citep{2016ApJ...821...82O}, have been proposed as responsible for the observed ringed structures as well. A correlation between the position of dust substructures and the location of the snowlines of major volatiles has been observed in some protoplanetary disks, supporting this scenario \citep[e.g.,][]{2015ApJ...806L...7Z, 2019ApJ...883...71C, 2021ApJS..257...14S}. 

At the current stage, the question of the origin of the substructures observed in protoplanetary disks is still open. With the notable exception of PDS70, where two planets and corresponding circumplanetary disks were imaged within the disk cavity \citep{2018A&A...617L...2M,2019ApJ...879L..25I,2021ApJ...916L...2B}, no conclusive explanation can be linked to the observed morphologies, even within the single sources. The comparison with simulations is complicated by the complex radiative
processes at play, which make the inference of the gas and dust density structures from the emitted radiation nontrivial. 
Dust in particular, despite being the most accessible observational tracer in disks, is not so straightforward to account for in global hydrodynamic simulations as it is subject to a variety of processes such as radial drift, coagulation, and fragmentation \citep[e.g.,][]{Birnstiel2010}.
The presence of rings and gaps in disks implies that dust could accumulate in certain regions which could promote further dust growth. It has been shown that grain growth has strong influences on understanding the lifetime and appearance of rings and vortices in 2D disks \citep{D2019,Laune2020,Li2020}, and it can affect the optical depth and spectral index of mutiple ringed structures \citep{Li2019}.

In this paper we focus on 
the bright disk around the Herbig Ae star HD~163296, and we use multiwavelength observations in the millimeter and centimeter range to reveal the radial variation in the dust properties with a nominal spatial resolution of about 5--8\,au. 
At a distance of 101.2 $\pm$ 1.2 \,pc \citep{2018AJ....156...58B}, this disk is an ideal candidate for studying planet formation in progress: a ringed structure in the dust emission of this disk was revealed  by ALMA (Atacama Large Millimeter/submillimeter Array) \citep{Isella2016,2016ApJ...818L..16Z}, and the presence of multiple planets perturbing the dust and gas dynamics was proposed to explain the observations \citep{Isella2016}. 
Several additional studies estimated 0.3--1~M$_\mathrm{Jup}$ planets at separations of about 50, 80, and 140\,au \citep{2018ApJ...857...87L,2018ApJ...860L..12T} and a $\sim$2~M$_\mathrm{Jup}$ planet at 260\,au \citep{2018ApJ...860L..13P}. 
Direct imaging of the giant planets was attempted at infrared wavelengths \citep{2018MNRAS.479.1505G,2019MNRAS.488...37M}, but without any robust detection. 

In this work we take advantage of new high resolution observations from ALMA and VLA (Very Large Array), in addition to archival DSHARP data at 1.3\,mm \citep{2018ApJ...869L..41A}, 
covering a wide spectral range (from 0.8\,mm to 3\,cm) to reconstruct the dust distribution in the midplane of the HD~163296 disk. 
{
In Sect.~\ref{sec:obs} we describe in detail the observations and the calibration procedures, we show the continuum intensity maps and corresponding brightness temperature profiles. Section \ref{sec:met} describes the methods we use for the analysis: the extraction of the non dust component and the radial profiles, and the models setups for fitting the spectral energy distribution. 
In Sect.~\ref{sec:res} we present our results: the spectral index maps and the dust properties we derive from the different models we apply to our multiwavelength analysis. In Sect.~\ref{sec:discussion} we discuss the results in terms of grain size distribution and dust mass, we compare them to previous studies and discuss the implications for the origin of the ringed structure in HD~163296. 
Finally, in Sect.~\ref{sec:concl} we summarize our main findings and draw our conclusions. }

\section{Observations and data reduction}
\label{sec:obs}
\subsection{ALMA Observations}
\label{sec:b3}
HD~163296 (also known as MWC~275) was observed with ALMA at Band~3 on September 7 and 13, 2016 with an array configuration of 42 antennas covering baselines from 15\,m  to 2.5\,km, resulting in an angular resolution of  $\sim$0.3$^{\prime\prime}$ and a maximum recoverable scale of 40$^{\prime\prime}$. The correlator was set up with four spectral windows in dual polarization mode: three SPWs for the continuum detection were set to cover the total bandwidth of 1.875~GHz, and were centered at 91.142, 103.006 and 104.694~GHz. One SPW was set to have a higher spectral resolution of 61~kHz ($\sim$0.2km/s after Hanning smoothing) and was centered at 93.165~GHz to include the N$_2$H$^+$(1--0) emission line. 
J1924-2914, J1733-1304 and J1751-1950 were observed to calibrate for bandpass, flux and phase, respectively. J1753-1934 was additionally observed every 20 minutes as a check source for the complex gains calibration. 
Data were calibrated using the ALMA pipeline in the CASA software package (version \texttt{5.4.1}), then self-calibration was performed on the lower side band (LSB) 
and upper side band (USB) separately, as the wavelength separation between the two resulted in an appreciable difference in flux (average flux difference of 35\% across the uv-space). 
Phase only self-calibration was carried out for each dataset, amplitude calibration was avoided in order to preserve the flux for multiband analysis. The final S/N (signal-to-noise) was improved by 60\%. 

Additional observations in Band~3 at a higher angular resolution were taken by ALMA on 16, 19 and 21 September 2017 (PI Isella). 
The array configuration consisted of 45, 44 and 41 antennas respectively, covering baselines from 41\,m  to 12\,km. The same calibrators listed for the more compact configuration were used for these observations. 
These were combined with the shorter baselines observations, again separating LSB from USB. Because of the different measurements of the flux calibrators, a flux-scaling was performed before combining the datasets: the reference flux was chosen by looking at the measurements of the correspondent ALMA flux calibrators, selecting the most recent in relation to the date of the observations. 

Observations at Band~4 were taken by ALMA in two complementary configurations: on November 13 and 23, 2017, HD~613296 was observed in the more extended configuration (C43--8), covering baselines from about 100\,m to 12\,km, using 43 and 50 antennas, respectively. The corresponding angular resolution was of 0\farcs06, with maximum recoverable scale of 1\farcs7 . The total on-source time for HD~163296 was 90 minutes; J1924-2914 was used for flux and bandpass calibration, J1753-1843 for the phase calibration. The correlator was set with 3 spectral windows for the continuum centered at 133.987, 135.925 and  145.976\,GHz, with a channel width of 976\,kHz ($\sim$ 2\,km/s). One spectral window was centered around the CS (3 --2) line at 146.963\,GHz, with a spectral resolution of 30.5\,kHz ($\sim$ 60\,m/s). 
In order to recover the large-scale emission that is filtered out by the extended configuration, HD~163296 was also observed in the more compact configuration C43--4 on January 17, 2018. Covering baselines from 15\,m to 1.4\,km and making use of 44 antennas, the  target was observed for 15 minutes with final angular resolution of $\sim$0\farcs5. 
Phase self-calibration was performed first on the short baseline dataset (configuration C43--4), then the dataset were combined, flux scaled as for the 3~mm datasets, and self-calibrated together.

\begin{figure*}
 \centering
 \includegraphics[keepaspectratio=True,width=\textwidth]{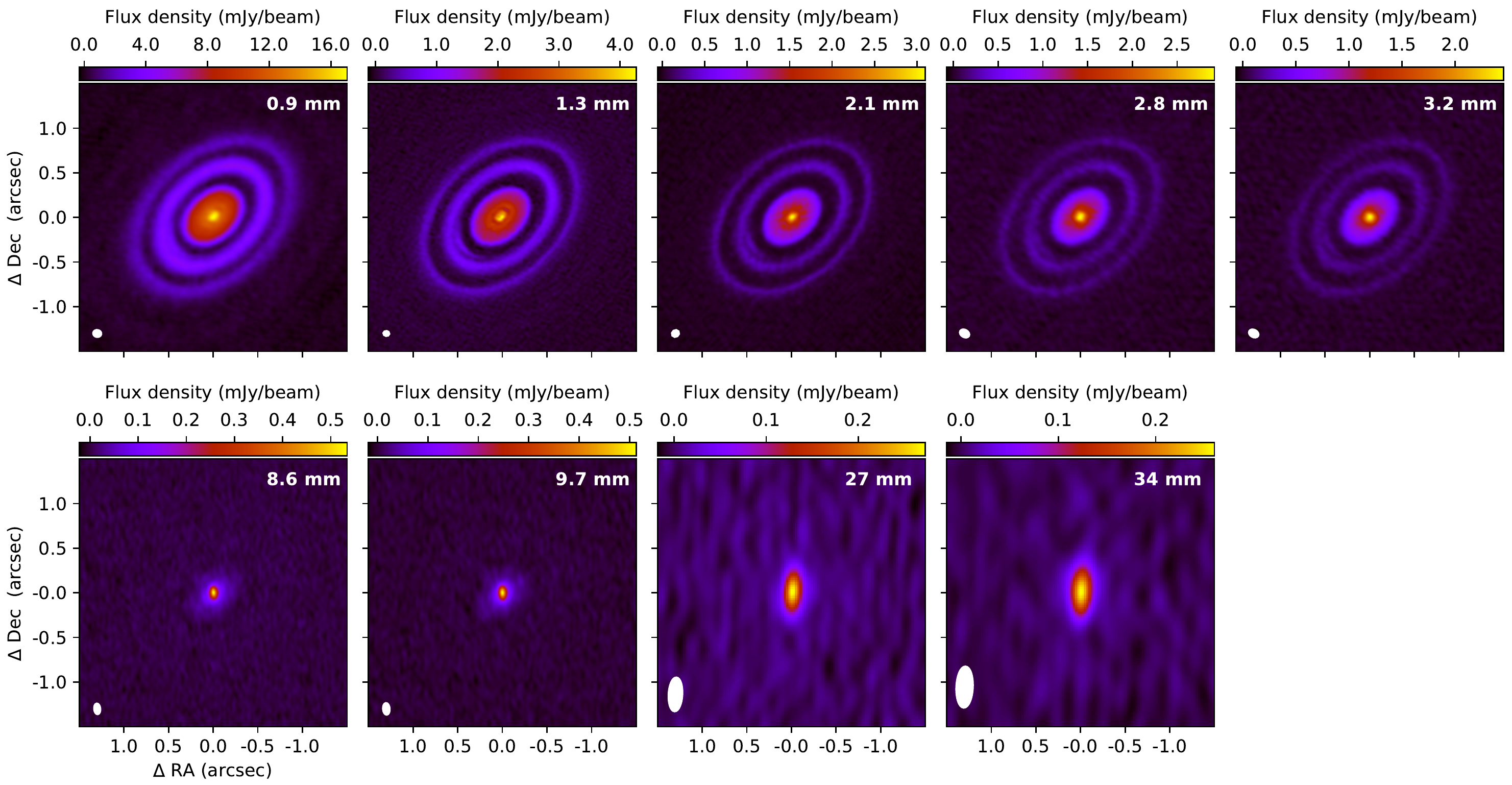}
 \caption{Continuum maps of HD~163296 from ALMA observations (top row) and VLA (bottom row). The synthesized beams of the final maps are drawn as white ellipse at the bottom left corner of each panel and are listed in Table \ref{tab:par}. }
 \label{fig:contmaps}
\end{figure*}

Observations at Band~7 were carried out on August 15, 2017 in the C40-7 configuration, using 42 antennas with a maximum recoverable scale of 0\farcs935 and longest baseline of 3.6\,Km. 
The on-source time on HD~163296 was 28 minutes, J1733-1304 was used as flux calibrator, J1751-1950 as phase calibrator and J1924-2914 as bandpass calibrator. 
The correlator was set with four spectral windows with 1.875\,GHz bandwidth, centered at 330.588\,GHz, 329.331\,GHz, 342.883\,GHz and 341.000\,GHz. 
In order to recover the flux at the shorter spatial frequencies, this dataset was combined with lower angular resolution data (project 2013.1.00053). 

{Finally, we used the calibrated Band~6 observation of HD~163296 made available by the DSHARP collaboration \citep{2018ApJ...869L..41A}, with three spectral windows for the continuum centered at 232.6, 245.0, and 246.9 GHz.
The high resolution DSHARP observations were combined with two additional datasets from previous ALMA programs, for a total baseline coverage spanning from 20\,m to 5.8\,Km (configurations C34-4, C34-7, C40-8), see also Table~2 and 3 in \citet{2018ApJ...869L..41A}. The final beam size is 0\farcs048$\times$0\farcs038 and maximum recoverable scale 13\arcsec. }

\subsection{VLA observations}
HD163296 was observed with  the VLA at Ka band in the A and B configurations during semesters 16B and 17B, respectively. The bandwidth covered the range between 29.04 and 36.96 GHz (corresponding to wavelengths of 8.1 -- 10.3 mm). 3C286 was used for absolute flux and bandpass calibration, J1820-2528 for pointing and J1755-2232 for complex gain (amplitude and phase) calibration. 
In the more compact configuration the maximum baseline was 11\,km and the total time on source was 2h 40min. In the extended configuration the baselines ranged over up to 36.6 km, with a total time on the science target of 7.5 hours. 
The datasets from the two configurations were first calibrated using the VLA pipeline script in CASA 5.6.1, epoch 17B was then self-calibrated with one round of phase calibration improving the signal-to-noise by $\sim$30\%. We combined the two configurations after scaling the flux of the A configuration, that resulted about 10\% lower than the B configuration (this was expected as the disk emission is likely larger than the maximum recoverable scale of $\sim$1\farcs6 in the extended array, so that spatial filtering can occur). 

Additional observations were carried out in the X band (frequencies from 8 to 10 GHz, corresponding to $\sim$3~cm) in the A configuration with maximum baseline 36.6~km, on October 1, 2016 and January 19, 2017 with a total of 1h 50min on the science target. The flux and bandpass calibrator was 3C286, the complex gain calibrator was J1820-2528 and After calibration using the CASA pipeline, we applied one round of self phase calibration improving the S/N of about 20\%.

\subsection{Bandwidth and time smearing}
For each of the ALMA and VLA datasets, a partial averaging in frequency and time was applied after the calibration. In order to avoid significant effects of chromatic aberration (also called bandwidth smearing, that is a radial smearing of the intensity distribution before it is convolved with the beam), we calculated the frequency bins corresponding to a reduction on the peak response of 1\% at the edge of the primary beam for a point source, as derived in Mangum 2016\footnote{https://safe.nrao.edu/wiki/pub/Main/RadioTutorial/\\BandwidthSmearing.pdf} (public NRAO documentation). We then applied the corresponding frequency averaging in each dataset with the task \texttt{mstransform} and using the closest integer number of channel bins that corresponded to the derived $\Delta\nu$. 

Similarly, an excessive time binning can produce smearing of the intensity, but in the azimuthal direction. We estimated the time bins to apply in order to have a reduction of the intensity peak up to 1\% anywhere in the image, writing the reduction of the peak response as 
\begin{equation*}
    R_a \simeq 1-\frac{1}{3} \left( \frac{0.832 \omega_e \tau_a}{\theta_b} \right)^2 (l_1^2 +m_1^2sin^2\delta_0)
\end{equation*}
as derived in Sect.~6.4 of \citet{2001isra.book.....T}. We chose a time bin $\tau_a$ to have a peak reduction of 1\% (R$_a$ = 0.99), using the synthesized beamwidth $\theta_b$, angular velocity of Earth $\omega_e$, image plane coordinates (l$_1$,m$_1$) and declination of the source $\delta_0$. 

\subsection{Systematic errors}
\label{sec:obserr}
The final uncertainty of the flux density measurements is mostly given by the accuracy of flux calibrators measurements: as they are in most cases quasar type objects, their intensity is intrinsically variable and this makes the evaluation of a flux density more difficult. Measurements of such calibrators are taken periodically by ALMA, but they are observed more frequently in certain bands (e.g. Band 3 and Band 6), while for other frequencies often an extrapolation of the flux from another wavelength is necessary (it is important to note that not only the intensity but also the spectral slope of a quasar can vary with time). Therefore the resulting calibration uncertainty will vary between ALMA Bands, and will depend on how close in time the calibrator measurement that is taken as reference was performed. 
Often a conservative calibration error of 10\% is used for ALMA observations, but the analysis of ALMA calibrator measurements spanning several years has shown that a value of 5\% is consistent with flux differences measured for short time spans in Band 3 and Band 6 \citep{2018MNRAS.478.1512B}. More recently, \citet{2021arXiv210205079F} compared ALMA and Planck observations and suggested that the calibration accuracy for ALMA Band\,3 should be taken as 2.4\% rather than the nominal 5\%. 

As it was pointed out that ALMA flux calibration accuracy could be poorer than the nominal value when the calibrator catalog is not up-to-date \citep{2020AJ....160..270F}, we check the individual calibrator measurements that were used as a reference during the calibration of our ALMA datasets. We find that the all the flux calibrations rely on recent measurements (taken within a few days of the science observations) of the corresponding calibrators, with deviations lower than than 3-4\% (see Appendix \ref{ap:calib}). Therefore we adopt the nominal flux uncertainties for the ALMA bands, corresponding to 5\% at Band~3 and Band~6, and to 10\% at Band 4 and Band~7.

For the VLA observations at the Ka and the X~band, 3C286 was used as flux calibrator, known to have flux densities and spectral index constant in time. 
Based on the indications by NRAO\footnote{https://science.nrao.edu/facilities/vla/docs/manuals/oss/performance/\\fdscale}, the single-epoch absolute accuracy is 10\% for the Ka band and 5\% for the X~band. 

\subsection{Continuum maps}
\label{sec:maps}
\begin{table}
	\caption{Parameters derived from the continuum images displayed in Figure \ref{fig:contmaps}.}
	\label{tab:par}
	\centering
	\begin{tabular*}{\columnwidth}{@{\extracolsep{\stretch{1}}}*{6}{@{}c@{}}}
		\hline \hline
		$\lambda$ & $F_{\rm{int}}$ & $F_{\rm{Peak}}$ & rms & beam & $R_{3\sigma}$\\
		{[mm]} & [mJy] & [mJy/beam] & [mJy/beam] & [arcsec $\times$ arcsec] & [arcsec] \\
		\hline
		0.88 & 1.68 $\cdot 10^{3}$ & 17.05 & 6.1 $\cdot 10^{-2}$ & 0.073 $\times$ 0.061 & 2.1 \\
		1.25 & 7.11 $\cdot 10^{2}$ & 4.26 & 2.3 $\cdot 10^{-2}$ & 0.048 $\times$ 0.038 & 2.2 \\
        2.14 & 1.76 $\cdot 10^{2}$ & 3.10 & 1.2 $\cdot 10^{-2}$  & 0.065 $\times$ 0.056 & 1.8\\
        2.82 & 8.30 $\cdot 10^{1}$ & 2.92 & 1.5 $\cdot 10^{-2}$ & 0.094 $\times$ 0.064 & 1.2\\
        3.17 & 5.82 $\cdot 10^{1}$ & 2.46 & 1.3 $\cdot 10^{-2}$& 0.095 $\times$ 0.065 & 1.2\\
        8.57 & 4.10 $\cdot 10^{0}$& 0.55 & 3.7 $\cdot 10^{-3}$ & 0.102 $\times$ 0.051 & 1.3\\
        9.67 & 3.03 $\cdot 10^{0}$& 0.51  & 3.4 $\cdot 10^{-3}$ & 0.113 $\times$ 0.057 & 1.2 \\
        27.4 & 5.76  $\cdot 10^{-1}$ & 0.27 & 4.3  $\cdot 10^{-3}$ & 0.362 $\times$ 0.136 & 1.2 \\
        33.6 & 4.30  $\cdot 10^{-1}$ & 0.26 & 4.1 $\cdot 10^{-3}$ & 0.444 $\times$ 0.167 & 1.2 \\
		\hline
	\end{tabular*}
	\tablefoot{The listed rms values represent the statistical uncertainties, that do not include the calibration error (see Sect.~\ref{sec:obserr}). 
	The last column lists the semi-major axis of the contours at a 3 times the rms level.}
\end{table}

In Figure \ref{fig:contmaps} we show the continuum maps of HD~163296 analyzed in this work, with wavelengths from 0.9~mm to 3.4~cm. All the observations come from new datasets at high resolution, with the exception of the image at 1.3\,mm taken from the DSHARP survey \citep{2018ApJ...869L..41A} and presented in \citet{isella2018}. The datasets at wavelength $\gtrsim$~3\,mm are spanning a larger bandwidth in terms of $\Delta \nu$/$\nu$, so they have been split into high and low frequencies as the fluxes can have a non-negligible variation within the baseband (an average difference across the spatial frequencies of 35\% is measure for ALMA Band~3 between Upper and Lower Side Band, 13\% between high and low frequencies of VLA Ka band, 10\% between high and low frequencies in VLA X~band). The images were then produced with the CASA task \texttt{tclean} in multifrequency mode with nterms = 2. 
The properties derived from each image are listed in Table \ref{tab:par}. 
The rms noise was derived as the standard deviation of the flux in the signal-free regions of the images. 
The integrated fluxes where derived through aperture photometry inside ellipses of 46$\degree$ inclination and 133$\degree$ position angle with steps
corresponding to $\sim$1 beam FWHM. The final flux correspond to the aperture at which successive variations of the flux remain within a 3 $\sigma$ level, where $\sigma$ is calculated as the standard deviation of the flux inside signal-free regions of the same area as the aperture.

\section{Methods}
\label{sec:met}
The main scope of this work is to constrain the dust properties in the HD~163296 disk using the high-resolution observations of the continuum emission presented in Sect.~\ref{sec:obs}. We describe in this section the methodology used for the determination of the non dust contribution, the extraction of the flux profiles, and the modeling setup for the spectral analysis.

\subsection{Contamination from ionized gas emission}
\label{sec:ff}
At radio frequencies, the emission from young stellar objects contains not only the thermal continuum from dust grains in the disk/envelope, but includes additional contributions that are often associated with free-free emission from ionized gas in the close surrounding of the star \citep[e.g.][]{2012ApJ...751L..42P}.  

The exact origin and therefore spatial extent of such emission in Herbig and T-Tauri stars is still not clear: free-free electrons can be generated in different environments, such as an envelope produced by either mass loss or mass accretion, ionized wind from the disk's atmosphere (i.e., photoevaporation) or collimated jets/ouflows from the star. 
An additional nonthermal process that produces a radio continuum is gyrosynchrotron emission from flares in the corona of magnetically active stars \citep{2002ASPC..277..491G}. 
The spectral slope at cm-wavelengths can help in identifying the responsible mechanism: emission from an optically thick, expanding shell at constant velocity is characterized by a slope of 0.6 \citep{panagia}. Free-free from disk atmospheres is expected to have a spectral index $\alpha$ between -0.1 and 2 for the optically thin and optically thick case respectively \citep[][]{2017MNRAS.466.4083U}. 
Nonthermal emission from corona flares have been observed to have a slope of around 2 from solar and stellar studies \citep{2002ASPC..277..491G}. 
\begin{figure}
 \centering
 \includegraphics[keepaspectratio=True,width=0.5\textwidth]{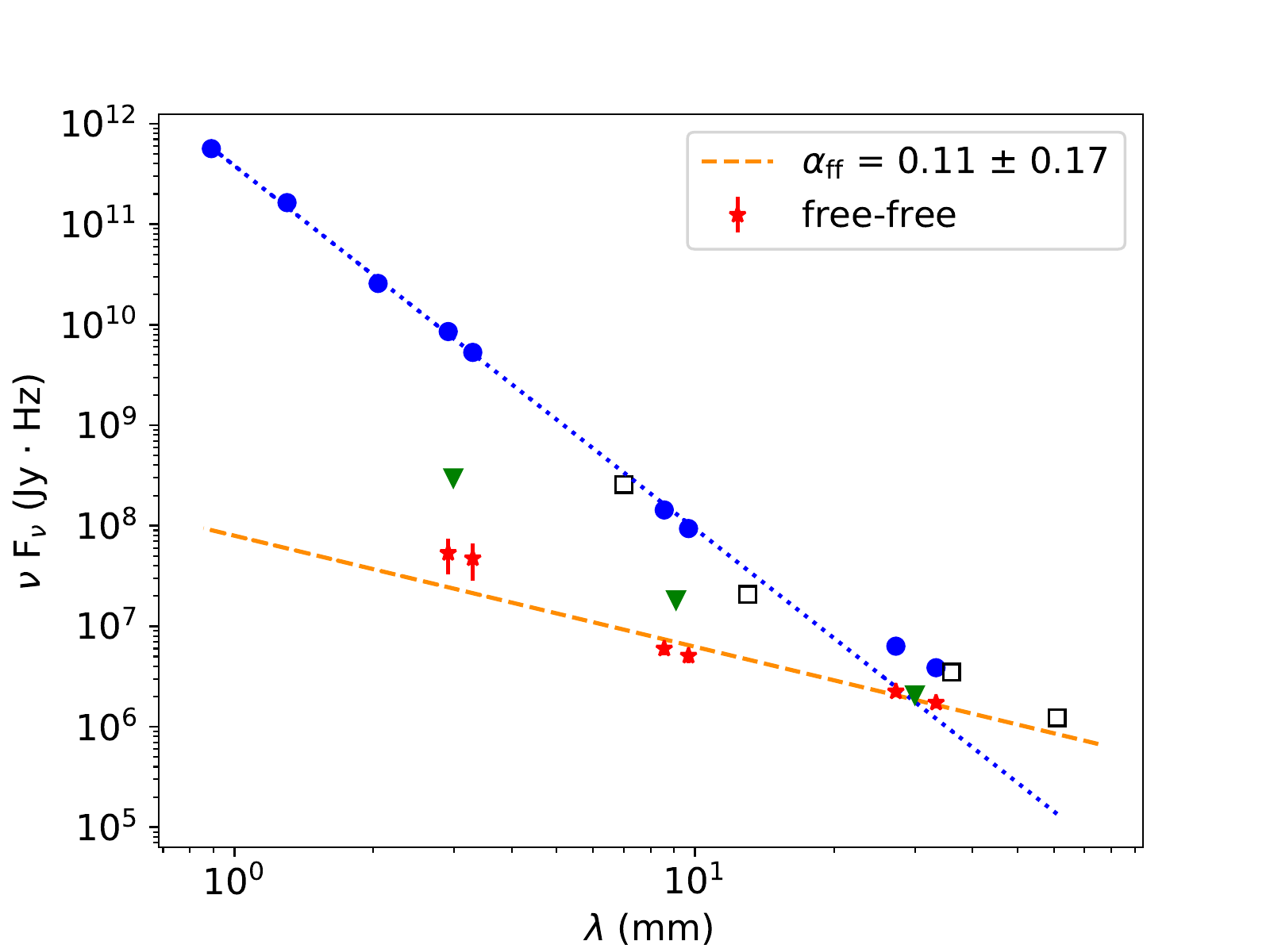}
 \caption{Spectral energy distribution of HD~163296. The blue circles represent the integrated fluxes from the data analyzed in this work, while empty squares are values from the literature \citep{isella2007,natta04,Guidi2016}. The blue dotted curve corresponds to a curve with spectral index of 2.6. Red stars are the contamination from a central compact emission, determined by the asymptotic values of the visibilities as described in Sect.~\ref{sec:ff}, and that are best fitted by a power law with index 0.11. Green triangles are upper limits for the free-free emission (see Sect.~\ref{sec:ff}), and correspond to the emission within a radius of 5~au. 
 }
 \label{fig:sed}
\end{figure}
Surveys at cm-wavelengths in nearby star-forming regions seem to indicate a correlation of the spectral index $\alpha$ with the evolutionary stage, consistent with free-free emission dominating in early type YSOs (Class~0 to Class~II, getting more optically thin toward the Class~II stage), and gyrosynchrotron emission dominating in Class IIIs \citep{2013ApJ...775...63D,2014ApJ...780..155L}.

This contamination in the HD~163296 system was estimated in previous works, and the slope of the free-free emission was found to be $\sim$0.6, using integrated fluxes at long wavelengths \citep{natta04} and $\sim$-0.2 using VLA resolved observations 
\citep{Guidi2016}. Both studies indicated that the contribution is negligible at wavelengths shorter than $\sim$7~mm. 

In Figure \ref{fig:sed} we plot the Spectral energy distribution in the mm/cm range for HD~163296: the change of slope of the SED when approaching cm wavelengths is clearly visible, and hints to the transition between different mechanisms responsible for the emission. 

To get an estimate of the dust contribution at the long wavelengths, we can fit only the ALMA integrated fluxes ($\lambda \le$3.3\,mm) with a single slope, obtaining an integrated spectral index $\alpha$ = 2.6 $\pm$ 0.1 (blue dotted curve in Figure \ref{fig:sed}). We note that such functional form seems to overpredict the fluxes at $\lambda \sim$ 9\,mm, that we know contain a fraction of non dust emission. We demonstrate later that a single $\alpha$ slope is not a good description of the dust emission, as the spectral index tends to get artificially lowered at shorter wavelengths by the high percentage of optically thick emission. 
For this reason, and given the higher resolution of the datasets presented in this work, we get an estimate of the nonthermal contribution based on the resolved data themselves, instead of extrapolating the dust flux from the integrated SED. 

We analyze the datasets at wavelengths as short as 3\,mm, and we fit for a constant emission in the visibility plane using the highest spatial frequencies (smaller spatial scales), in the assumption that the free-free is a point-like source. 
The fits for each dataset are shown is Appendix \ref{ap:ff}, and the estimated free-free contributions are displayed in Table \ref{tab:ff}.

As a further step, we employ a different method to get upper limits on the non dust emission and check the consistency with the point-source approach. We produce images from the VLA datasets at the highest possible resolution, and extract the flux of the deconvolved models within a radius of 5\,au. These measurements will contain both the entire free-free contamination and the dust emission, and are displayed as green triangles in Figure \ref{fig:sed}. 

\begin{table}
	\caption{Contamination from non dust emission ( F$_{\mathrm{c}}$) and associated error as estimated from the visibility profiles.}
	\label{tab:ff}
	\centering
	\begin{tabular*}{\columnwidth}{@{\extracolsep{\stretch{1}}}*{4}{c}@{}}
		\hline \hline
		$\lambda$ & F$_{\mathrm{c}}$ & $\Delta$ F$_{\mathrm{c}}$ & \% total Flux \\
		{[mm]} & [mJy] & [mJy] &\\
		\hline \hline
		2.91 & 0.52 & 0.20 & 0.6 \\
		3.29 & 0.52 & 0.21 & 0.9 \\
		8.57 & 0.17 & 0.03 & 4.1 \\
		9.67 & 0.16 & 0.03 & 5.2 \\
		27.3 & 0.20 & 0.02 & 35\\
		33.3 & 0.19 & 0.02 & 44 \\
		\hline
	\end{tabular*}
	\tablefoot{The last column is the ratio of the free-free over the total flux for each wavelength, multiplied by 100.}
\end{table}
The procedure and the motivation of the choice of 5~au is presented in detail in Appendix \ref{ap:ff}, along with the model images. 

\begin{figure*}[t]
 \centering
 \includegraphics[keepaspectratio=True,width=\textwidth]{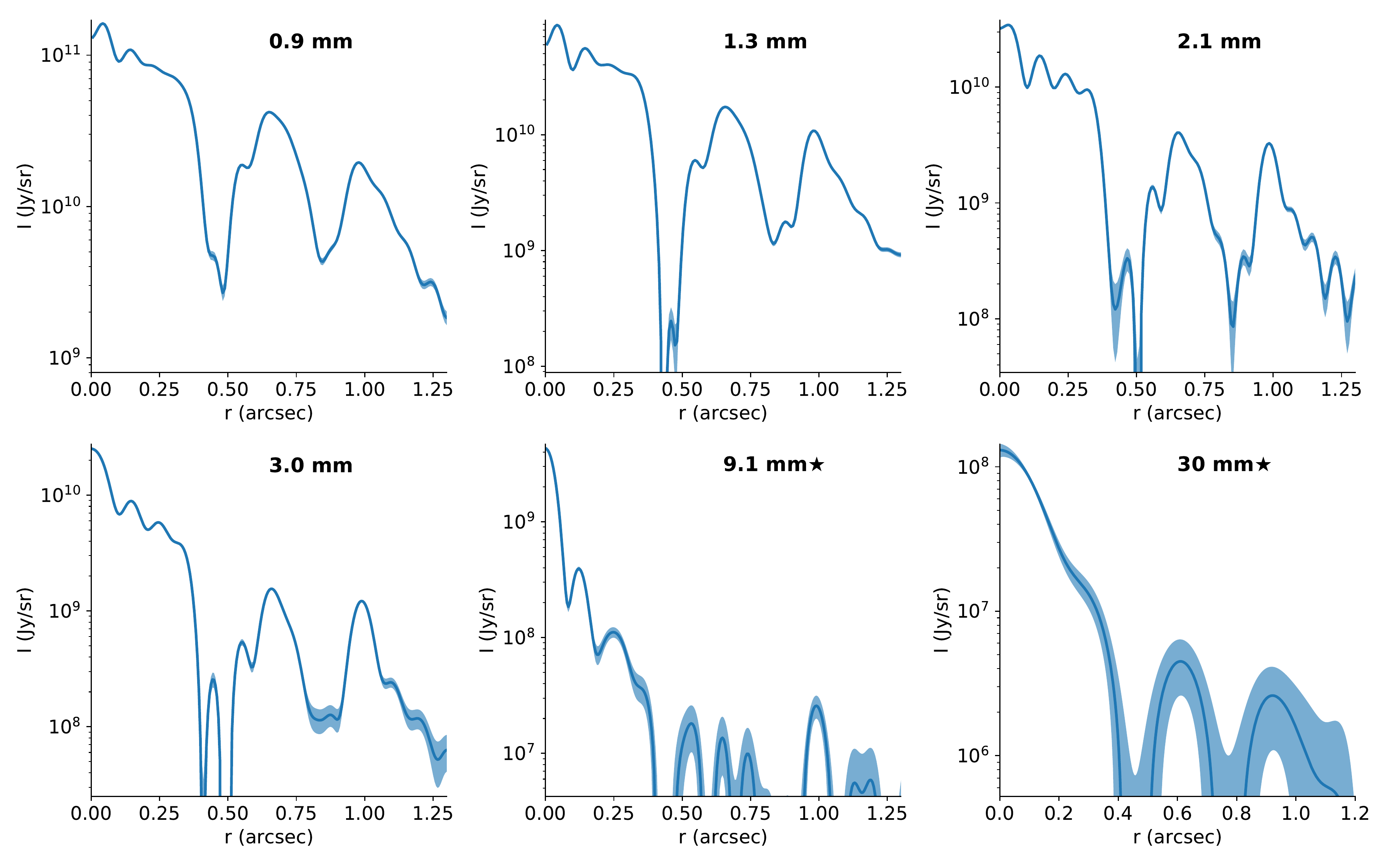}
 \caption{Radial profiles of the continuum emission derived with \texttt{frank} for each dataset, the shaded regions represent the statistical error computed by the \texttt{frank} fit. The star symbol in the plots relative to the VLA datasets indicates that these have been first corrected for the free-free/centrally peaked contamination, as described in Appendix \ref{app:vis}.}
 \label{fig:bff}
\end{figure*}

Based on our estimates in Table \ref{tab:ff}, we derive a free-free spectrum as a power-law as a function of frequency, that results $\nu^{0.11 \pm 0.17}$ (dashed line in Figure \ref{fig:sed}). The fit is done with a least-squares method (using the \texttt{python} routine \texttt{scipy.optimize.curve\_fit}) and the error on $\alpha$ is defined as the square root of the variance as estimated from the fit. 
We can observe that the value of 0.11 for the spectral slope is consistent with optically thin free-free emission from a stellar or disk wind \citep[][and references therein]{2014ApJ...795....1P}, while it seems to rule out the hypothesis of gyrosynchrotron emission for which an $\alpha_\mathrm{cm}$ around 2 is predicted. This would be also 
in agreement with what reported for Class~II systems in nearby star forming regions \citep{2013ApJ...775...63D}. 

We note that the assumption of a constant spectrum across cm wavelengths can be incorrect: models of free-free emission from disk winds ionized by X-rays or EUV predict for example a variation of the spectral index with frequency in the 1 to 100 GHZ range \citep{2013MNRAS.434.3378O}. Indeed, the intra-band spectral index that we measure at the peak of our VLA multifrequency images, where the emission is likely dominated by the free-free(Appendix \ref{ap:ff}) is different for 33\,GHz ($\alpha$ = 0.92 $\pm$ 0.07) and 10\,GHz ($\alpha$ = 0.24 $\pm$ 0.03), where the low uncertainties are due to the fact that they consist of a relative measurement within the same frequency band and therefore not affected by absolute amplitude calibration errors. 

While it is not possible at the moment to link such spectral indexes to a specific underlying mechanisms of the free-free emission, they can provide some useful reference for future studies of photoevaporation/accretion mechanisms in this system, subject that is beyond the scope of this paper.

\subsection{Extraction of radial profiles}
\label{sec:vis}
The more complex the sky brightness distribution, the less trivial is the image reconstruction from interferometric measurements through aperture synthesis, in our case by means of the CLEAN algorithms in CASA. Besides requiring some initial assumptions on the clean components model (e.g. multiscale cleaning), the image reconstruction requires a deconvolution, which is a nonlinear operation that can introduce similar nonlinear alteration on the flux distribution. In order to mitigate the uncertainties of aperture synthesis and bypass the deconvolution operation, we additionally analyze the observations in the visibility (or Fourier) plane. 
It has been recently showed that the resolution of interferometric datasets can be ``pushed'' beyond the limit obtained with synthesized imaging, by analysing the data in the visibility space \citep[see e.g.][]{2018MNRAS.476.4527T,2021arXiv210302392J}. 
By fitting the visibilities of the DSHARP observations at 1.3~mm  with an axisymmetric model, \citet{isella2018} characterized the continuum emission in HD~163296 with 5 gaussian rings and highlighted two minor asymmetric features: one at about 4~au and a crescent in the south-east direction at about 55~au separation. This latter is partially visible also in our images at 0.9, 2 and 3~mm displayed in Figure \ref{fig:contmaps}.

\begin{table}
\caption{Positions of the peaks in the brightness profile for each wavelength.}
\label{tab:frings}
	\centering
	\begin{tabular*}{\columnwidth}{@{\extracolsep{\stretch{1}}}*{1}{l}@{\extracolsep{\stretch{1}}}*{6}{c}}
		\hline \hline
	$\lambda$ & R1	 & R2 & R3 & R4 & R5 & R6   \\
		\vspace{3pt}
	[mm] & [au] & [au] & [au] & [au] & [au] & [au]  \\
0.9 & 4.3 & 14.4 &-- &-- & 65.9 & 99.0 \\  
1.3 & 4.4 & 14.7 & 22.9 &-- & 66.5 & 99.3  \\  
2.1 & 3.1 & 14.8 & 24.3 & 32.3 & 66.0 & 99.7 \\   
3.0 & -- & 14.8 & 24.7 &-- & 66.6 & 100.0 \\  
9.1 & -- & 12.3 & 25.1 &-- & -- & 100.2 \\  
30 & -- & -- &-- &-- & 62.7 & 93.2 \\ 
		\hline
	\end{tabular*}
\end{table}
We perform a similar analysis in this work, but we use the python package \texttt{frank} \citep{2020MNRAS.495.3209J}, that allows the recovery of brightness profiles without assuming any functional form for the intensity as a function of the radius. The main assumption is always that the brightness is axisymmetric, therefore only the real part of the visibilities is modeled and the Fourier transform can be simplified as an Hankel transform. We refer to \citet{2020MNRAS.495.3209J} for an exhaustive description of the metodology and convergence criteria of this tool, and to Appendix \ref{app:vis} for the details of our modeling. 
The extracted brightness profile for each dataset is displayed in Figure \ref{fig:bff}. 

The VLA datasets have been corrected for the strong central emission likely due to free electrons (see Sect.~\ref{sec:ff}), to avoid artificial oscillations in the brightness profiles (see Appendix \ref{app:vis}). 

In Table \ref{tab:frings} we show the position of the peaks in the radial brightness profiles at each wavelength ranging from 0 to 1\farcs25, as found with the \texttt{python} routine \texttt{scipy.signal.find\_peaks}. 
{The most pronounced ringed-structure is shown by the ALMA Band~4 dataset (2.1\,mm) where we can clearly identify 6 pronounced peaks, i.e. one more than what found by previous studies \citep{isella2018}. This likely results from this datasets having the best combination of angular resolution and intrinsic width of the peaked emission. In fact, the DSHARP Band\,6 dataset that has the best resolution shows less pronounced peaks in the inner disk with respect to Band\,4. This can be either an optical depth effect, or could reflect the different radial distribution of smaller dust particles.} 
We use the radii identified for Band~4 as reference for the ring positions in the analysis carried out in this paper. 

Worth noticing is also the difference in the relative peak intensity of the outer rings: while the intensity at the 67~au ring is much higher than the one at the 100~au for the shortest wavelengths, this difference tends to disappear as the wavelength increases. In the next section, we show that this is due to the higher optical depth in the outer ring, that is also increasing with frequency. 

{For the purpose of the multiwavelength analysis described in Sect.~\ref{sec:mwle}, we need to compare the emission at the same resolution at all wavelengths. To make sure this is verified, 
we perform a second round of fits with \texttt{frank}, where we truncate the visibility distribution for the ALMA tables at shorter spatial frequencies, in order to obtain the same accuracy B$_{80}$ = (2100 $\pm$ 100)~k$\lambda$ from 0.9\,mm up to 9\,mm (see Appendix \ref{app:vis} for the details). The 30\,mm dataset has a considerably lower resolution (see Table \ref{tab:frankfits}), and since degrading the spatial resolution at the shorter wavelengths to match the $\sim$0\farcs3 of 30\,mm would result in a consistent loss of information, we do not include the 30\,mm observation in the multiwavelength analysis described in the next section. }

\subsection{Spectral analysis setup}
\label{sec:mwle}
Resolving the disk structure at multiple wavelengths can help us constraining the dust properties as a function of the radius. 
We use the brightness profiles extracted with \texttt{frank} at a matched resolution (see Sect.~\ref{sec:vis}) 
to fit the Spectral energy distribution with three different analytic models for the dust emission. 
We start by using a simple parametric model where we describe the optical depth as a power law in function of the frequency. We then employ a physical model that includes only the absorption opacity for the dust, and finally we introduce the contribution from scattering. 

In the parametric model, we assume that the intensity emitted from dust is regulated by an opacity with a power-law dependency from the frequency. Under the assumption of LTE (Local Thermodynamic Equilibrium), the dust thermal emission can be written as 
\begin{equation}
\label{eq:pl}
I_{\nu}(r) = B_{\nu}(T(r)) \cdot (1 - e^{-\tau_{\nu}(r)/\mu}),
\end{equation}
with the optical depth given by $\tau_{\nu} = \tau_0 (\nu/\nu_0)^{\beta}$, and the inclination parameter $\mu$=cos($i$). 
At a given radius, the intensity in eq. \ref{eq:pl} depends on three 
parameters: the midplane temperature T, the optical depth $\tau_0$ (at a reference frequency $\nu_0$), and the spectral index $\beta$.  

We use the python package \texttt{UltraNest} \citep{2021arXiv210109604B} a Monte Carlo Nested Sampling tool based on the MLFriend method \citep{2014arXiv1407.5459B}, that computes both the posterior probability of the three free parameters and the marginal likelihood of the model.  
The convergence and termination criteria relative to this tool are described in detail in \citet{2021arXiv210109604B}, and references therein. 
We fit eq.\ref{eq:pl} independently at each separation, i.e. without assuming any trend as a function of radius, with bins of 2~au starting from  0 up to 120~au. We use a flat prior for all parameters, corresponding to log$_{10} \tau_0$: [-4,3], $\beta$: [0,5] and T: [3, T$_\mathrm{up}$], where T$_\mathrm{up}$ is an upper limit variable with radius (see Appendix \ref{app:sedfit} for the details). 
The likelihood function in this Bayesian framework is defined at each radius as the sum of the $\chi^2$ values at the different wavelengths, calculated weighting the squared residuals by 1/$\sigma^2$ as $\chi^2_r = \sum_{\nu_i} (F_{obs,r,\nu_i}-F_{mod, r, \nu_i})^2/ \sigma_{r, \nu_i}^2$, where $\sigma$ is the error on the fluxes given by the 
statistical errors estimated in Sect.~\ref{sec:vis}. 
As the measurements are affected by a systematic error on the order of 5--10\% related to the flux calibration, we perform 30 independent fits where we scale the flux densities at all radii and at each wavelength by a random offset, generated from a normal distribution with standard deviation corresponding to the flux calibration uncertainty (10\% for ALMA Band 7, Band~4 and VLA Ka band, and 5\% for ALMA Band~3 and Band~6).
The resulting posterior probabilities at each radius are obtained by merging the posteriors of the 30 fits.  
\begin{figure}
  \includegraphics[keepaspectratio=True, width = 0.5\textwidth]{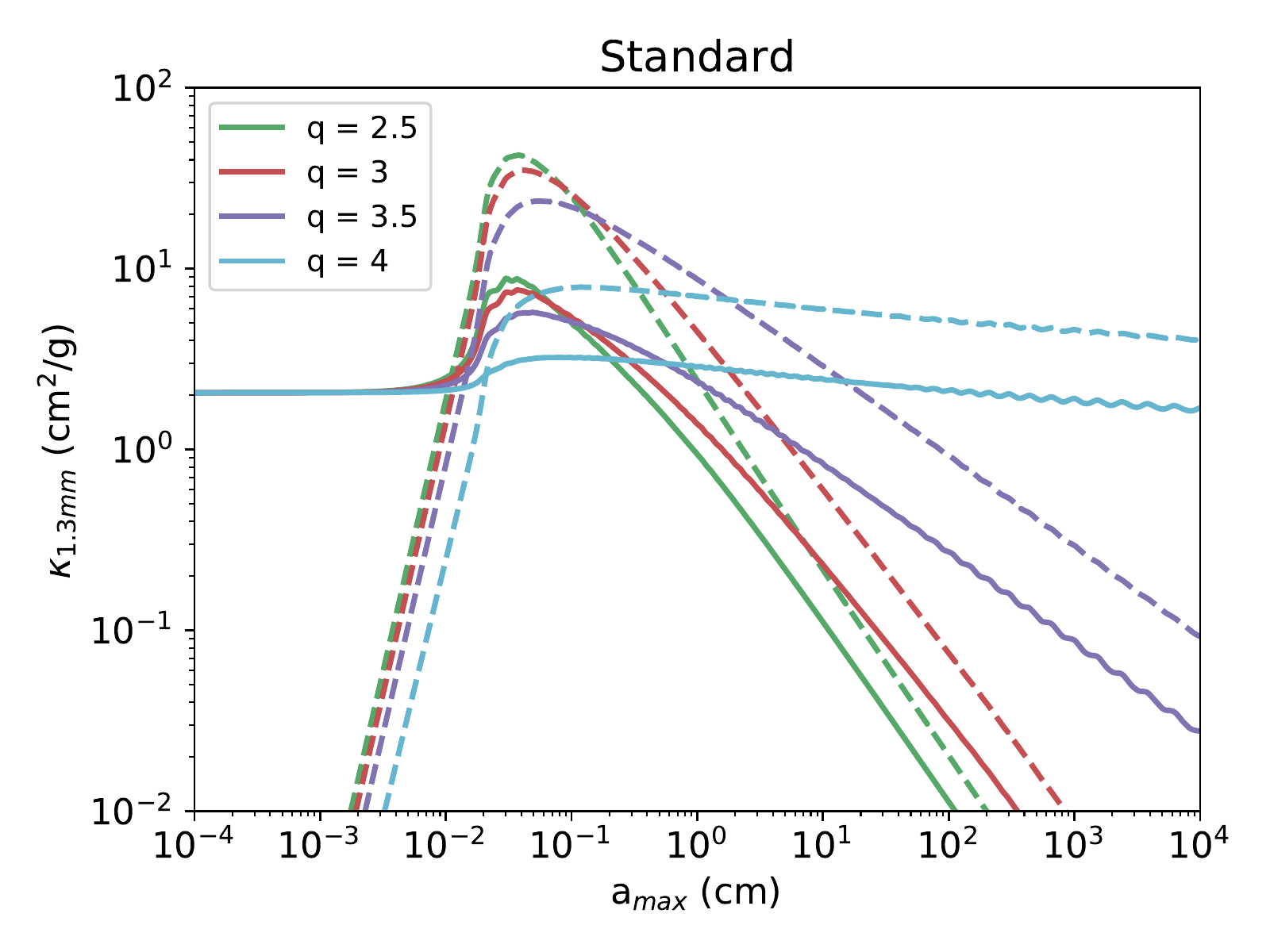}
 \includegraphics[keepaspectratio=True, width = 0.5\textwidth]{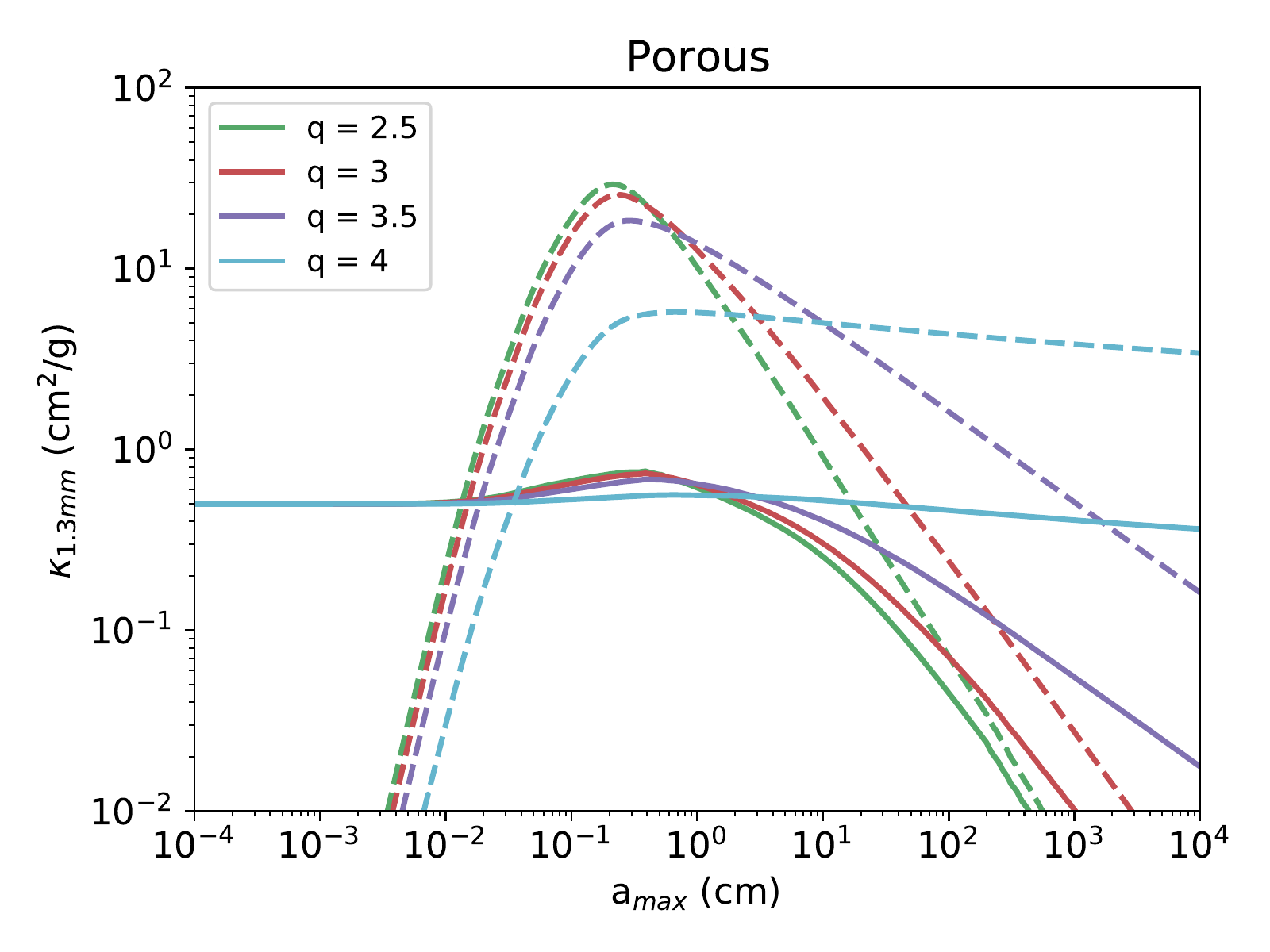}
 \caption{Dust opacity at 1.3\,mm as a function of the maximum grain size. \textit{Upper panel}: absorption (solid lines) and scattering (dashed lines) opacities for the standard dust composition and different size distribution $q$. \textit{Lower panel}: same as the upper panel but for grains with a porosity of 80\%.}
 \label{fig:kappa}
\end{figure}
\\
\\
\indent
To obtain dust physical properties such as the surface density and grain size, we need to employ a physical model that relates these quantities to the observed intensity. 
A necessary step in this direction involves the computation of the dust opacity as a function of the grain size. 

The main caveat is that the exact constituents of the dust grains are not known, and different properties (especially composition and porosity) determine dramatic differences in terms of opacity and albedo of the dust \citep[e.g.][]{2016A&A...585A..13M}.
An overview of the effects of different compositions assumed for grains in protoplanetary disks on their optical properties is given in \citet{2018ApJ...869L..45B}. The result is that each analysis can lead to very different conclusions in terms of the quantities of interest (opacity, mass, grain size), depending on the initial assumption of dust composition/porosity. 
Furthermore, the dust in protoplanetary disks does not consist of a single-size population, but rather an ensemble of grains of different sizes. This is usually described with a power-law distribution $d n(a)/da \propto a^{-q}$, with $a$ representing the particle size and $n$ the number of particles with a size $a$, between a minimum size $a_{min}$ and a maximum $a_{max}$. The opacity of such an ensemble is mostly sensitive to the maximum grain size and it will depend on the size distribution power-law $q$. This latter is estimated by theoretical and experimental studies of collisional and coagulation processes in dust grains \citep{Testi2014}, resulting in typical values of $q \sim$ 2 -- 4. Often a value of $q$ = 3.5 is assumed, following studies characterizing the interstellar dust \citep[see e.g.][and references therein]{2006ApJ...636.1114D, Testi2014}.  Ultimately, the value of $q$ for protoplanetary disks is not known as it is expected to vary with time evolution \citep[see e.g.][]{Testi2014} and location within the disk.

With these uncertainties in mind, we consider a set of different grain populations for a total of 8 initial models. We compute a SED fit for each model and we compare the evidence to assess which one is more representative of the dust population at each radius. 
We compute the opacities from the DIANA project \citep{2016A&A...585A..13M}, through the DIANA OpacityTool Fortran package. The authors define a standard grain composition for protoplanetary disks as a mixture of amorphous silicates \citep{1995A&A...300..503D} and amorphous carbonaceous materials \citep{1996MNRAS.282.1321Z} in a volume fraction of 60\% and 15\%, respectively, with a remaining 25\% of vacuum. In the opacity calculations, grains are modeled as distributions of hollow spheres \citep{2005A&A...432..909M}, overcoming the assumption of spherical grains used in the Mie scattering theory \citep{1908AnP...330..377M}. 
We generate a series of opacities for a range of wavelengths covering our observations, varying the maximum grain size from 10$^{-4}$ to 10$^4$\,cm and keeping a minimum size of 0.05\,$\mu$m. This procedure is repeated for 4 different slopes $q$ of the size distribution, with values of 2.5, 3, 3.5, 4. 
The absorption and scattering opacities as function of the maximum grain size at 1.3\,mm are plotted in Figure \ref{fig:kappa}, upper panel, and labelled as ``standard''. We compute a further set of opacities enhancing the grain porosity to 80\%, while keeping the same relative fraction of silicate and carbons as in the standard composition (Figure \ref{fig:kappa}, lower panel labelled ``porous''). 
This results in 8 different dust populations: 2 compositions (compact and porous) with 4 size distribution each. 

{The optical depth in equation \ref{eq:pl} can then be written explicitly as function of the opacity and surface density: 
\begin{equation}
\label{eq:noscat}
    I_{\nu}(r) = B_{\nu}(T_{mid}(r)) [1 - exp(-\kappa_{\nu, abs}(a_\mathrm{max}(r))\cdot \Sigma_d(r)/\mu)]
\end{equation}

where $\kappa_{\mathrm{abs}}$ is the absorption opacity and $\Sigma_d$ the dust surface density. For a given dust composition and size distribution with spectrum $q$, we have that $\kappa_{\nu,\mathrm{abs}}$ depends only on the maximum grain size a$_{\mathrm{max}}$ (see Figure \ref{fig:kappa}) . At a given radius $r$ we still assume that the temperature is the same for the emission at all wavelengths. 
As for the parametric fit, we use the Monte Carlo nested sampling algorithm with the UltraNest software to fit eq.\ref{eq:noscat} independently at radial bins of 2~au from the star, and using a flat prior of [3, T$_\mathrm{up}$] for T, [-4,3] for log$_{10}$ a$_\mathrm{max}$ and [0.0001,10] for $\Sigma_d$.} 
\\
Finally, to include the possible effects of dust self-scattering, we use the analytic expression for the emergent intensity given in  \citet{2019ApJ...877L..18Z}, that is valid in the assumption of isotropic scattering from an isothermal slab:

\begin{multline}
\label{eq:scattering}
 I (r)= B_{\nu}(T(r)) \cdot(1.-exp(-\tau_d(r) /\mu)) \\
 \resizebox{0.45\textwidth}{!}{$\left( 1 - \omega_{\nu}(r) \frac{exp(-\sqrt{3 (1 - \omega_{\nu}(r))} \tau_d(r)) + exp(\sqrt{3 (1 - \omega_{\nu}(r))} (\tau(r) - \tau_d(r)))}{exp(-\sqrt{3(1-\omega_{\nu}(r))}\tau_d(r))\cdot (1 - \sqrt{1 - \omega_{\nu}(r)}) + (\sqrt{1 - \omega_{\nu}(r)} + 1) }   \right) $.}
 \end{multline}

The deprojected optical depth in this description is calculated as $\tau = 2 \mu \tau_d/ (3 \tau_d + 1)$, from the total optical depth $\tau_d$ that includes both absorption and scattering, given by $\tau_d = \Sigma_d (\kappa_{\mathrm{abs}}+ \kappa_{\mathrm{sca, eff}})$  \citep{2019ApJ...877L..18Z}. Here $\Sigma_d$ represents the dust surface density, $\kappa_{\mathrm{abs}}$ is the absorption opacity. and the effective scattering opacity is obtained with a correction by the asymmetry parameter $g$ as $\kappa_{\mathrm{sca, eff}} = (1 - g) \kappa_{\mathrm{sca}}$. 
Finally, the albedo $\omega$ is the ratio between the scattering and the total opacity $\kappa_{\mathrm{sca, eff}}/(\kappa_{\mathrm{abs}}+\kappa_{\mathrm{sca, eff}})$. 

We can observe that for a given dust composition and size distribution with spectrum $q$, we have that $\kappa_{\mathrm{abs}}$, $\kappa_{\mathrm{sca}}$ and $g$ at a certain wavelength depend only on the maximum grain size a$_{\mathrm{max}}$. 
Therefore, as in the non-scattering case described by eq. \ref{eq:noscat}, equation \ref{eq:scattering} depends only on T(r), a$_{\mathrm{max}}$(r) and $\Sigma_d$(r). 
We recall here that the underlying assumption is that at a given radius $r$ the temperature at the emitting layer is the same at all wavelengths. 
Following the same procedure as for the first two models, we find the best-fit parameters and relative uncertainties with \texttt{ultranest}, and we perform 30 additional fits at each radius to account for the systematic calibration offset.  

\subsection{Model comparison with Bayes factor K}
\label{sec:bayesK}
{We take advantage of our Bayesian framework to compare the performances of the different analytical models we use to fit our observations of HD~163296. The Monte Carlo nested sampling routine we apply provides not only the posterior probabilities but also the Bayesian evidence (or marginal likelihood) Z. This corresponds to the normalization factor in the Bayes equation, or the integral over the whole parameter space of the likelihood times the prior density $Z = \int L(D\vert \theta) \pi(\theta) d\theta $. }
{
It follows that we can compute the Bayesian factor K between different models, defined as the radio of the marginal likelihoods, and assess which model is more compatible with our datasets. 
To interpret the Bayes factors, we refer to the scale proposed originally by \citet{1939thpr.book.....J}, that associates different ranges of the K factor to a strength of a change in evidence. Defining K$_{12}$ as Z$_1$/Z$_2$, where Z$_1$ and Z$_2$ are the marginal likelihood of model 1 and 2, respectively, we use the following scale:}
\begin{table}[h!]
    \centering
    \begin{tabular}{r r r}
      K$_{12}$   & >100 & Decisive evidence for model 1 \\
         & 30 -- 100 & Very strong evidence for model 1\\
          & 10 -- 30 & Strong evidence for model 1\\
       & 3 -- 10 & Moderate evidence for model 1\\
      & 1 -- 3 & Anecdotal evidence for model 1\\ 
       & = 1 & No change in evidence\\ 
    \end{tabular}
\end{table}

\section{Results}
\label{sec:res}
\subsection{Spectral index}
\label{sec:spindex}
\begin{figure*}
 \centering
 \includegraphics[keepaspectratio=True, width = \textwidth]{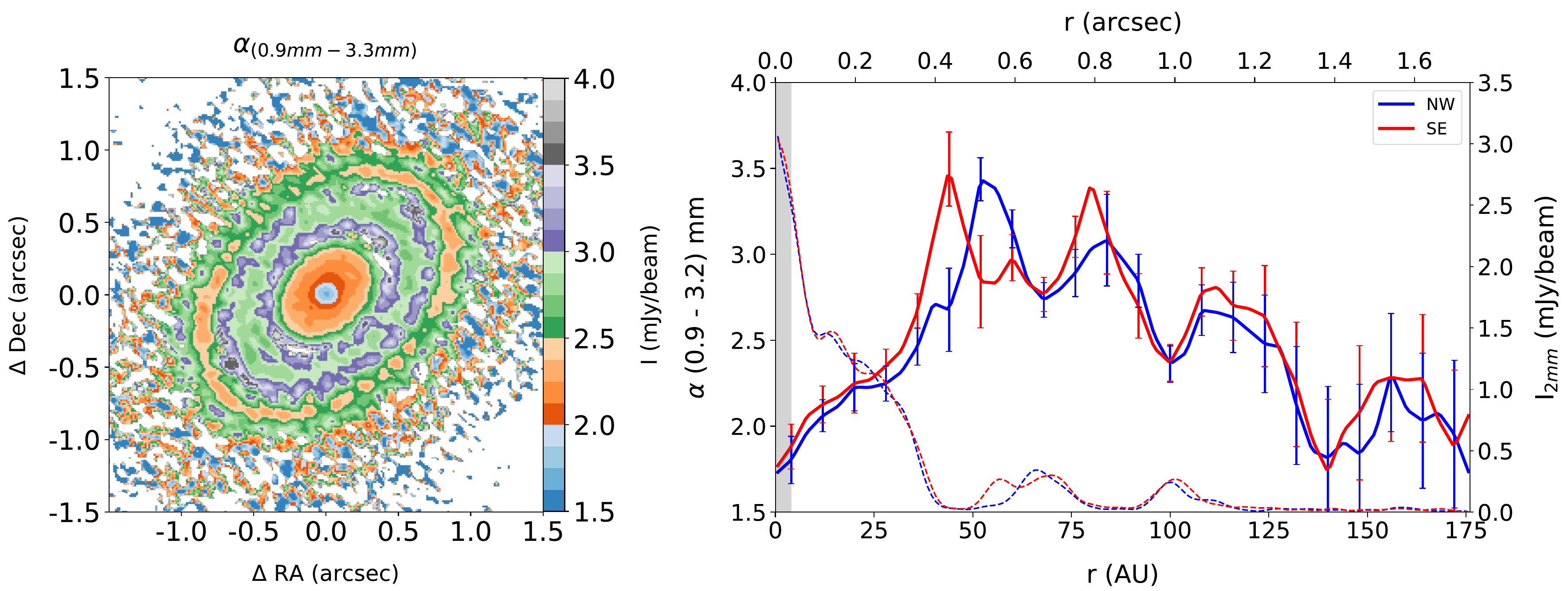}
 \caption{Flux density spectral index. \textit{Left:} Map of the spectral index computed from the ALMA continuum images at 0.9\,mm, 1.3\,mm, 2.1\,mm, 2.8\,mm and 3.2\,mm with a matching beam of 0\farcs095x0\farcs065 and position angle of 61.24\degree. 
 \textit{Right:} Radial profile of the spectral index 
 in the northwest and southeast sides (blue and red curve, respectively), error bars 
 are shown for each bin and calculated as described in the text.  
 Overplotted with dashed lines is the corresponding flux density profiles in Band~4 (right $y$ axis). The grey shaded region on the left denotes the angular resolution of the map as half of the average beam FWHM. }
 \label{fig:alp}
\end{figure*}
The measure of the flux spectral index of the dust emission at millimeter wavelengths has been the primary tool for deriving 
information on the grain properties, through its relation with the dust opacity spectral index $\beta$ (where $\kappa_{\nu} \propto \nu^{\beta}$). 
Multiple surveys of protoplanetary disks allowed in the past to estimate integrated values of $\beta$ - in the Rayleigh-Jeans and optically thin assumptions - that resulted systematically lower than the values measured for the insterstellar medium $\beta_\mathrm{ISM} \simeq 1.7$ \citep[e.g.][]{1990AJ.....99..924B,2003A&A...403..323T,natta04,2010A&A...512A..15R}. This trend was confirmed by more recent ALMA surveys targeting disks in nearby star forming regions \citep[e.g.][]{2018ApJ...859...21A,2021MNRAS.506.5117T}. 

A common explanation for these measurements (typically $\beta <$1) was found in the growth of solids, since other mechanisms such as chemical composition of the grains and 
porosity, are expected to have only moderate effects on the total opacity, within certain limits of the grain size distribution. Specifically, when such distribution follows a power law of the form $dn/da \propto a^{-q}$, 
a size distribution with $q = 3.5$ will have $\beta \lesssim$ 1 for $a_{max} \gtrsim 3 \lambda$ \citep{2006ApJ...636.1114D}. 

In the ALMA era, more accurate measurement of the spectral index are possible: by spatially resolving the continuum emission from the disk, the radial variation of the spectral index can be computed. Studies employing medium-resolutions ALMA observations showed for different disks a monotonically increasing spectral index from small to larger radii: this was interpreted as a signature of larger grains in the inner regions, as expected from the differential action of radial drift \citep[e.g.][]{2012ApJ...760L..17P,2015ApJ...813...41P,Tazzari2015,Guidi2016}. 
More recently and thanks to higher spatial resolution observations ($\leq$20~au) it has been possible to measure the spectral index variations on smaller scales and show that it deviates from a pure monotonic behavior in several disks, such as HLTau \citep{2017A&A...607A..74L,2019ApJ...883...71C}, TWHya \citep{2021A&A...648A..33M}, HD~169142 \citep{2019ApJ...881..159M}, GMAur \citep{2020ApJ...891...48H} HD~163296 \citep{2019MNRAS.482L..29D,2019ApJ...886..103O} and a few disks in the Taurus association \citep{2020ApJ...898...36L}. These improved measurements are showing that a low $\alpha$ in the millimeter range does not necessarily coincide with the presence of large (millimeter) grains, but it is in some cases artificially lowered by the high optical depth of the continuum emission. 

With the extended and improved set of ALMA observations of HD~163296 we present in this work, we can build a detailed map of the spectral index between 0.9\,mm and 3.2\,mm and measure the spectral index with higher resolution and smaller uncertainties compared to previous works.  

After producing images with a matching beam of $\sim$0\farcs08 and centered on the same pixels, corresponding to the lowest resolution available (in this case the 3\,mm observations), we can compute the spectral index $\alpha$ of the flux (assuming $F_{\nu} \propto \nu^{\alpha}$), with the least-squares method for each pixel. 
The resulting $\alpha$ map and corresponding radial profile are shown in Figure \ref{fig:alp}: 
the profiles are computed inside segments within a 45\degree (deprojected) angle centered on the disk major axis (PA = 133\degree) with bins taken every half-beam (4~au) and treating the NE and SW side separately. For each bin we computed the weighted mean of the values, with weights given by the inverse squared error of each pixel $\sigma_{\alpha}$, derived analytically from the linear least-squares regression. The uncertainties on the flux densities at each pixel were taken as $\sigma_{F} = \sqrt{rms^2 + (\Delta F_\mathrm{cal})^2}$, i.e. the sum in quadrature of the rms of the image and the flux calibration error, corresponding to 5\% or 10\% depending on the wavelength (see Sect.~\ref{sec:obserr}). 
The error on each bin is then computed as the standard error of the weighted mean, accounting for the fact that the pixels are not independent. 

The spectral index map appears overall azimuthally symmetrical and the profiles do not show strong deviations between the south-east and north-west sides; some  appreciable differences are observed inside the gaps - where we are dominated by the noise - and at the location of the crescent at 55~au on the south-east side (red curve in Figure \ref{fig:alp}): here the spectral index is consistent with what is measured in the adjacent ring at 67~au ($\alpha$=2.8 $\pm$ 0.1). 
Interestingly, the outermost ring at 100~au shows a value of $\alpha$ = 2.4$\pm$ 0.1, lower with respect to the ring at 67~au. {This is in agreement with previous recent measurements of the millimeter spectral index at lower resolution \citep{2019MNRAS.482L..29D,2021ApJS..257...14S}.} 
\begin{figure}[h!]
 \centering
 \includegraphics[keepaspectratio=True,width=9cm]{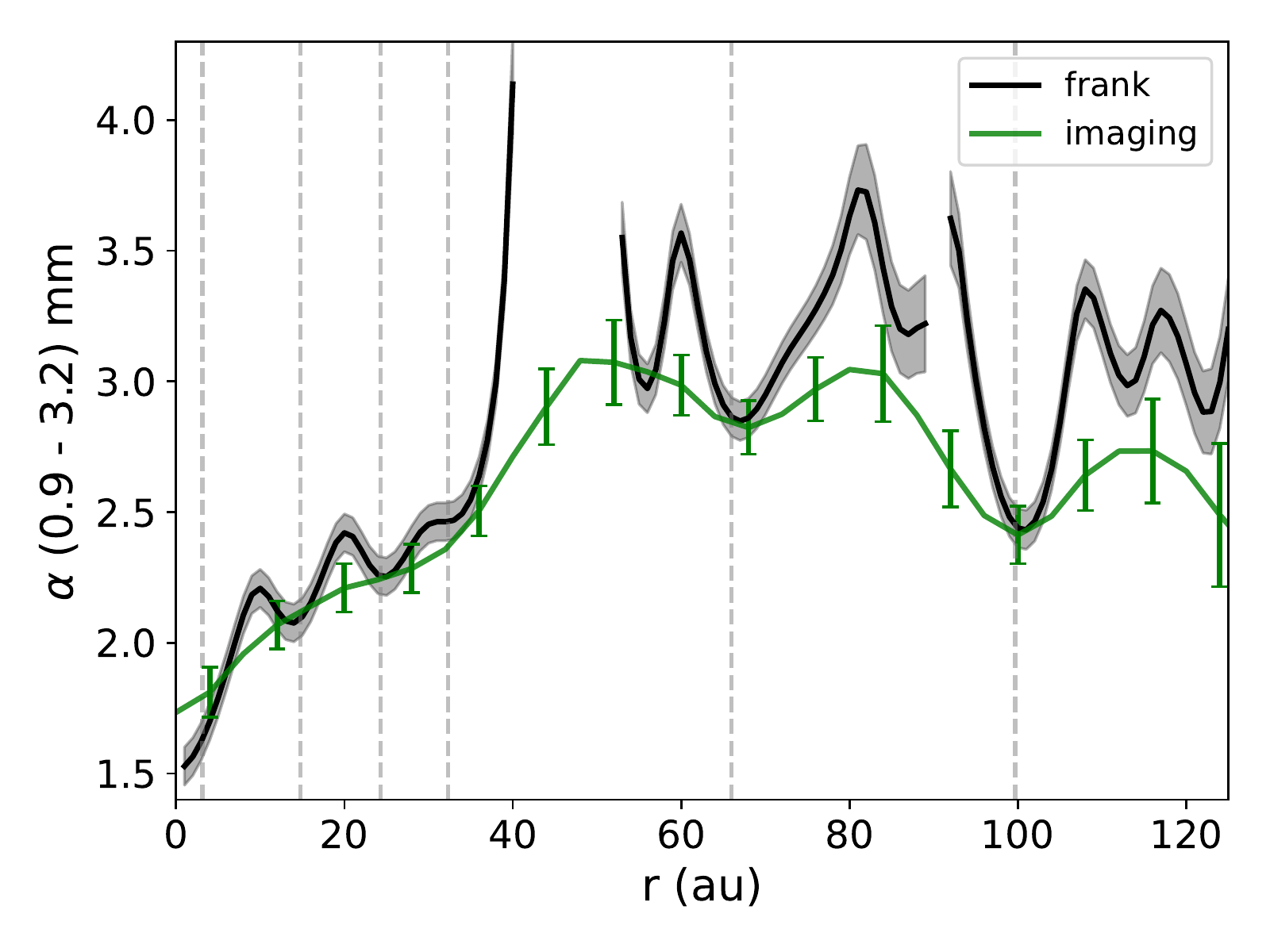}
 \caption{Flux spectral index between 0.9\,mm and 3.2\,mm computed from the radial profiles extracted with \texttt{frank} (\ref{sec:vis}, black curve), with shaded areas representing the errors derived from the linear least-squares regression (with the uncertainties for the single fluxes defined as the sum in quadrature of the statistical and calibration errors). The profile is masked at the locations where the flux S/N < 3 for at least one wavelength. Overplotted is the azimuthal average of the map in Figure \ref{fig:alp} (green curve), calculated with bins of 1 beam ($\sim$ 0\farcs08) and with errorbars computed as for the profiles in Figure \ref{fig:alp}, right panel.}
 \label{fig:alp_frank}
\end{figure}
{We compare in Figure \ref{fig:alp_frank} the spectral index obtained from the images with the one computed from the radial profiles (see Sect.~\ref{sec:vis}). While there is generally a very good agreement between the two profiles, we note how the fluxes extracted with \texttt{frank} allow us to reveal with more detail the variation of the $\alpha$ index in the inner region of the disk (inside $\sim$40\,au).} 

Finally, we note that $\alpha$ reaches values lower than 2 in the innermost regions (r$\lesssim$10~au), i.e. below the black-body limit for an optically thick emission. This has been already observed in HD~163296 using a smaller set of observations \citep{2019MNRAS.482L..29D}, and in other sources \citep[e.g. TWHya][]{2018ApJ...852..122H}. This feature 
can be due to multiple effects, such as the self-scattering reducing the emission at shorter wavelength in the optically thick inner regions, or the free-free emission increasing the contribution of longer wavelength in the central flux (e.g. the contamination calculated in Sect.~\ref{sec:ff} accounts for $\sim$20\% of the flux at 3\,mm in the central beam, and reduces to about 1\% at 0.9\,mm).

\subsection{Parametric model}
\label{sec:parmodel}
\begin{figure}
 \centering
 \includegraphics[keepaspectratio=True, width = 0.5\textwidth]{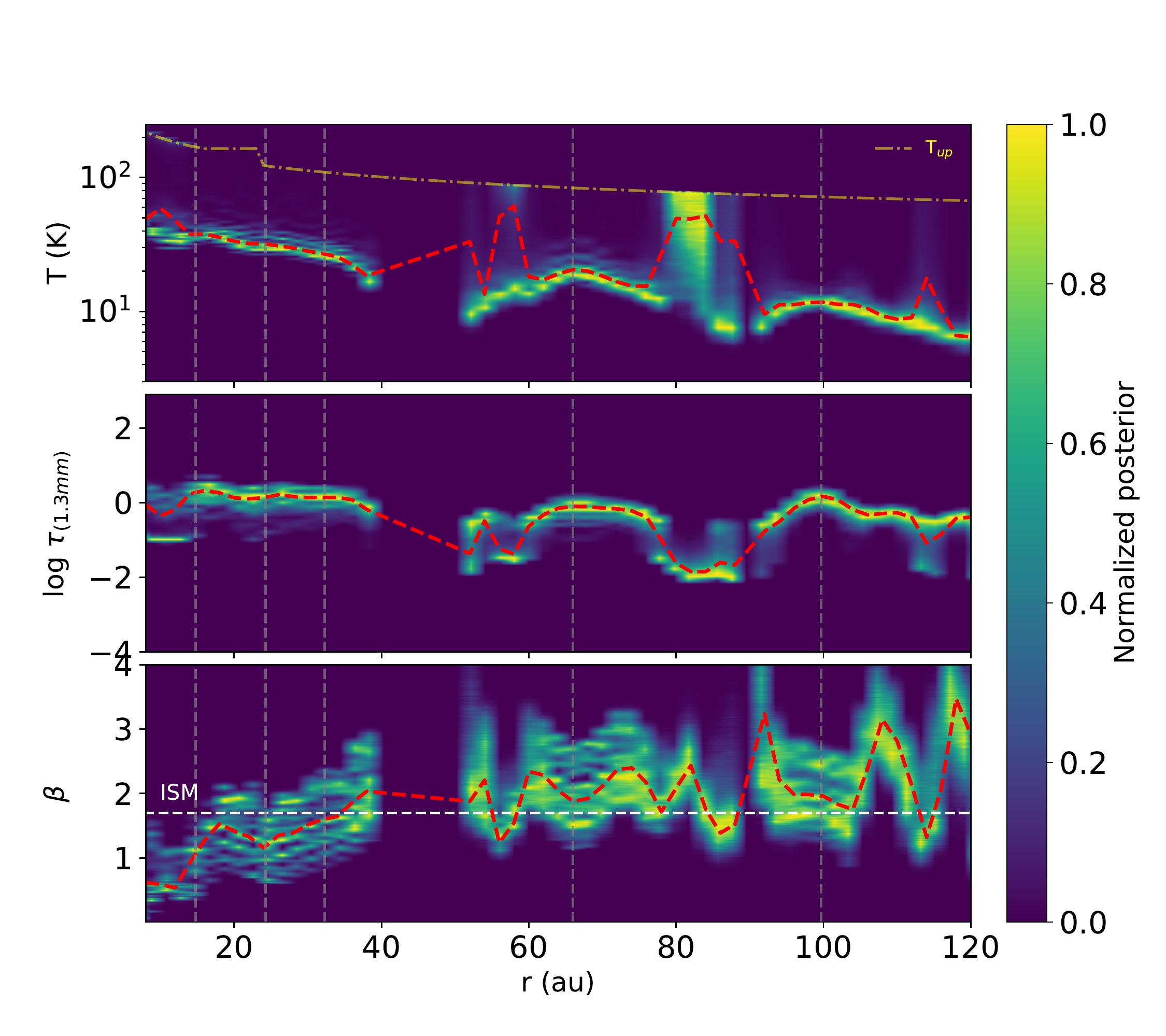}
 \caption{Best fit parameters for the power law model: temperature, optical depth, and opacity spectral index $\beta$ as function of the radius. The dashed red curve represents the estimates obtained as the mean values of the posterior distributions from the Monte Carlo nested sampling fit. The color map is the normalized probability after merging the 30 posteriors obtained introducing a random offset in the fluxes according to their flux calibration accuracy. The dashed-dotted curve in the upper panel shows the upper limit of the Temperature prior for each radius used in the fit (see Appendix \ref{app:sedfit}). The vertical dashed lines correspond to the dust ring found in Sect.~\ref{sec:vis}, while the white dashed line in the bottom panel is drawn for $\beta$=1.7, corresponding to ISM dust grains.}
 \label{fig:fitA}
\end{figure}

We show in Figure \ref{fig:fitA} the best-fit parameters of our parametric model described in Sect.~\ref{sec:mwle} as a function of the radius; the results are plotted starting from 8~au as for lower separations some of the parameters resulted highly unconstrained (see Appendix \ref{app:sedfit}). {The parameters estimates from our nested sampling method that includes the statistical error are shown with a dashed red line, and we overplot the total normalized posterior obtained merging the 30 fits with a random calibration offset, as described in Sect.~\ref{sec:mwle}.}

The results indicate that the 1.3\,mm emission is moderately optically thick: while the average value in the inner disk inside 35\,au is $\tau_{1.3mm} \simeq$ 1.3, we find  
$\tau_{1.3mm} = 0.8^{+0.2}_{-0.2}$ at the 66\,au ring, and $\tau_{1.3mm} = 1.5^{+0.3}_{-0.3}$ in the outer ring at 100\,au, where the uncertainties are given as the 16th and 84th percentile of the total posterior distribution. The temperature is well-constrained across the disk except in the dust gaps, where on the contrary the optical depth is lower and the temperature uncertainties are higher. 
Nevertheless, the temperature profile obtained with this approach deviates from a smooth decreasing power-law (the functional form that is generally assumed for the radial dependence of T when analysing medium-resolution observations), and in particular it shows enhanced values in correspondence of the dust gaps. This temperature increase in the dust gaps of HD~163296 was already pointed out by \citet{2018ApJ...867L..14V} and \citet{2020A&A...642A.165R}, and is typically explained with the higher penetration of the scattered light from the disk surface into the midplane, because of the dust depletion. This is also consistent with the effect of a planet on the disk temperature structure as shown by hydrodynamic simulations \citep[e.g.][]{2018ApJ...860...27I}. 
From our modeling, an increase in temperature is appreciable in the large dust gaps at about 50 and 85 and 115 au, with temperature peaks corresponding to 2.5, 2.7 and 3.7 times the values at the closest inner rings (32, 66 and 100\,au , respectively). A tentative increase in T is also observed in the innermost gap at $\sim$10\,au. 

Beyond 30 au, the $\beta$ index seems consistent or larger than the ISM value, and no difference in $\beta$ is measured between the ring at 66~au ($\beta$=1.9 $\pm 0.3$) and the ring at 100~au ($\beta$=2.0 $\pm 0.3$), i.e. no indication of larger grains at 100~au, as the spectral index shown in Sect.~\ref{sec:spindex} might have suggested. The lower $\alpha$ can be explained by the higher optical depth of the outer ring as mentioned above. 

\subsection{Physical models}
\label{sec:pm}
\begin{figure}
 \centering
 \includegraphics[keepaspectratio=True, width = 0.5\textwidth]{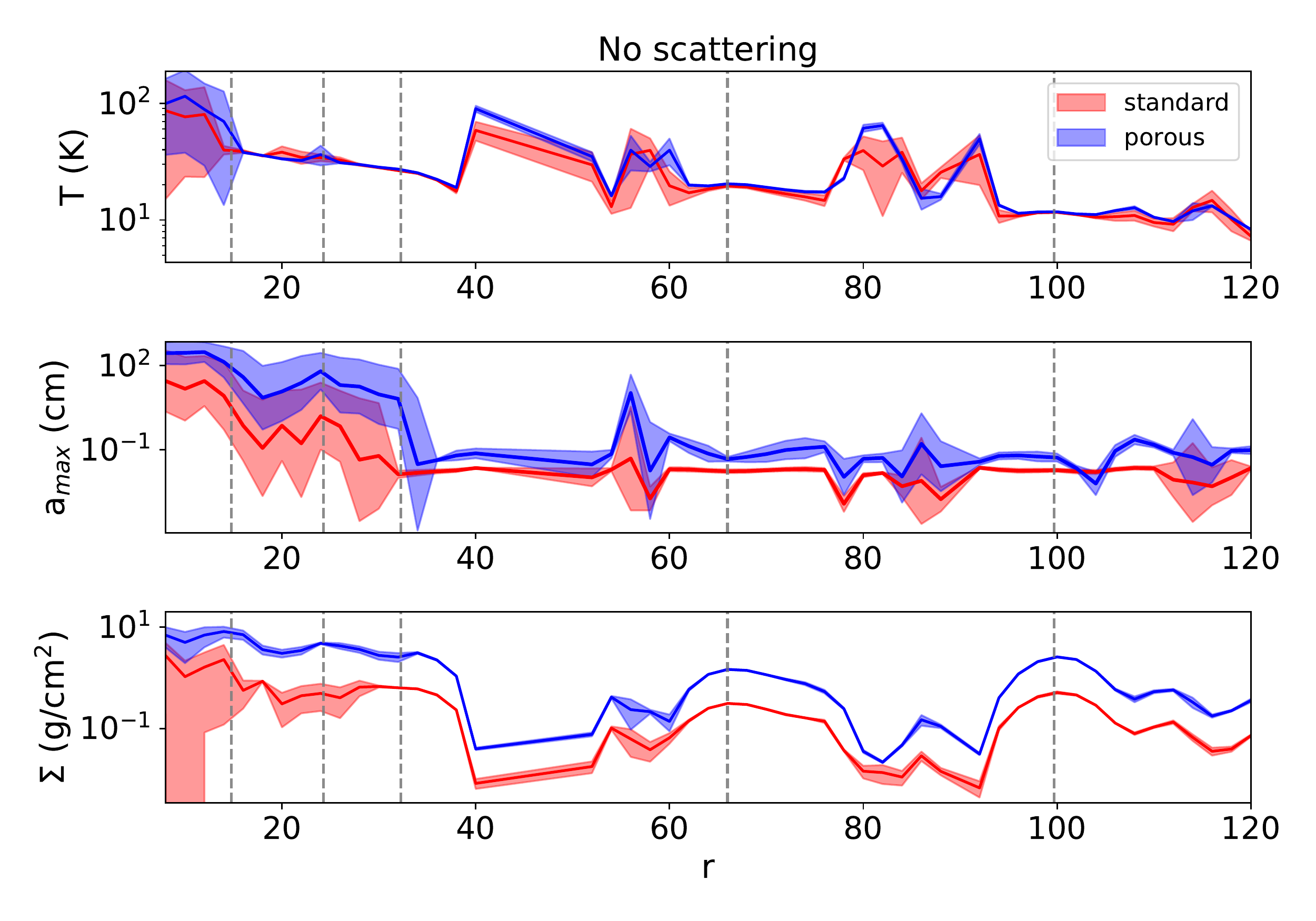}
 \caption{{Dispersion of the best-fit temperature, maximum grain size and surface density in the non-scattering model, calculated as the mean value at each radius of the four different size distributions (the single best-fit models are show in Appendix \ref{app:sedfit}), considering standard and porous grains separately. The shaded regions correspond to the standard deviation of the four best-fit values at each radius.}}
 \label{fig:stan_por_noscat}
\end{figure}
\begin{figure}
 \centering
 \includegraphics[keepaspectratio=True, width = 0.5\textwidth]{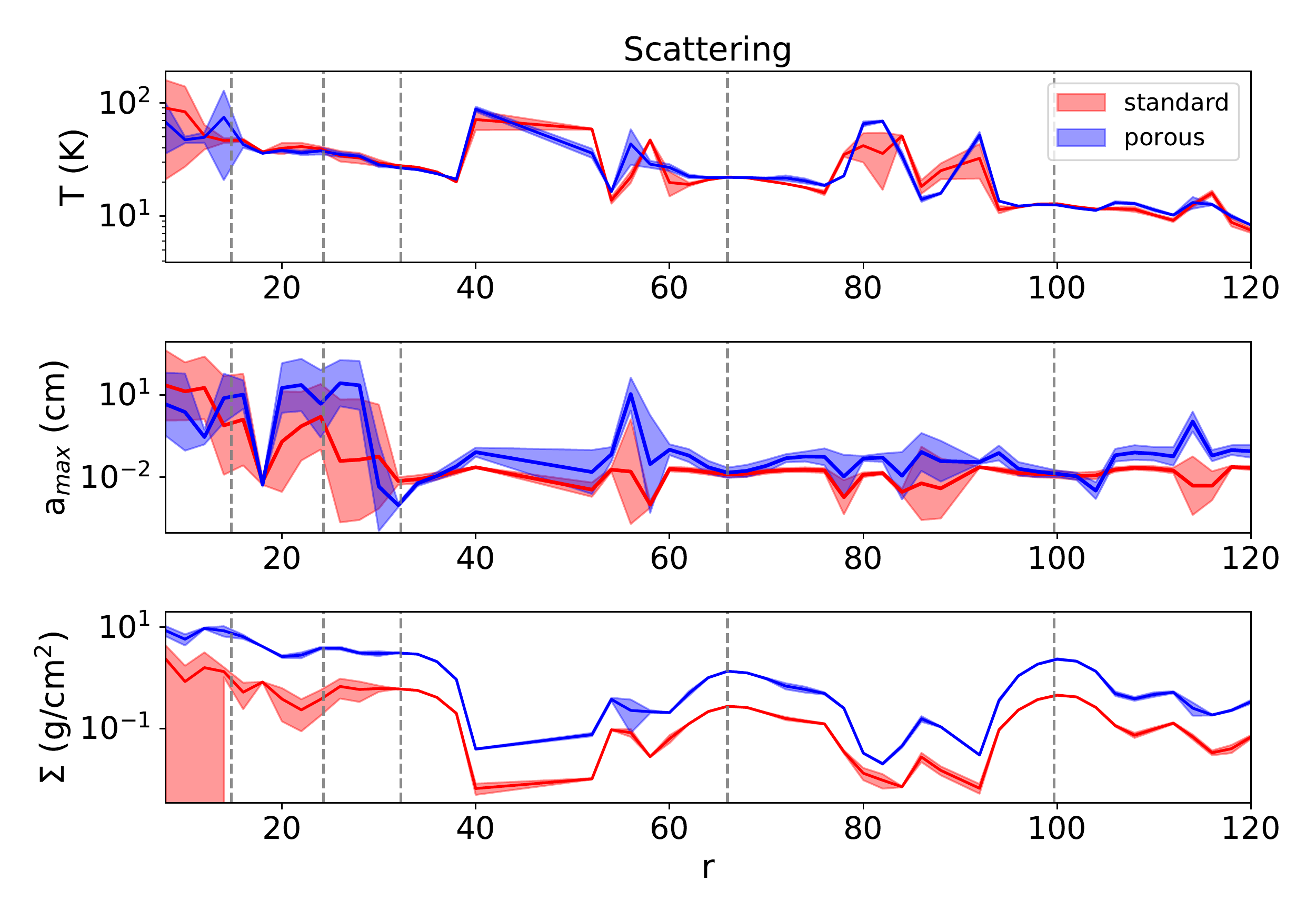}
 \caption{Same as Figure \ref{fig:stan_por_noscat}, but for the model that includes scattering (the single best-fit models are show in Appendix \ref{app:sedfit}).}
 \label{fig:stan_por}
\end{figure}

{We fit equation \ref{eq:noscat} at each radius for all the 8 dust models, and we show in Figure \ref{fig:stan_por_noscat} the best-fit values averaged on the four size distributions ($q$ ranging from 2.5 to 4) for both the standard and porous dust grains. 
In Appendix \ref{app:sedfit} we show the results for all the single models. We observe that while the temperature profile is consistent between the 2 sets of dust, the maximum grain size for porous grains is on average a factor of 3--6 higher (with a ratio increasing with the size distribution spectral index $q$), and the surface density a factor of 5 higher (roughly consistent across the size distributions). 
Similarly, in the model that includes scattering (eq. \ref{eq:scattering}), assuming porous grains we predict a $a_\mathrm{max}$ between 2 and 4 times larger (again increasing with the $q$ spectral index) and a surface density that is on average $\approx 5$ times larger (see Figure \ref{fig:stan_por}).} 

\begin{figure}
 \centering
 \includegraphics[keepaspectratio=True, width = 0.5\textwidth]{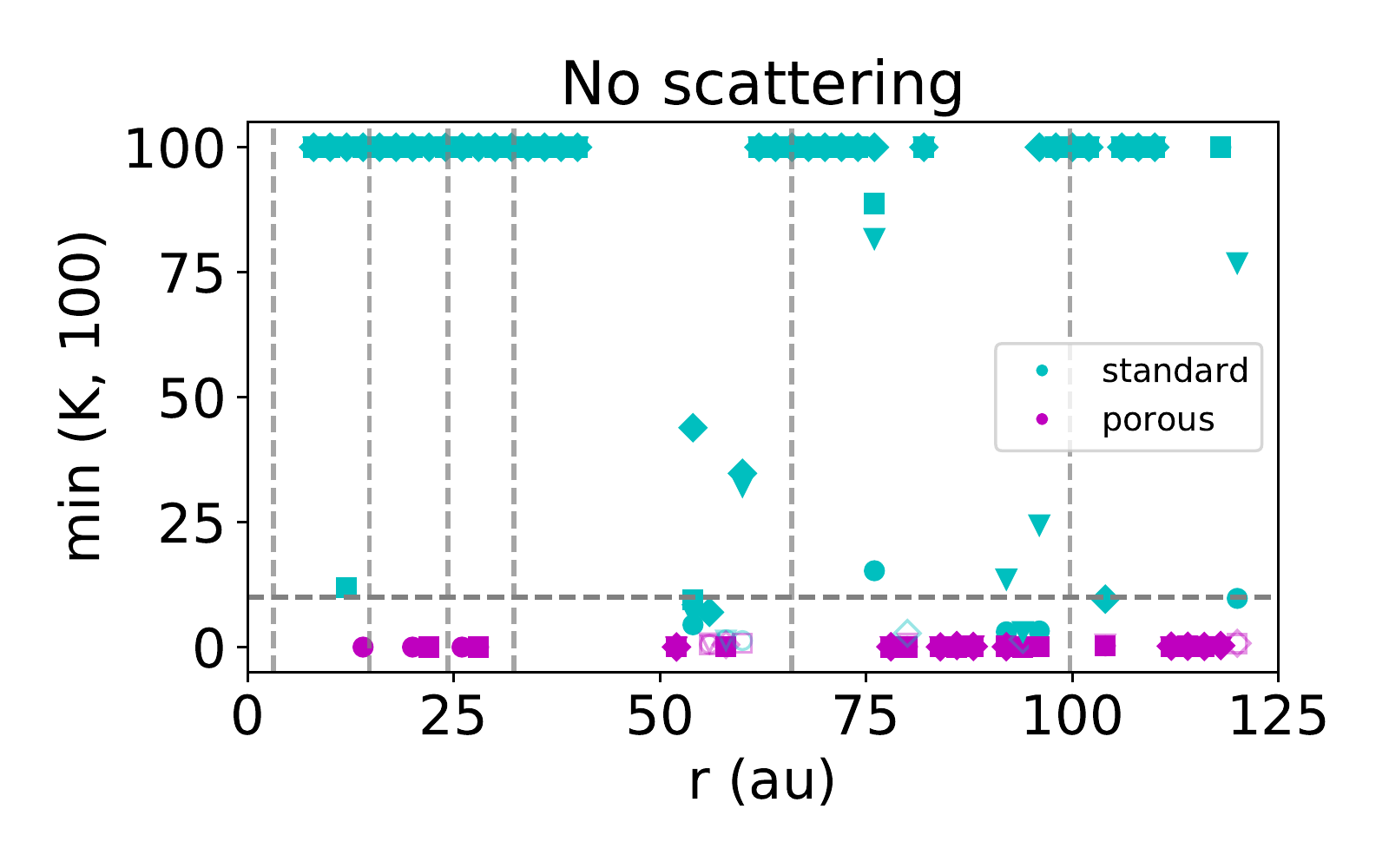}
 \includegraphics[keepaspectratio=True, width = 0.5\textwidth]{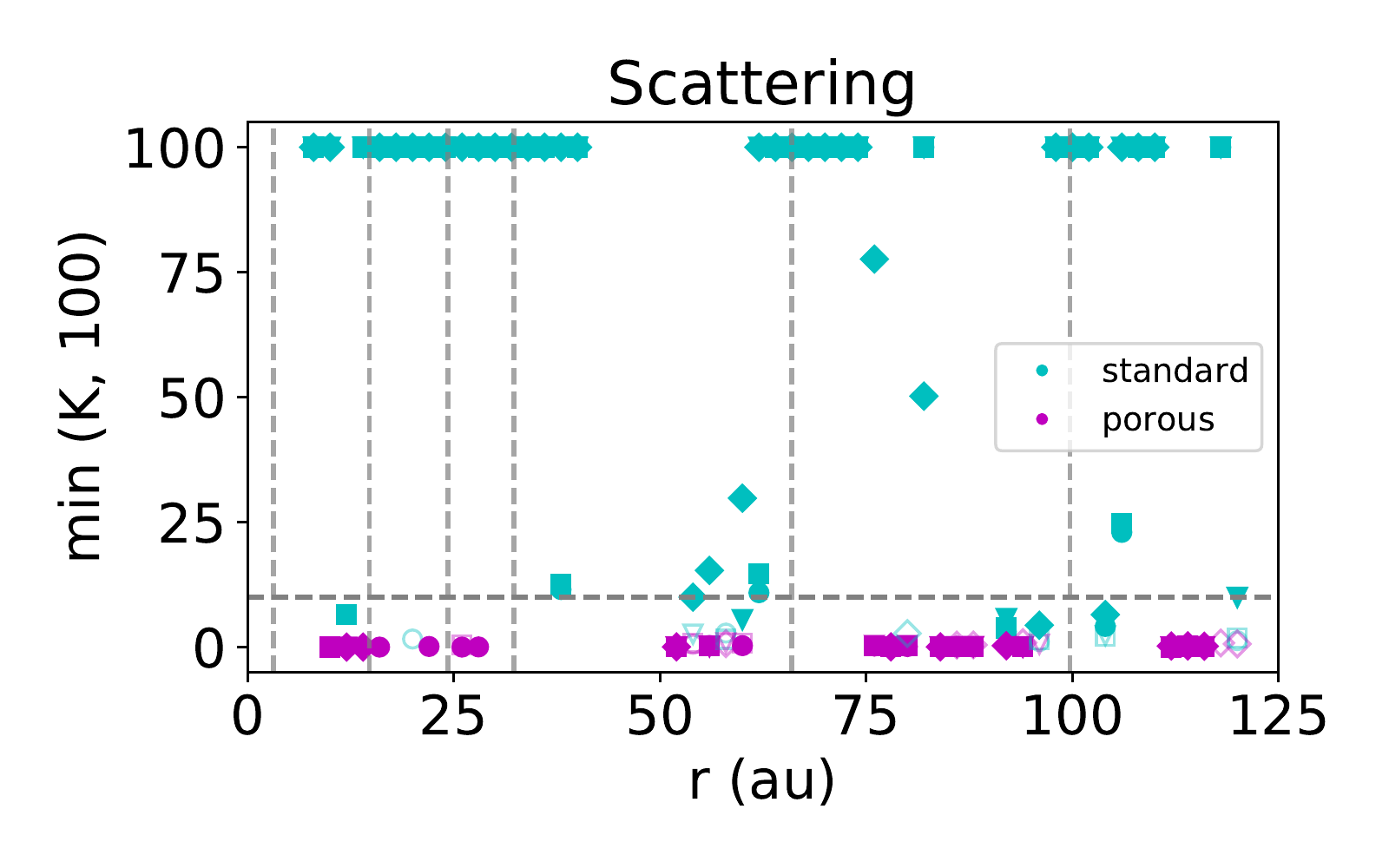}
 \caption{Bayes factor K calculated between the standard and the porous models as Z$_\mathrm{standard}$/Z$_\mathrm{porous}$, for the nonscattering model (upper panel) and scattering model (lower panel). The different markers correspond to the different size distribution coefficients $q$. The points are color-coded according to their values: K>1 (larger evidence for the standard composition) are drawn in cyan and K<1 (larger evidence for porous composition)in magenta. Empty markers correspond to 1<K<3 or 1/3<K<1, full markers to K>3 or K<1/3. The horizontal dashed line is drawn at K = 10, so that all points above this line correspond to a strong evidence in favor of standard grains. The points larger than 100 are drawn at the location of 100, since for K$\geq$100 the interpretation in terms of evidence strength does not change (see Sect.~\ref{sec:bayesK}).}
 \label{fig:K_stan_por}
\end{figure}

{To determine the most probable model between the two different dust compositions (standard and porous), we look at the Bayesian evidence of the Monte Carlo nested sampling fits and identify the models with the highest evidence. We illustrate the results for the non-scattering and scattering case in Figure \ref{fig:K_stan_por}, showing the Bayes factors between the standard and porous models for each value of $q$. Referring to the scale reported in Sect.~\ref{sec:bayesK}, we find that overall standard grains seem to better reproduce our data, with K > 10 (strong evidence for standard grains) in 61\% of the cases (cases corresponding to the number of $q$ values times the number of separations), while for porous grains we have K<1/10 only in 16\% of the cases. Moreover, standard grains seem to have a decisive superior evidence (K>100) at 57\% of the radii, while this goes down to 5\% for porous grains. These values are calculated over all the 4 size distributions, but they present similar values when looking at the single $q$ models.  
In the scattering case we find similar results, with a strong evidence for standard grains 55\% of the times (decisive 50\% of the times), and 15\% for porous grains (decisive for 3\%). 
Interestingly, porous grains result a better model only in correspondence of the gaps in the dust distribution (although for the few radii in the inner disk where the rings are not resolved, this is less clear).} 

\begin{figure}
 \centering
  \includegraphics[keepaspectratio=True, width = 0.5\textwidth]{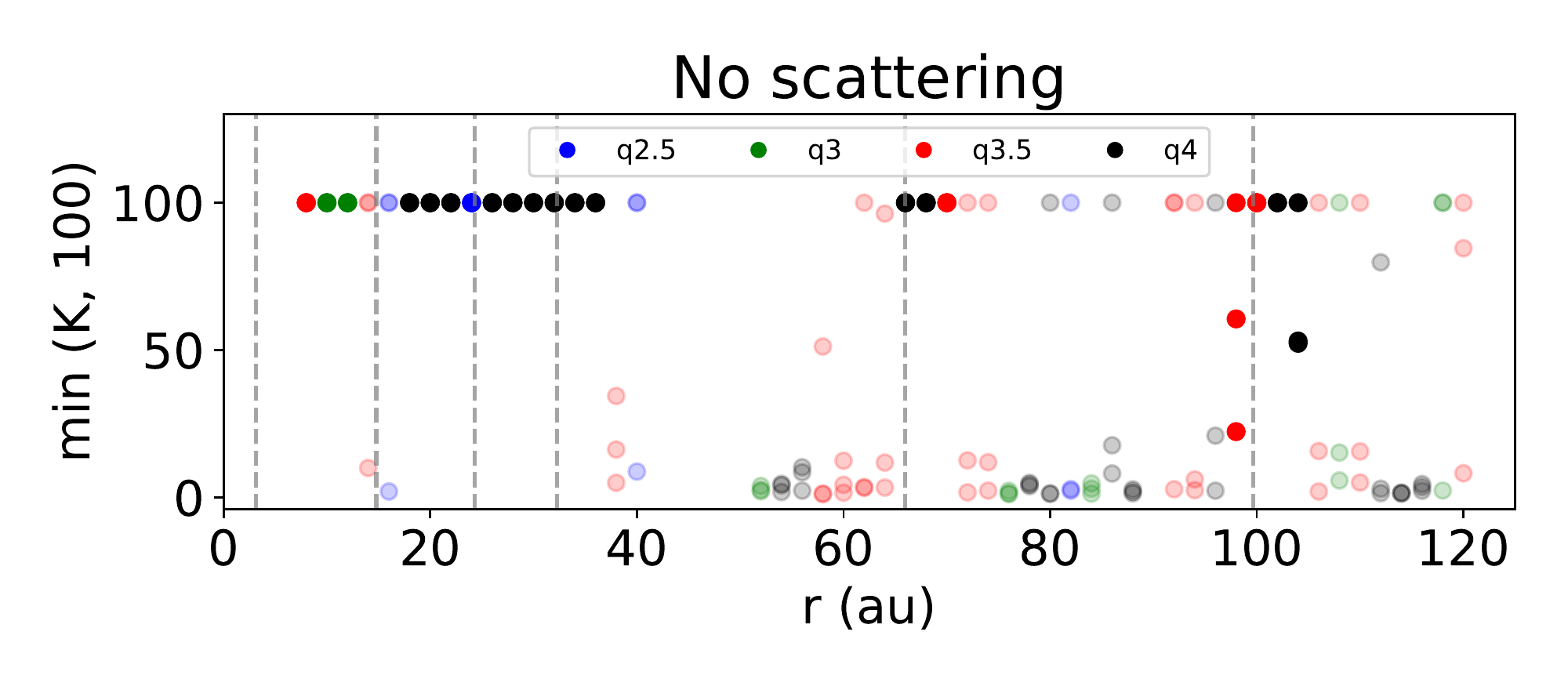}
 \includegraphics[keepaspectratio=True, width = 0.5\textwidth]{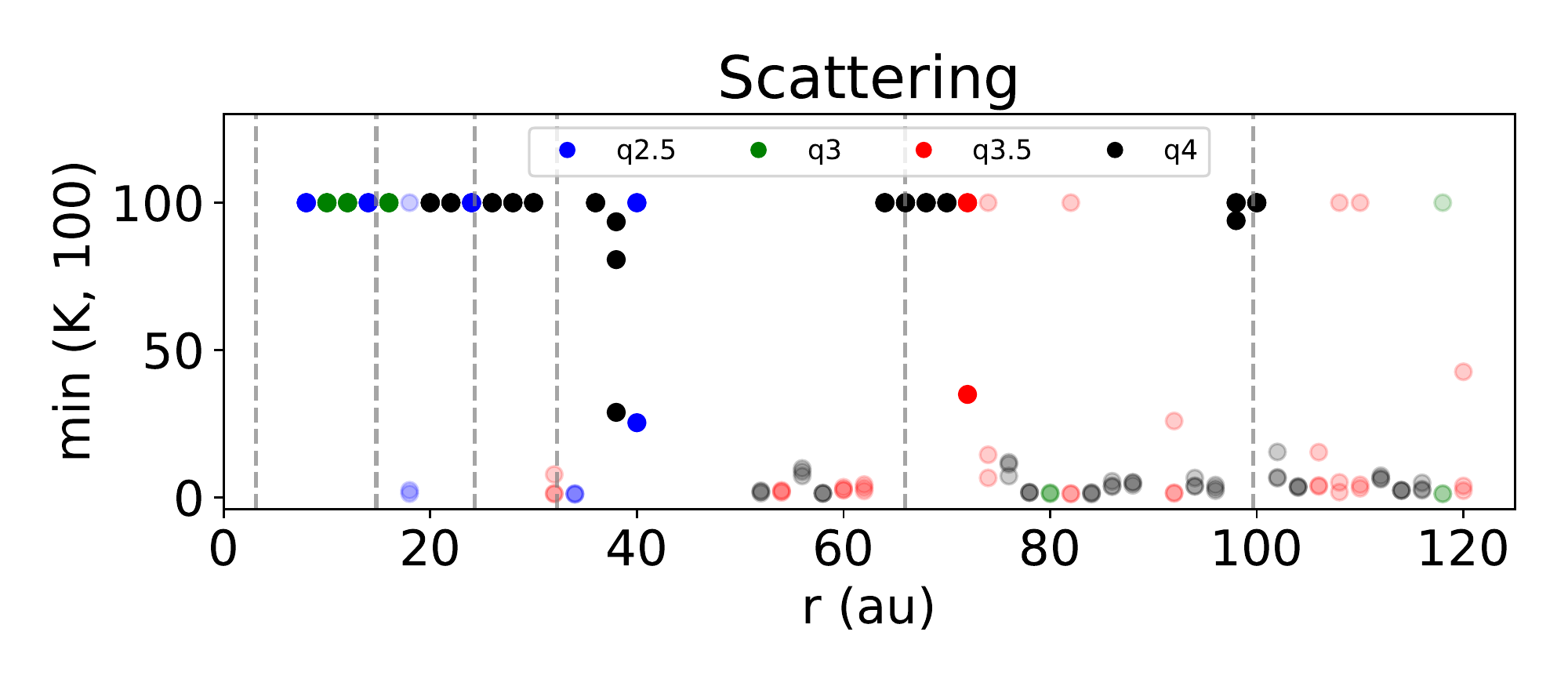}
 \caption{Bayes factor K computed between the model with the highest evidence (identified by the color of the marker) and the remaining 3 size distributions at each radius, for standard grains in the non-scattering (top panel) and scattering (bottom panel) case. The full color indicates that the Bayes factors K at that specific radius are all > 10, i.e. the is a strong evidence for that size distribution compared to the other three discrete values.}
 \label{fig:bestq}
\end{figure}

\begin{figure*}
 \centering
  \includegraphics[keepaspectratio=True, width = \textwidth]{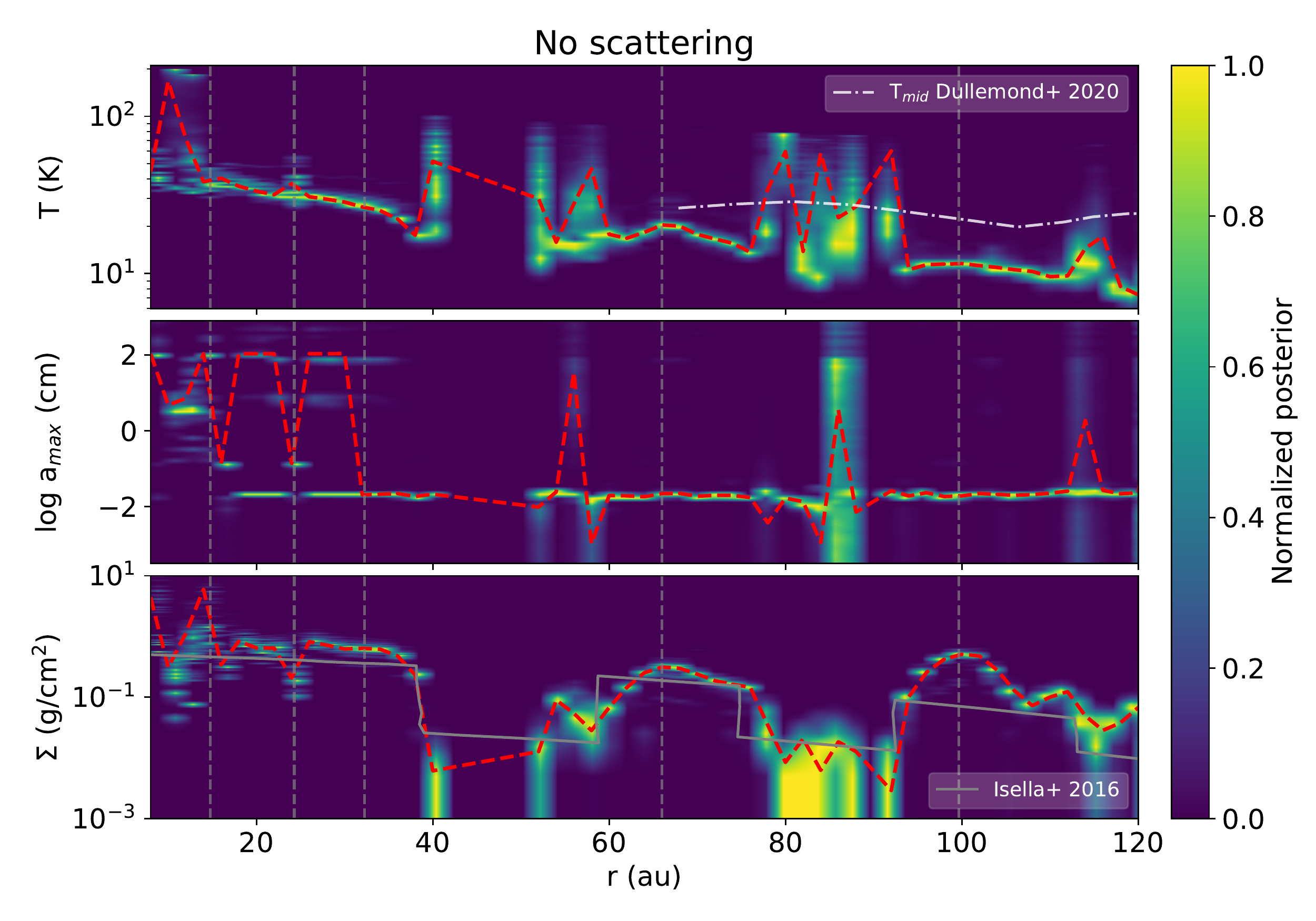}
 \caption{Temperature, maximum grain size and surface density as a function of the radius from the SED fitting with a model without scattering. The dashed red curve represents the estimates from the Monte Carlo fit including the statistical error only. The color map shows the normalized posterior distribution obtained merging 30 additional fits after introducing a random offset in the fluxes according to their flux calibration accuracy. The resonance in the opacity as function of a$_\mathrm{max}$ (see Figure \ref{fig:kappa}) can result in a degeneracy of this parameter, e.g. in the inner disk inside $\sim$40\,au. In the top and bottom panes we overplot the temperature and surface density from previous studies.}
 \label{fig:fitnoscat}
\end{figure*}

\begin{figure*}
 \centering
  \includegraphics[keepaspectratio=True, width = \textwidth]{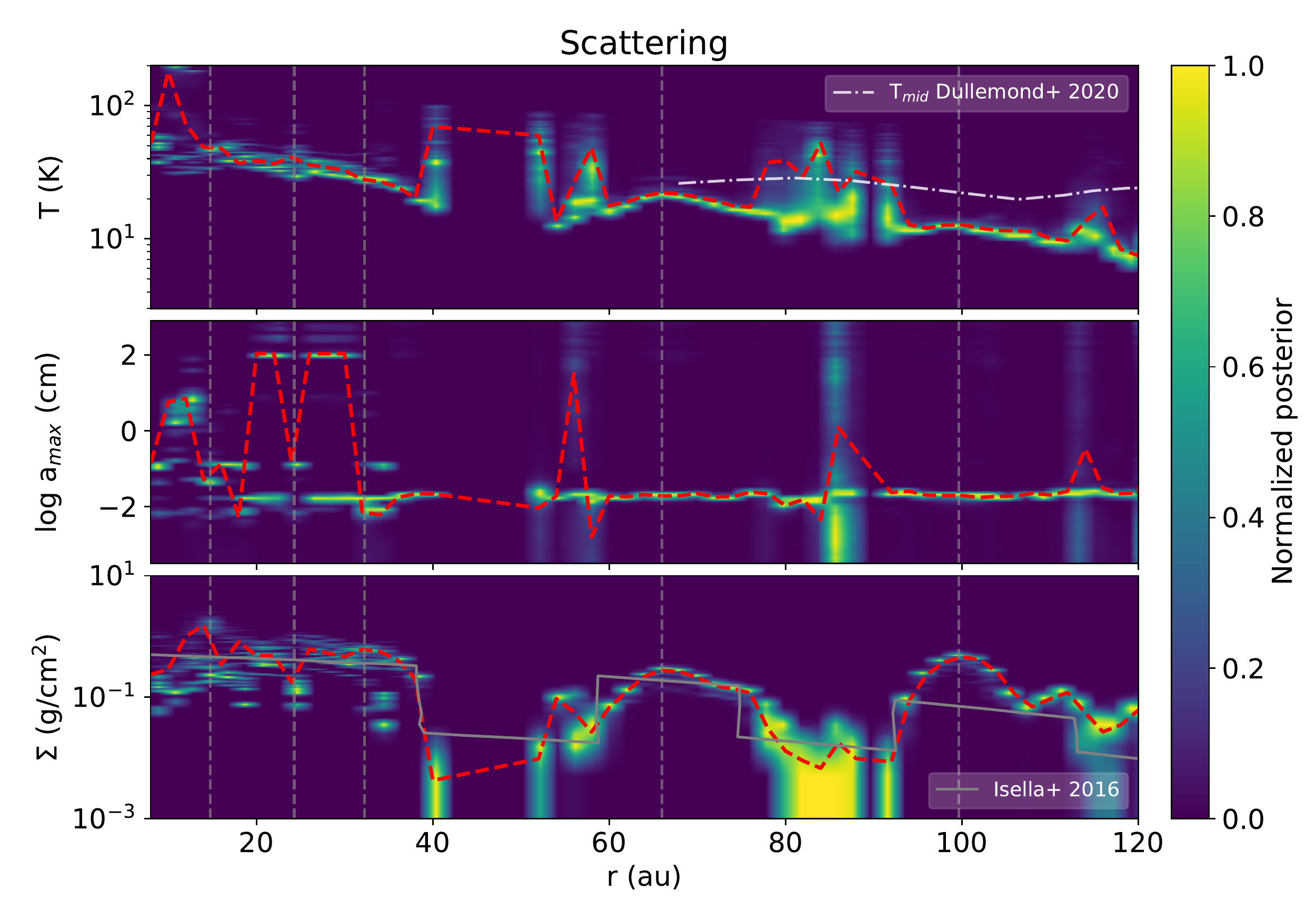}
 \caption{Same as Figure \ref{fig:fitnoscat}, but for the model that includes scattering.}
 \label{fig:fitscat}
\end{figure*}

{Since we determined that the models with standard grains are preferred to the ones with porous grains, we focus on the former to infer the final estimates of our best-fit parameters for the HD~163296 disk. The further step consists in estimating the best size distribution as function of the radius, following the same procedure based on the K factor.  
In Figure \ref{fig:bestq} we plot the Bayes factor K relative to the $q$ model with the larger evidence: this shows simultaneously the preferred size distribution at each radius and the strength of the evidence ratios. 
We show the corresponding final best-fit parameters, relative to the best size distribution at each radius, in Figure \ref{fig:fitnoscat} and \ref{fig:fitscat} for the nonscattering and scattering model, respectively. 
We draw with a red curve  the best-fit values for the Temperature, maximum grain size and surface density obtained from the Monte Carlo fit with statistical error only, and the total posterior distribution obtained with the 30 additional fits to account for the calibration error as a color map. }

\begin{figure*}
 \centering
 \includegraphics[keepaspectratio=True, width = \textwidth]{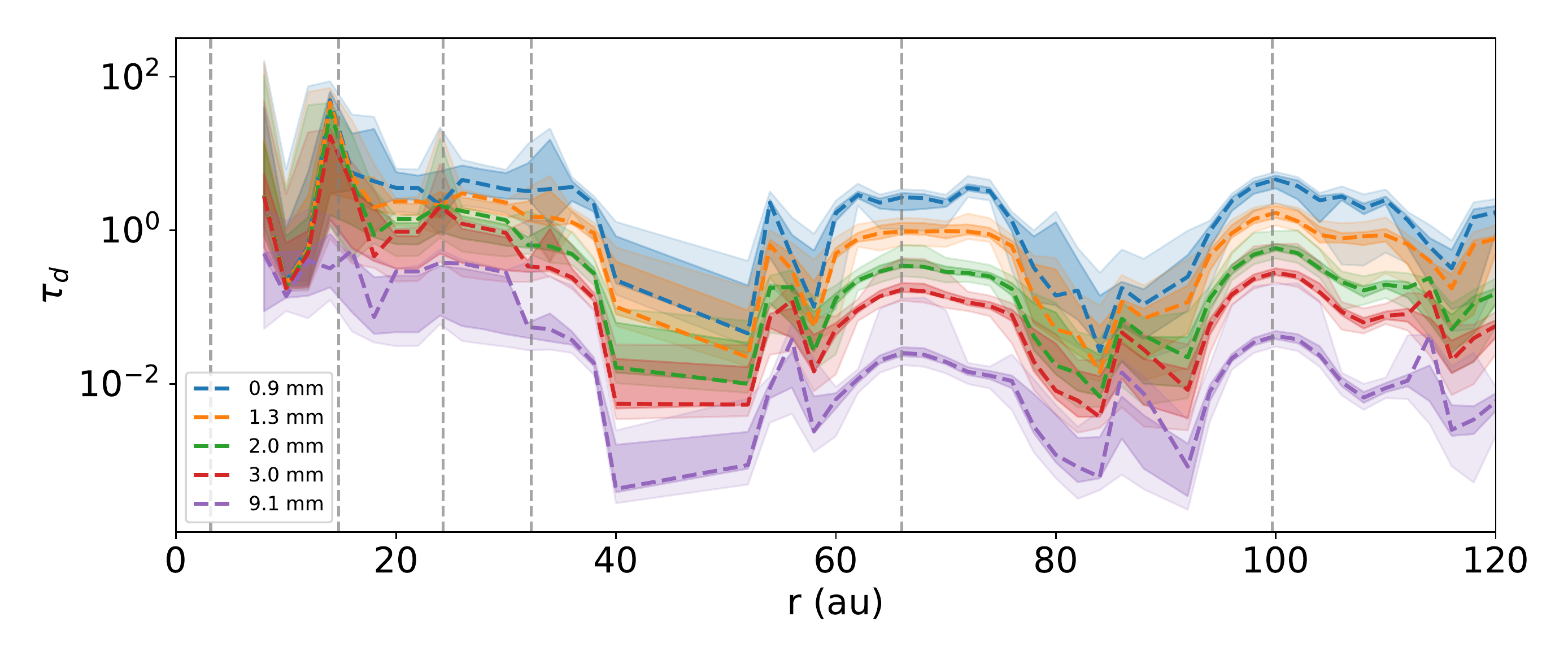}
 \caption{Optical depth at the five different wavelength, resulting from the best fit model shown as a dashed curve in Figure \ref{fig:fitscat}. The uncertainties (shaded regions) correspond to 1 and 2 $\sigma$ (16th/84th and 2.5th/97.5th percentiles, respectively) of the distribution at each radius (see main text). }
 \label{fig:tau}
\end{figure*}

The best-fit temperature is consistent with the one derived from the simple power-law model (Figure \ref{fig:fitA}), with hints of higher values of T in the dust gaps at about 50, 85 and 115\,au. 
In addition, in both the nonscattering and scattering model we see a 
steep increase in the temperature in the innermost gap, with T $\simeq$ 167\,K and 180\,K respectively. 
{If we compare our midplane temperature profile with the one derived by \citet{2020A&A...633A.137D} using CO emission lines (gray curve in the upper panels of Figures \ref{fig:fitnoscat} and \ref{fig:fitscat}), we note that while at the 66\,au ring there is only a small difference between the two studies, at the 100\,au ring we get a significantly lower T ($\simeq$12\,K) in both the nonscattering and scattering case. Since this is below its condensation temperature, the CO would be frozen-out onto dust grains at this location in the midplane. This invalidates the assumption made by \citet{2020A&A...633A.137D}, that would find a higher temperature as their measured CO emission is coming from higher vertical layers at this separation.} 

The maximum grain size radial profile is interestingly flat outside $\sim$40\,au, roughly consistent with a constant a$_{\mathrm{max}} \simeq$ 200\,$\mu$m from 40 to 120\,au with no significant difference between gaps and rings. 
{While the grain size is well constrained in the outer rings (at r$\geq$40\,au), in the inner disk (in particular between 15 and 30\,au) a degeneracy in $a_\mathrm{max}$ results in higher uncertainties on this parameters, with values ranging from 10$^{-2}$ to 10$^2$\,cm. }
The surface density drops, i.e. indicates dust depletion, at the same locations where the temperature increases (see previous paragraph). 

Comparing our result with the work by \citet{Isella2016}, that performed radiative transfer modeling using a smooth surface density profile with rectangular gaps, we find that outside 20\,au the two surface densities are consistent within a factor of 2. A significant exception is the ring at 100\,au, where we find a higher $\Sigma_\mathrm{dust}$ by a factor of 7. 
The optical depth at the five wavelengths {from our scattering model} is shown in Figure \ref{fig:tau}. The errors are estimated by taking the 2.5/16th and 97.5/84th percentiles of the distribution of $\tau$ obtained drawing 1000 random  samples from a gaussian distribution of $a_\mathrm{max}$ and $\Sigma_d$ (with $\sigma$ corresponding to the standard deviation of their total posterior distribution at each radius). 
In the outer rings at 66 and 100\,au, the emission is still moderately optically thick at a wavelength of 1.3\,mm, with $\tau$ of order unity. 
It is worth noticing that the optical depth in the outer ring at 100\,au results larger that the one at the 66\,au ring at all wavelengths, with $\tau_{\mathrm{100}} \sim 1.7 \tau_\mathrm{66}$. At the higher frequencies this is even comparable with the optical depth of the inner rings, with 
$\tau_{\mathrm{0.9 mm}}$ = 4.6$^{+0.6}_{-1.1}$ and $\tau_{\mathrm{1.3 mm}}$ = 1.7$^{+0.4}_{-0.3}$ at 100\,au, with uncertainties corresponding to $\sim$1$\sigma$ (16th and 84th percentiles) of the posterior distribution and showed as shaded areas in Figure \ref{fig:tau}. 

\begin{figure}
 \centering
 \includegraphics[keepaspectratio=True, width = 0.5\textwidth]{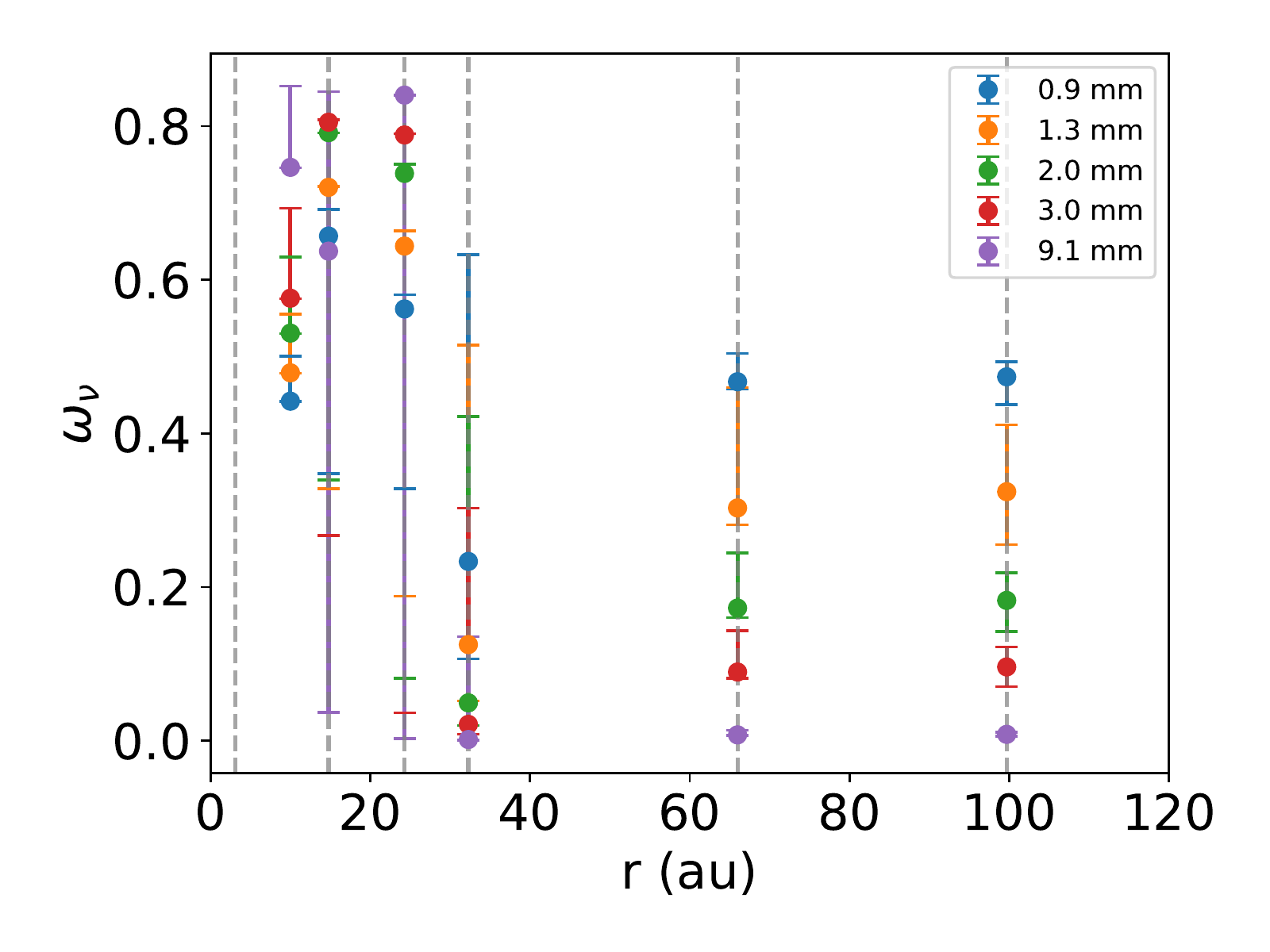}
 \caption{Albedo at the different wavelengths from our best-fit physical model, shown at specific locations in the disk. The error bars correspond to the 16th and 84th percentile of the set models, the vertical dashed line are drawn at the position of the flux peaks at 2\,mm.}
 \label{fig:albedo_r}
\end{figure}

In Figure \ref{fig:albedo_r} we plot the albedo at the different wavelengths in correspondence of the rings {and at the innermost gap at 10\,au}: while in the outer disk the albedo decreases with wavelengths, in the inner disk 
(r $\sim$ 10\,au) it shows the opposite trend. Across the intermediate rings (at 14, 24 and 32\,au) the trend appears similar to the innermost part of the disk, but the uncertainties on the albedo are too high to draw a robust conclusion. 
This ``spectral inversion'' of the albedo from the outer to the inner disk is related to the transition from small grains in the outer rings ($\omega_{\mathrm{1 mm}} \sim$ 0.3) to large grains inside ($\omega_{\mathrm{1 mm}} \sim$ 0.7), with the peak of the albedo falling at wavelengths $\lambda \sim$ 2$\pi$a$_{\mathrm{max}}$.  

\section{Discussion}
\label{sec:discussion}
\subsection{Grain size}
Previous works that analyzed mm-observations of HD~163296 at low and moderate spatial resolution indicated the presence of $\sim$millimeter/centimeter sized grains \citep{1990AJ.....99..924B,natta04, isella2007,Guilloteau2011, Guidi2016} based on the measured millimeter flux spectral index. However, in the recent years the robustness of a straightforward interpretation of this parameter as a direct proxy for the grain size has been disputed: observational evidence suggest that the assumption of optically thin emission at millimeter wavelengths is no longer justified for protoplanetary disks, and that dust self-scattering can therefore play a significant role in regulating the emitted intensity \citep[e.g.][]{2016ApJ...831L..12K,2019ApJ...877L..22L,2019ApJ...877L..18Z,2020ApJ...893..125U}. Accounting for this effect, several multiwavelength studies have been carried out for other bright disks \citep{2020ApJ...891...48H,2021A&A...648A..33M,2021ApJS..257...14S}, and indicate that in fact the optical depth and albedo can assume high values especially in the inner disk, but also at large separation from the central star \citep[e.g. HLTau][]{2019ApJ...883...71C}. 

Similarly to the studies reported above, in our multiwavelength analysis of the continuum emission we include self-scattering from an isothermal slab within an analytical formulation of radiative transfer. Using this physical model we could successfully fit the observed profiles of HD~163296 disk deriving the dust temperature, maximum grain size and surface density at separations larger than 8~au. 
At smaller separations, the surface density could not be constrained, likely because of the high optical depth that results in the Planck term in Equation \ref{eq:scattering} dominating the emitted intensity. 
The fact that at shorter radii we could not constrain the surface density and maximum grain size 
could indicate that the underlying assumption of emission from an isothermal surface at all wavelengths does not hold in the very inner disk. 
The best-fit model 
indicates that the outer rings (at 66 and 100~au) are composed by grains on the order of 200 $\mu$m, with no significant variations in grain size between these rings and the adjacent gaps. 
In the inner disk (r $\lesssim$ 40\,au) the a$_\mathrm{max}$ is less well constrained: {the total posterior shows a degeneracy in the grain size, that can take values from $\sim$200\,$\mu$m to $\sim$100\,cm, and the best estimates in this region are more dependent from the size distribution spectral index $q$ (see Fig. \ref{fig:allbf_scat}). Despite these localized degeneracy, we note that a$_\mathrm{max}$ overall shows larger values in the inner disk (e.g. $\geq$millimeter-size in the rings at $\sim$16 and 24\,au, see Figure \ref{fig:fitscat}). 
We note that this degeneracy is not present in the outer disk where we spatially resolve the rings at all wavelengths.}

{In the outer rings at 66 and 100~au, the solutions are strongly dominated by the steeper size distributions (q = 4) in both the non-scattering and the scattering model, which corresponds to the bulk of the dust mass contained in the small grains. On the contrary, a prevalence of flatter distribution with q = 2.5/3 is found at small separations ($\leq$40\,au), but only for the scattering model. Although the explored grid of values is very limited and therefore we cannot determine the size distribution slope with high accuracy, we note that the steeper size distributions in the outer rings of HD~163296 are consistent with what found in the outer disks of HD~169142 \citep{2019ApJ...881..159M} and TW~Hya disk \citep{2021A&A...648A..33M}, by fitting multiwavelength observations with $q$ as a free parameter.}

{In a recent study, \citet{2021ApJS..257...14S} analyzed ALMA observations of HD~163296 at medium resolution ($\sim$19\,au) to constrain the dust properties across this disk. Using a physical model that includes scattering (as in eq. \ref{eq:scattering} in this work) and assuming a power-law size distribution with q = 2.5, they find two possible families of solution for the a$_{max}$ parameter in the inner disk (r $\leq$40\,au): one with grains on the order of 100\,$\mu$m and one with millimeter grains, similarly to what found in this work. At larger separation they find a maximum grain size on the order of millimeter, with a local increase at the 100\,au ring. These latter values differ by about one order of magnitude with our findings of $\sim$200\,$\mu$m grains outside 40\,au. This is not entirely surprising, as we showed in this work how the initial assumptions, such as the dust composition and size distribution, can affect the derived parameters in these analysis. 
We note that these two studies differ for the choice of dust opacities and the fixed temperature profile that \citet{2021ApJS..257...14S} assume while fitting only for the maximum grain size and surface density. 
To understand how each of these factors affects the final results, we performed some tests varying the dust opacities and introducing a fixed temperature profile. We describe the procedure and results in Appendix \ref{app:dsharp}, where we find that both the chosen composition and the fixed temperature profile significantly affect the final estimates of maximum grain size. }

{At this stage we still lack information on the typical dust composition in protoplanetary disks, therefore it is important to test a wider range of dust properties when fitting disk observations. In this direction, we note that  \citet{2022arXiv220201241Z} recently found that dust opacities including such amorphous carbons from \citet[][the same used in this study]{1996MNRAS.282.1321Z} can better reproduce the size-luminosity relation observed in nearby star forming regions \citep{2017ApJ...845...44T}, with respect to the DSHARP opacities.} 

Another important measure of the dust properties in disks comes from polarization studies: ALMA polarimetric observations of HD~163296 indicated that the grain size across this disk is smaller than 100-150 $\mu$m, when interpreting the polarized emission in terms of dust self-scattering \citep{2019MNRAS.482L..29D,2019arXiv191210012L,2019ApJ...886..103O}. 
While our results are in agreement with dust polarization measurements in the outer disk (r $\geq$ 40\,au), we find evidence for larger grain sizes in the inner disk. However, it was recently pointed out that a mix of dust scattering and magnetic alignment could be responsible for the detected polarized signal in the disk of HL~Tau \citep{2021ApJ...908..153M}. If this is the case for HD~163296, this would loosen the constraint on the maximum grain size of 100~$\mu$m in the inner disk and mitigate the discrepancy with our results. 

The bimodal distribution of the maximum grain size between the inner and outer disk could have multiple explanations. 
As a consequence of the aerodynamical friction with the surrounding gas moving at sub-Keplerian velocities, dust grains lose angular momentum and drift toward smaller radii, with larger particles drifting inward more efficiently than smaller particles \citep{Weidenschilling77}. As a result we expect to find 
larger particles in the inner disk and smaller outside. If pressure traps are present in the disk, they could stop or slow down radial drift and retain some large particles within localized structures. {Because of the poor constrains on a$_\mathrm{max}$ in the inner disk, we cannot tell whether this is the case for the inner rings (r$\leq$40\,au). On the opposite, we find a more robust evidence of no differential trapping (no change in grain size between rings and gaps) in the outer rings at 66 and 100\,au.} 
This could mean that the timescales of radial drift are shorter than the ones relative to the mechanism that created the rings, or the rings could have form recently and not have enough time to trap the particles. {Another reason could lie in the sticking properties of dust grains, hindering their growth  outside the water snowline: new laboratory measurements are showing that water ice-coated grains have a lower sticking force than previously reported and especially at the low temperatures of protoplanetary disks \citep[e.g.][]{2018MNRAS.479.1273G}, whereas dry grains (bare silicate or refractory carbonaceous) have an increase in sticking force at high temperatures \citep[e.g.]{2015ApJ...812...67K}, leading to a ``sweet spot'' for grain growth around 1200--1400\,K \citep{2020A&A...638A.151B,2021A&A...652A.106P}. These predictions are also consistent with a recent multiwavelength study of FU Ori \citep{2021ApJ...923..270L}, showing mm-sized grains in the hot inner disk and grain size $\leq$200\,$\mu$m at larger separations.} 

Finally, another scenario involves the replacement of the original dust particles by a second-generation of dust: N-body simulations presented in \citet{2019ApJ...877...50T} show that within an evolved disk - where dust has already evolved into larger bodies and planetesimals - the formation of giant planets in a disk can generate dynamical perturbations and  create a highly collisional environment in its surroundings. Depending on the mass of the planets and on the original size distribution of the planetesimals, these collisions would generate a populations of rejuvenated dust (with sizes from 100 $\mu$m to centimeters) that could account for a large fraction of the dust that is measured in evolved disk (50\% to 100\% of the dust mass in the case of HD~163296 as estimated in \citet{2019ApJ...877...50T}). {The same scenario is proposed to explain the trend of dust mass as function of age observed in nearby star forming regions in \citet{2022arXiv220104079T}: after an initial decrease with $\sim$1/t the solid mass increases again at 2--3\,Myr, which could be a sign of early planet formation and production of reprocessed dust.} 
Recently, \citet{2021ApJ...912..164D} claimed that the ring at 68\,au exhibits an increased dust-to-gas scale height $h_d/h_g$ with respect to the inner disk and to the outer ring at 100\,au. A large scale height of smaller (micron-sized) dust in the two outer rings (corresponding to aspect ratios $h/r \sim 0.25-0.3$) was suggested as well in \citet{2018MNRAS.479.1505G} to interpret the scattered light emission from HD~163296 in the thermal infrared.
This hints to the presence of dynamically excited dust at this location, that we recall is predicted by planet-disk interaction simulations \citep{2019ApJ...877...50T,2021ApJ...912..107B,2021MNRAS.tmp.1852B}. This could also reconcile the dust temperature at 66\,au derived in this work being closer to the gas temperature measured by \citet{2020A&A...633A.137D} compared to the 100\,au ring (see Figure \ref{fig:fitscat}). 
We note that the DSHARP observations of HD~163296 at 1.3\,mm were carefully analyzed to search for localized emission from circumplanetary material in the main gaps at 48 and 86\,au in \citet{2021ApJ...916...51A}, who did not find any detection for this or other DSHARP disks included in the study. One of the possible causes could be the aforementioned higher scale height of the dust rings in inclined disk such as HD~163296, that would increase the extinction of the circumplanetary disk emission along the line of sight (e.g. see Figure 4 in \citet{2018MNRAS.479.1505G} in relation to the small grains).

\subsection{Dust mass}
\label{sec:dmass}
We calculate the dust mass from our derived surface density profile as \(M_d = \int_\mathrm{{r_i}}^\mathrm{{r_f}} \Sigma(r) 2 \pi r dr\). 
{With r$_\mathrm{i}$ = 8\,au and r$_\mathrm{f}$ = 120\,au this is equal to 8.0 $\cdot$ 10$^{-4}$ M$_{\odot}$ or 265 M$_{\oplus}$ for the scattering model, and 1 $\cdot$ 10$^{-4}$ M$_{\odot}$ 337~M$_{\oplus}$ for the nonscattering model.} We note that since we are pushing the spatial resolution of our observations to characterize in detail the inner structure of the HD~163296 disk, we have typically a low signal-to-noise in the outer disk, that was not included in our SED fitting. Therefore we are not sensitive to the dust mass contained at radii $>$120~au, that should anyway represent only a few percent of the total dust mass (see next paragraph). 

We can compare our result with previous studies considering the same separation range: integrating the surface density from \citet{Isella2016} between 8 and 120~au, we find a dust mass of 0.47 $\cdot$ 10$^{-3}$ M$_{\odot}$, about 1.7 times smaller than what found in this work. We note that the dust surface density in \citet{Isella2016} was derived using RADMC3D \citep{2012ascl.soft02015D} without including scattering opacity on a single wavelength ALMA dataset at 1\,mm. The difference can therefore be due to the fact that we fit for the opacity at each radius and we include dust self-scattering, although with a simplified analytical description. We note that integrating the surface density function found in \citet{Isella2016} up to radii of 400\,au, we find that the remaining mass is only a small fraction of the total, specifically M$_\mathrm{dust}$\,[8--120]\,au $\simeq$ 97\% M$_\mathrm{dust}$\,[0.5--400]\,au. 

We can estimate the mass of the spatially resolved rings at 66 and 100\,au, where we define the radial limits for each ring as the closest minima to the ring peaks, calculated with \texttt{scipy.signal.argrelmin} on the surface density profile obtained in Sect.~\ref{sec:pm}. The values are displayed in Table \ref{tab:mrings}: we note that the mass of the outer ring at 100\,au is larger than the one at 66\,au and accounts for about a third of the total mass. This is opposite to what found in previous studies, for example, \citet{2018ApJ...869L..46D} found 56.0 M$_{\oplus}$ for the 66\,au ring (consistent with our result), but only 43.6 M$_{\oplus}$ for the 100\,au ring, using only the DSHARP dataset at 1.3\,mm. The discrepancy at 100\,au remains if we compute the ring masses using the same radius intervals as in the cited paper, as we find 57 M$_{\oplus}$ in the limits 52--82 au and 84 M$_{\oplus}$ between 94--104 au. 
\begin{table}
	\caption{Dust mass and temperature from the nonscattering and scattering models with standard grains.}
	\label{tab:mrings}
	\centering
		Nonscattering model\\
		\vspace{2pt}
		\begin{tabular*}{\columnwidth}{@{\extracolsep{\stretch{1}}}*{4}{c}@{}}
		\hline \hline
		 & R &  M$_d$ & T$_d$ \\
	     & [au] & [M$_{\oplus}$] & [K] \\
		\hline
		R5 & 58--84 & 61$^{+9}_{-11}$ & 18.0 \\
		\vspace{3pt}
		R6 & 92--108 & 107$^{+11}_{-14}$ & 11.2 \\
		\vspace{2pt}\\
		disk & 8--120 & 337$^{+71}_{-88}$ & 20.4 \\
		\vspace{0.2pt}\\
		\hline
	\end{tabular*}
	\vspace{4pt}\\
	Scattering model\\
	\vspace{2pt}
		\begin{tabular*}{\columnwidth}{@{\extracolsep{\stretch{1}}}*{4}{c}@{}}
		\hline \hline
		 & R &  M$_d$ & T$_d$ \\
	     & [au] & [M$_{\oplus}$] & [K] \\
		\hline
		R5 & 58--84 & 53$^{+11}_{-9}$ & 21.4 \\
		\vspace{3pt}
		R6 & 92--108 & 96$^{+13}_{-16}$ & 12.1 \\
		\vspace{2pt}\\
		disk & 8--120 & 265$^{+75}_{-76}$ & 22.5 \\
		\vspace{0.2pt}\\
		\hline
	\end{tabular*}
	\tablefoot{Dust mass and median dust temperature are computed at the outer rings and across the portion of the disk where the spectral analysis was carried out. The confidence intervals are calculated using the upper and lower estimates of the surface density from our best-fit model shown in Figure \ref{fig:fitscat}. The numbering of the rings is taken from Sect.~\ref{sec:vis}.}
\end{table} 

In the context of disk demographic studies, the dust mass is typically calculated 
using a simple scaling of the flux density with a fixed dust opacity, as originally described by \citet{1983QJRAS..24..267H}:
\begin{equation}
    M_d = \frac{F_{\nu} d^2}{\kappa_{\nu} B_{\nu}(T)}.
\end{equation}
If we use this relation to derive the dust mass in HD~163296 from the integrated fluxes from the ALMA observations (Table \ref{tab:par}), using T=21\,K as the median disk temperature for the Black Body term, and an opacity of $\kappa_{\nu}$ = 10 cm$^2$/g ($\nu$/1000\,GHz)$^\beta$ with $\beta$ = 1, we obtain dust masses on the order of 60-80\% of the one derived in our multiwavelength analysis. Specifically, we obtain M$_{\mathrm{d, 0.9mm}}$ = 63\% M$_\mathrm{d,mwle}$, M$_{\mathrm{d, 1.3mm}}$ = 73\% M$_\mathrm{d,mwle}$, M$_{\mathrm{d, 2.1mm}}$ = 64\% M$_\mathrm{d,mwle}$ and M$_{\mathrm{d, 3.0mm}}$ = 75\% M$_\mathrm{d,mwle}$. 

We must stress here that the value for the dust mass we obtain from our spectral fit are heavily dependent from the choice of the dust composition. Within the different models we explored, we showed that a difference of a factor of 5 in surface density - that corresponds to the same factor in terms of dust mass - is already present varying only the porosity of the grains. Our dust mass estimates from the porous grains models are listed in Appendix \ref{app:sedfit} and indicate a total dust mass of 4.26$\cdot10^{-3}$\,M$_\odot$ or 1417 M$_{\oplus}$, 
i.e. about 5 times larger than our reference model with standard grains. Changing the dust constituents would also cause a difference: for example, the amorphous carbons from the DIANA project that we employ in this study have absorption opacities that are a factor of a few larger than the standard opacities used in the DSHARP modeling \citep{2018ApJ...869L..45B}. {We showed in Appendix \ref{app:dsharp} how for the case of HD~163286 this translates in surface densities (i.e. dust mass) larger by a factor of $\sim$3.} 

In relation to demographic studies that found a linear size-luminosity relationship for protoplanetary disks at the frequency of 340\,GHz \citep{2017ApJ...845...44T,2018ApJ...865..157A}, we recall here that one of the scenarios proposed to explain such scaling relation invoked the presence of sub-structures in the dust that would generate optically thick emission with filling factors of a few tens percent \citep[see][]{2012A&A...540A...6R}. In the case of HD~163296 this is verified, as we measure a filling factor flux$_\mathrm{thick}$/flux$_\mathrm{tot}$ of 82\% at Band~7 in the range 8--115\,au, with the upper limit in separation  taken as the effective radius R$_\mathrm{eff}$ given in \citet{2017ApJ...845...44T}. 

\subsection{Origin of the ringed structure}
{Revealing the grain size and distribution across a structured-disk can provide useful information on the mechanisms that generate such substructures. Theoretical studies show that 
various forms of instabilities, including the presence of embedded planets, generate pressure traps in the gas distribution of different shape and intensity; these, in turn, cause different levels of segregations of dust grains depending on their size \citep[e.g.][]{2012A&A...545A..81P,2019MNRAS.485.5914N}. 
So far, several studies have proposed the presence of planets in the disk of HD~163296, based on 
the observed depletion in the dust and gas surface density  \citep{Isella2016, 2018ApJ...857...87L} and signatures in the gas kinematics \citep{2018ApJ...860L..13P,2018ApJ...860L..12T}.}  

\citet{2021ApJS..257...14S} found evidence of dust trapping in a pressure bump at the 100\,au ring (hint by the higher grain size at this ring). We do not find such an indication in this study, but we showed how this result is highly dependent on the assumptions in the dust composition (see Appendix \ref{app:dsharp}). 
Even if we do not find evidence of such a differential trapping of dust grains in the rings, we note that we see an increase in the surface density at the 100\,au ring, with respect to the inner 66\,au ring. This is generally predicted by disk-planet interaction simulations, where the presence of a planet results in an accumulation of dust in the ring external to the planet, and a starving of material in the internal ring \citep[e.g.][]{2012ApJ...755....6Z,2016MNRAS.459.2790R,2021MNRAS.tmp.1852B}. 
Finally, as discussed in Sect.~\ref{sec:dmass} the relatively small grain size in the outer rings could indicate the effect of collisions excited by an embedded planet. In this case, we note that the size distribution of the particles could be significantly different than the power-law assumed in this study, and this can in turn affect the derived maximum size of the  distribution.

\subsection{Model comparison}
\label{sec:mcomp}
\begin{figure}
 \centering
 \includegraphics[keepaspectratio=True,width=0.5\textwidth]{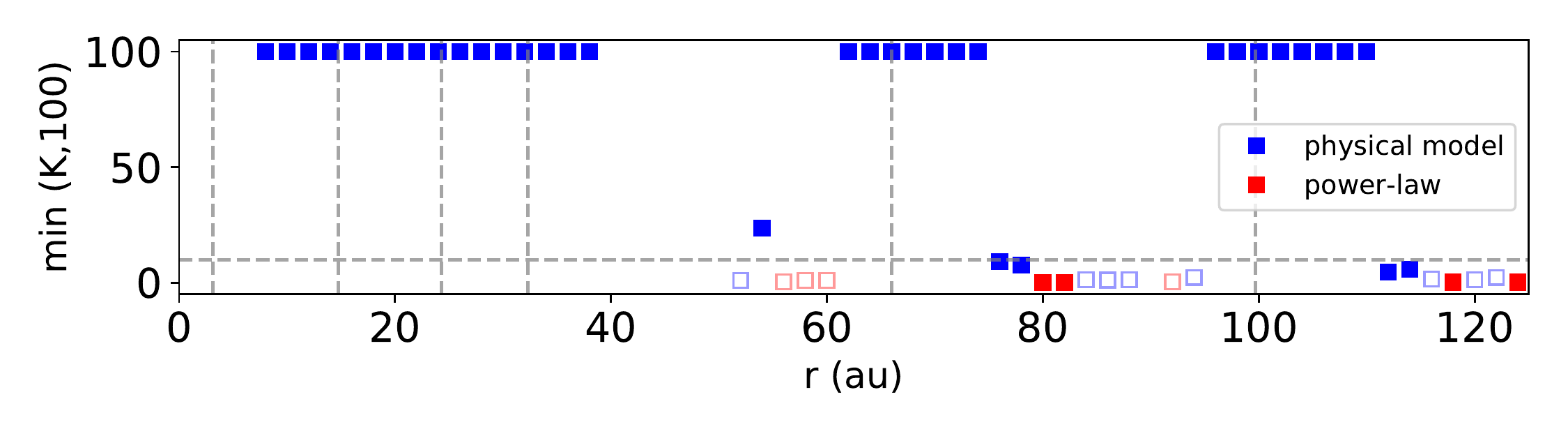}
  \includegraphics[keepaspectratio=True,width=0.5\textwidth]{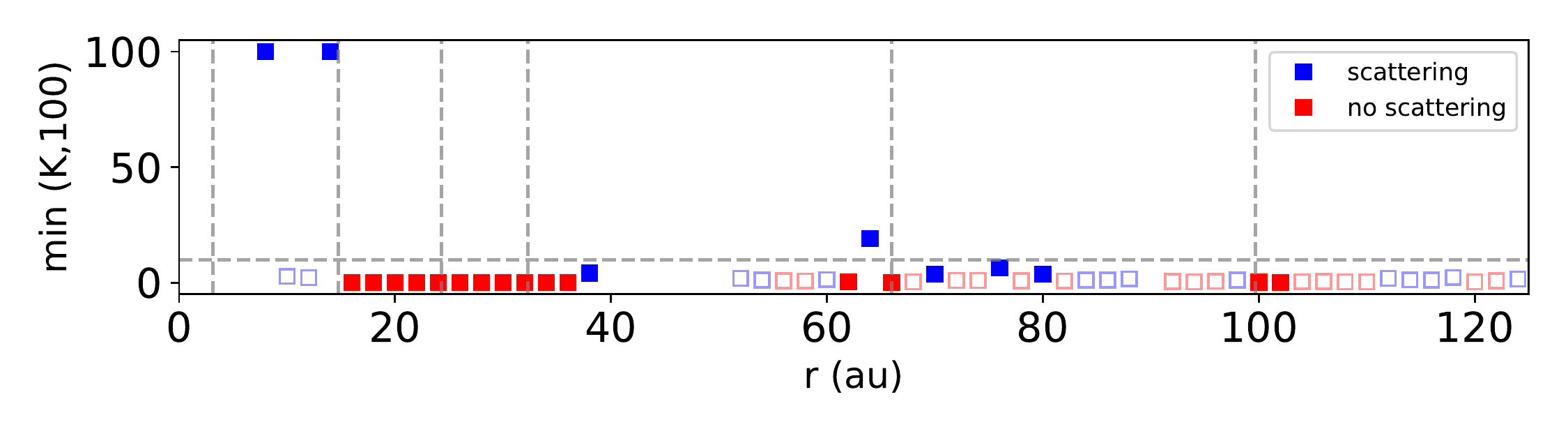}
 \caption{Model comparison. \textit{Top panel}: Bayes factors computed between the evidence of the fits with the physical model without scattering (eq. \ref{eq:noscat}) and the parametric model (eq. \ref{eq:pl}). \textit{Bottom panel}: Bayes factor between the scattering (eq. \ref{eq:scattering}) and nonscattering model. }
 \label{fig:model_comp}
\end{figure}
In this study we fit observations of HD~163296 at 5 wavelengths using a parametric model and a physical model (with and without scattering). To determine the relative strength of each model in predicting our datasets, we use again the Bayes factor K (see Sect.~\ref{sec:bayesK}). In the top panel of figure \ref{fig:model_comp} we confirm the expected outcome that a physical model is a better description of our submillimeter data with respect to a simple power-law. In particular, there is decisive evidence (K>100) in favor of the physical model at 62\% of the radii, a strong evidence (K>10) at 64\% or the radii and a moderate evidence (K>3) at 72\% of the radii. On the opposite, strong evidence goes down to only 4\% and moderate evidence at 6\% of the radii for the power-law model. 
This indicates that a simple prescription of the optical depth as a power-law function of the frequency is less-likely to predict the observed emitted intensity (see also Figure \ref{fig:kappa_plaw} in the Appendix), compared to the analytic physical model described in equation \ref{eq:noscat}. 

When comparing the physical models with and without scattering, we do not find a strong change in evidence in favor of the former (Figure \ref{fig:model_comp}, bottom panel). We also find that the relative evidence between scattering and nonscattering model varies significantly among the single size distribution models. We can only conclude that a nonscattering model can equally reproduce the observations, generally predicting a lower temperature with respect of the scattering model.

\section{Conclusions}
\label{sec:concl}
We analyzed new ALMA and VLA observations to study the dust properties in the rings of HD~163296. We employ parametric and physical descriptions of the flux density to reproduce the Spectral energy distribution as function of the radius, and compare the performances of the different models in a Bayesian framework. 
We summarize here our main results:
\begin{itemize}
    \item we estimate a non dust contribution from the inner disk (r$\lesssim$5~au) that accounts for about 5\% of the total flux at 9~mm and 40\% of the total flux at 3\,cm. Based on its mm-cm spectrum, its origin is consistent with free-free emission from a disk wind. 
\item We extract the radial brightness profiles at all wavelengths with the python tool \texttt{frank} fitting the original visibilities. We recover a total of six bright peaks in the flux distribution of HD~163296 within a separation of 120~au, i.e. one more ring than what derived from the convolved ALMA images at the highest available resolution \citep[e.g.][]{2018ApJ...869L..42H}. 
\item We fit a simple opacity power-law prescription to the extracted radial profiles at wavelengths from 0.88\,mm to 9\,mm and independently at each radius. We find that the dust temperature increases in the gaps at average separations of 10, 50 and 83\,au, and the opacity spectral index $\beta$ is consistent with the ISM values (micron-sized grains) at most radii, while it decreases to values below 1.7 in the inner disk (r $\lesssim$ 30\,au)
\item We follow the same procedure using a physical model with an analytical expression for radiative transfer with and without dust scattering. We consider a grid of 8 different dust populations, where we vary the size distribution slope and the composition. The best-fit models indicate the presence of 200\,$\mu$m sized grains in the outer disk at r$\gtrsim$40\,au. The grain size is less constrained in the inner disk, but there are local indications of larger grains ($\gtrsim$millimeter) at separation smaller than $\sim$30\,au. 
\item We find that our observations are generally better described by compact grains (porosity of 25\%) than by porous grains (80\% porosity). We also observe that a steeper size distribution (with q = 3.5/4) better describes the outer rings at 66 and 100 au.  
\item By comparing the evidence of the parametric and physical model, we confirm that the latter is better at predicting our datasets with respect to the simple power-law.
\item Finally, we note that different choices of dust opacities change the estimates of the dust mass of a factor of a few (a factor of 5 just varying the dust porosity), and locally affect the derived grain size in a less systematic but significant manner. 
\end{itemize}
 
Our results confirm that the presence of optically thick structures can artificially lower the millimeter spectral index, and therefore in general effects of dust self scattering should not be neglected when interpreting the continuum millimeter emission in terms of dust properties. 
This factors, combined with the highly non homogeneous spatial distribution of dust grains, accentuates the necessity of both high spatial resolution and spectral information, as they are crucial for resolving the small-scale structures and removing the degeneracies between parameters such as dust density and grain size.  It is not always possible to fully break such  degeneracy, despite the large  spectral information available (wavelengths from 0.9 to 9 mm), especially where the substructures are not resolved. 

We stress that a major caveat is still represented by the unknown composition of the dust grains, that can significantly affect the final dust parameters and make the comparison with theoretical prediction more challenging. 

Regarding the origins of the observed dust rings in the HD~163296 disk, the most accredited scenario invokes the presence of three to four giant planets perturbing the dust and gas structure. Our results are in general agreement with this scenario: the higher values of the dust temperature we derive in the optically thin gaps at 10, 50, 85 and 115\,au is expected if a planet is present in the gap as resulting from hydrodynamical calculations \citep{2018ApJ...860...27I}. A higher dust scale height in the 66\,au ring reported independently by \citet{2021ApJ...912..164D} would explain the temperature values we derive at this location being closer to the gas temperature, and would be consistent with vertical stirring of the dust caused by planets \citep[e.g.][]{2021MNRAS.tmp.1852B}. 

We do not detect a significant size difference between the grains in the two  outer rings and the adjacent gaps, as one could expect in presence of planets, as the pressure trap generated by the planets would result in larger grains being trapped at the pressure maxima in the gap edges \citep[e.g.][]{2012A&A...545A..81P,2016MNRAS.459.2790R,2019MNRAS.485.5914N}. However, the small 200\,$\mu$m grains found in the rings at 66 and 100\,au could belong to a second-generation dust population resulting from collisions of large km-sized bodies, that could account for a large fraction of the dust mass in the rings \citep{2019ApJ...877...50T}. Furthermore, we derive a higher surface density in the 100\,au ring with respect to the 66\,au ring, which is interestingly consistent with some theoretical predictions of disk with embedded planets and could be related to the protoplanet predicted at a radius of 80-90\,au  \citep[e.g.][]{Isella2016,2018ApJ...860L..12T,2022ApJ...928....2I}.
In this scenario the dust grains responsible for the (sub)millimeter emission could have a more complex three-dimensional structure rather than being distributed in a thin midplane. A future 3D Monte Carlo radiative transfer modeling (including full scattering) of multiwavelength observations will help in reconstructing the full structure of the solids in this disk.

\begin{acknowledgements}
This work has been carried out within the framework of the National Centre of Competence in Research PlanetS supported by the Swiss National Science Foundation. G.G. acknowledges the financial support of the SNSF. A.I.
acknowledges support from the National Science Foundation under grant No. AST-1715719 and from NASA under
grant No. 80NSSC18K0828. H.B.L. is supported by the Ministry of Science and Technology (MoST) of Taiwan (Grant Nos. 108-2112-M-001-002-MY3). GR acknowledges support from the Netherlands Organisation for Scientific Research (NWO, program number 016.Veni.192.233) and from an STFC Ernest Rutherford Fellowship (grant number ST/T003855/1). 
IdG-M is partially supported by MCIU-AEI (Spain) grant AYA2017-84390-C2-R (co-funded by FEDER). H.L. gratefully acknowledges the support by the LANL/CSES program and the NASA/ATP program. M.T. has been supported by the UK Science and Technology research Council (STFC) via the consolidated grant ST/S000623/1, and by the European Union’s Horizon 2020 research and innovation programme under the Marie Sklodowska-Curie grant agreement No. 823823 (RISE DUSTBUSTERS project).
This paper makes use of the following ALMA data: ADS/JAO.ALMA\#2017.1.01682.S, ADS/JAO.ALMA\#2015.1.00725.S, 
ADS/JAO.ALMA\#2016.1.01086.S. ALMA is a partnership of ESO (representing its member states), NSF (USA) and NINS (Japan), together with NRC (Canada), MOST and ASIAA (Taiwan), and KASI (Republic of Korea), in cooperation with the Republic of Chile. The Joint ALMA Observatory is operated by ESO, AUI/NRAO and NAOJ. 
The National Radio Astronomy Observatory is a facility of the National Science Foundation operated under cooperative agreement by Associated Universities, Inc.
\end{acknowledgements}



\bibliographystyle{aa}
\bibliography{biblio} 



\begin{appendix}
\section{ALMA flux calibration errors}
\label{ap:calib}
To get an estimate of the calibration error that affects our ALMA datasets, we compared the amplitudes of the flux calibrators that were set in the ALMA pipeline or calibration script with the measurements of the same calibrators taken before and after the science observations from the ALMA Calibrator Source Catalogue\footnote{https://almascience.eso.org/sc/}. We note that, as described in Sect.~\ref{sec:obs}, when combining multiple epochs and/or antenna configurations we scale the amplitudes to match the execution block that has been calibrated using the most recent calibrator measurements. Therefore for each ALMA band we plot only the fluxes from the epoch we used as reference, instead of plotting all the single scheduling blocks. 

\begin{figure}
 \centering
 \includegraphics[keepaspectratio=True,width=0.5\textwidth]{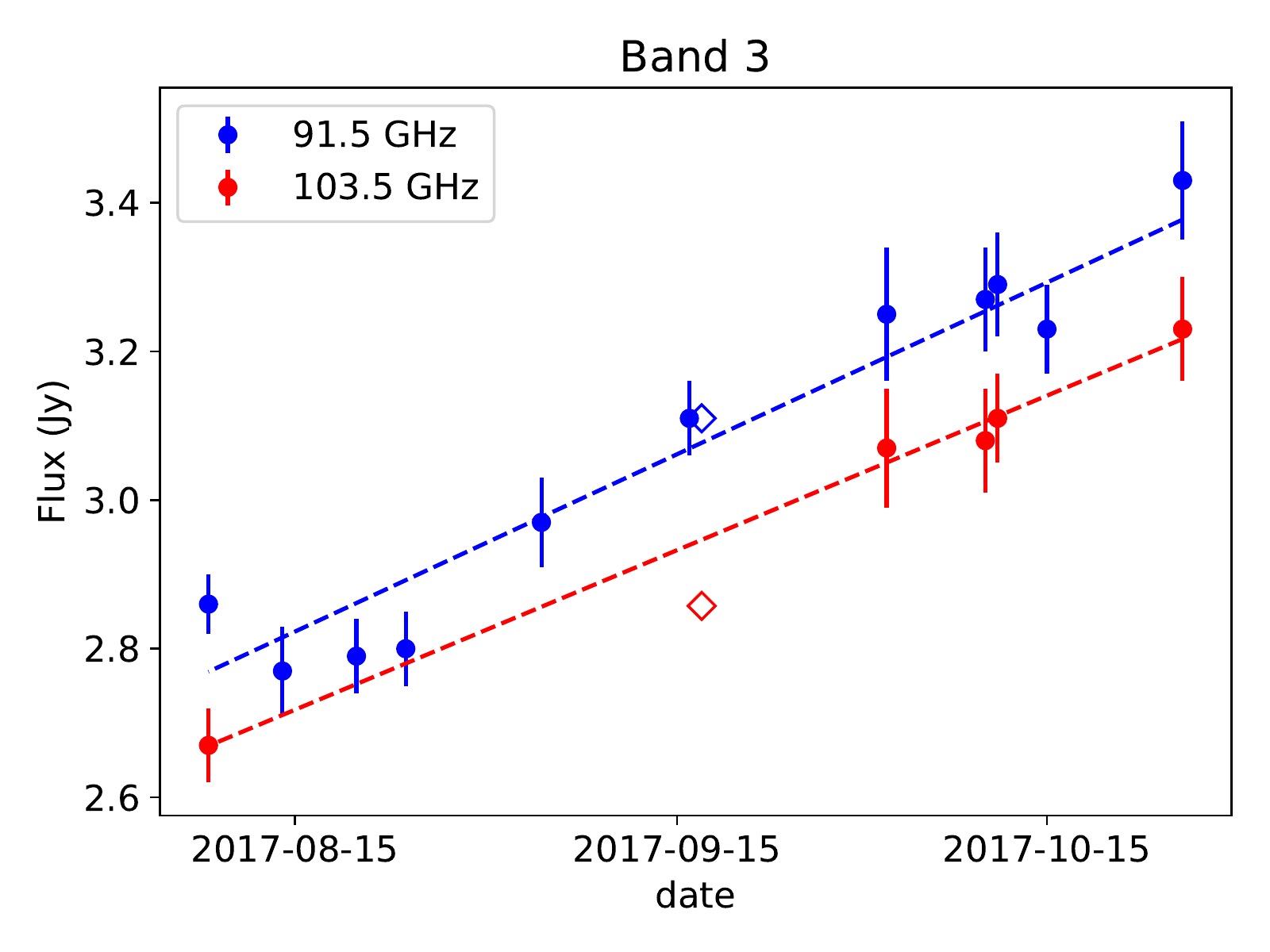}
  \includegraphics[keepaspectratio=True,width=0.5\textwidth]{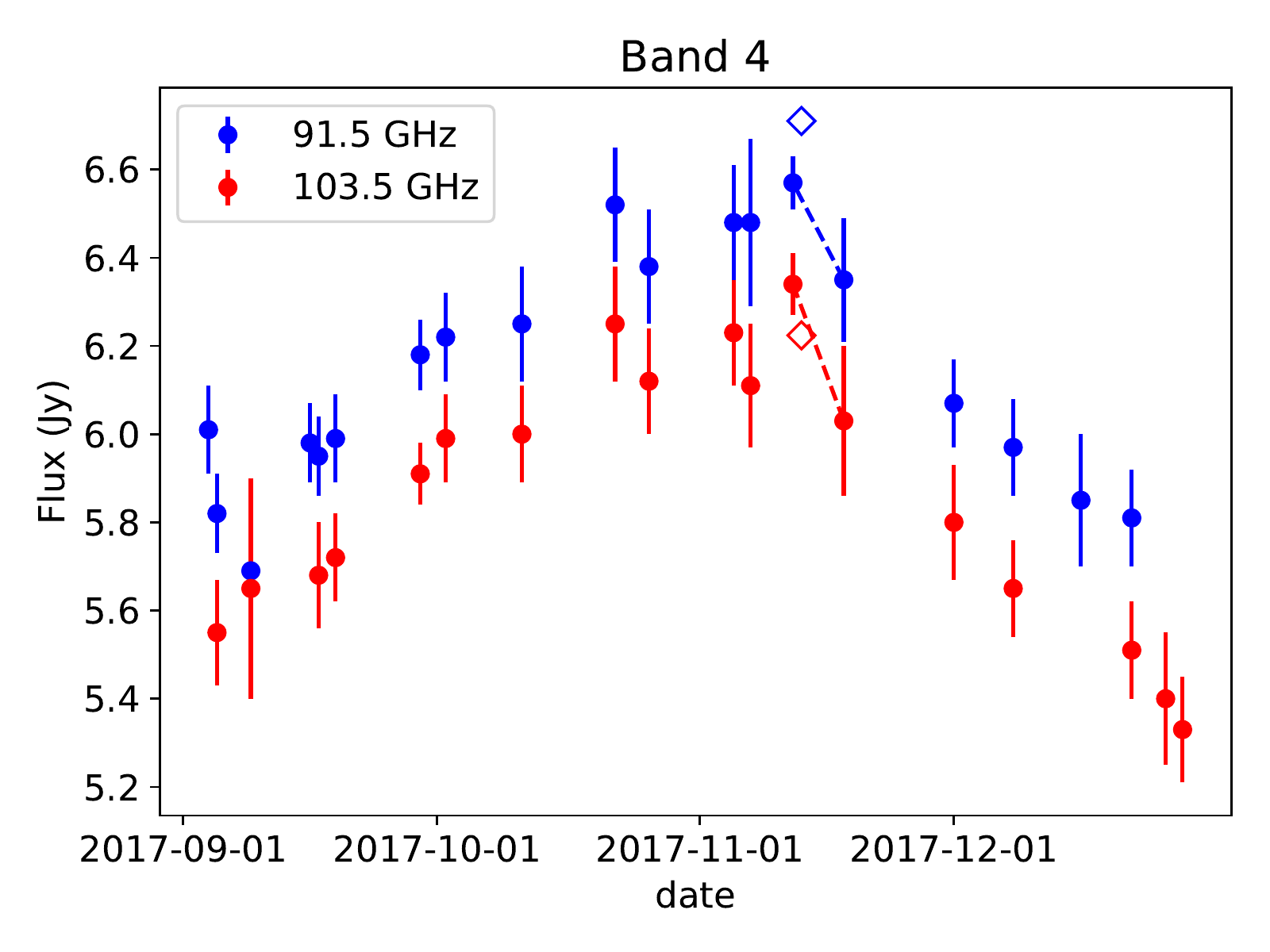}
  \includegraphics[keepaspectratio=True,width=0.5\textwidth]{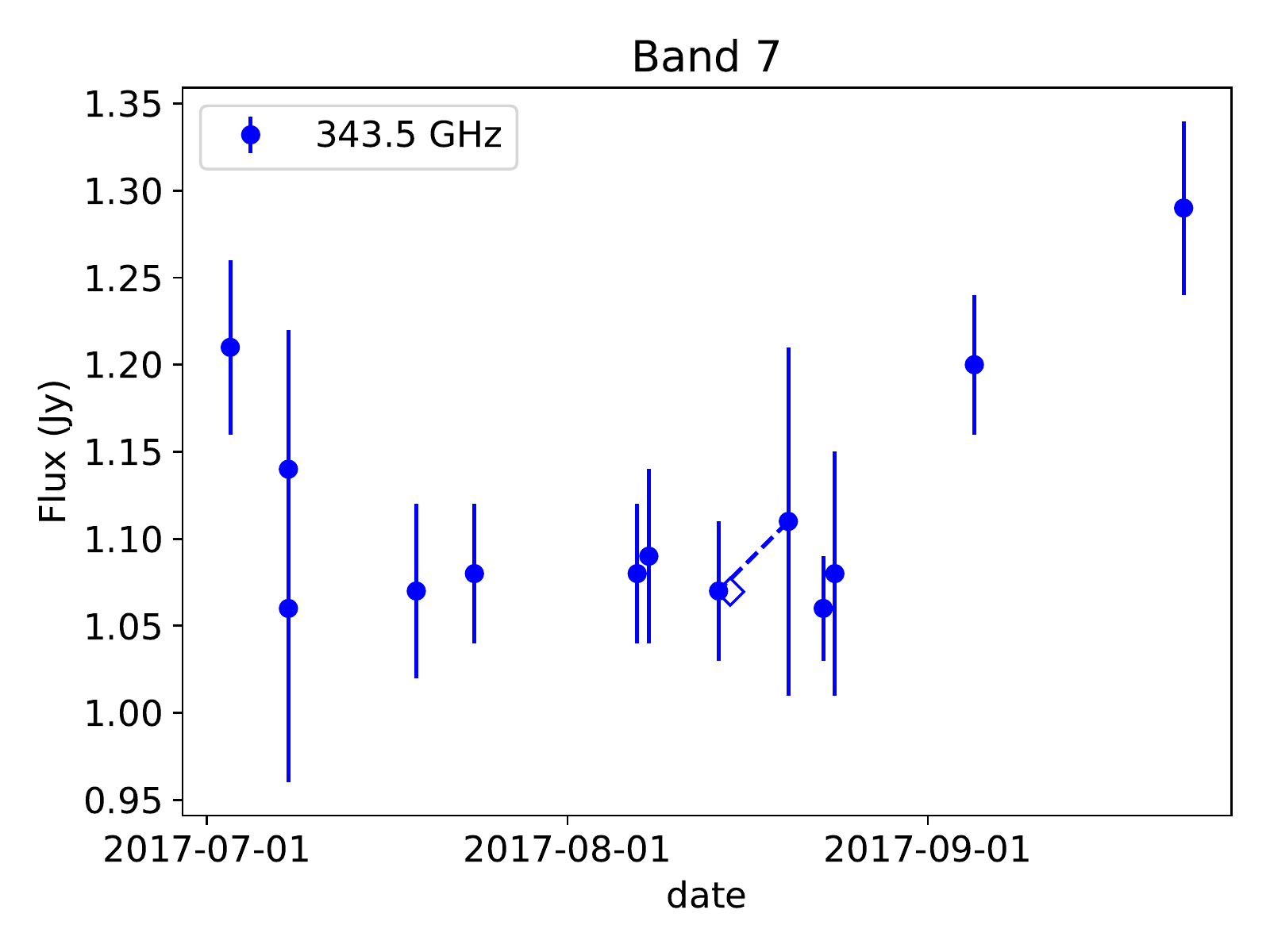}
 \caption{Calibrators measurements. \textit{Upper panel}: ALMA measurements of the J1733--1304 calibrator are plotted as blue and red circles, with corresponding error bars, at the two frequencies of Band 3, respectively. The diamond markers represent the amplitudes set for the flux calibrator in the CASA calibration script. Dashed lines are the linear regression between the plotted catalogue measurements.\textit{Middle panel}: same as for the upper panel, but with the values of the flux calibrator for Band 4 (J1924--2914), scaled at the frequencies of Band~3 with the spectral index used during calibration. \textit{Lower panel}: measurements of the J1733--1304 calibrator used for the observations at Band~7. The time range on the x-axis is three months in all the panels.}
 \label{fig:calib_b3}
\end{figure}

At Band~3 we find that frequent measurements of the flux calibrator J1733-1304 were taken around the date of our HD~163296 observations, as we show in Figure \ref{fig:calib_b3}, upper panel. The quasar shows a flux increasing with time within the selected interval (1.5 months before and after the science observations), so we can get an estimate of its flux in function of time by performing a linear fit on the data points. 
The error bars on the calibrator fluxes in the plotted time range are on the order of 1--3\%, and the deviation of the flux used during ALMA calibration (diamond markers in Figure \ref{fig:calib_b3}) lie within this range (1\% for the lower side band and 3\% for the upper side band). 

At Band 4 the amplitude calibration relied on measurements of the flux calibrator J1924-2914 at other frequencies: therefore we plot the observed data points at the two sidebands of Band 3 were close to the Band 4 observation date. We scale the flux used during calibration to the Band~3 frequencies using the spectral index specified in the corresponding CASA calibration script in the \texttt{setjy} task. We note that the measured fluxes present a rapid temporal variability within the selected time frame, so that a linear fitting is no longer adequate to interpolate the data points. If we estimate the calibrator fluxes at the time of our science observations by a simple linear regression between the precedent and following measurements (Figure \ref{fig:calib_b3}, middle panel), we find a deviation on the order of 1\% and 2.6\% in the upper and lower side-band, respectively. 

Band~6 calibrated data were taken from the DSHARP program\footnote{https://almascience.eso.org/almadata/lp/DSHARP/}: amplitude scaling to match the most recent catalog entries was performed \citep[as described in][]{2018ApJ...869L..41A} and the final dataset results from the combination of several datasets from multiple epochs and configurations, so that we can reasonably assume that the calibration error was partially attenuated from the nominal 5\% \citep{2018MNRAS.478.1512B}.  

Finally, we find several measurements of the J1733-1304 calibrator at Band~7, so we can compare directly the flux used during calibration with the measurements before and after (Figure \ref{fig:calib_b3}, lower panel). 
We find a deviation $<$1\% between the flux used during calibration and the estimated flux at that date, with the closest measurements taken only one day before the science observations and reported with an error of 3.7\%. 

With this exercise we do not aim to estimate accurately the calibration error at the different bands, but only to verify that there are no major issues affecting the absolute flux calibration, mostly out-of-date reference values for the calibrators. 
We note that using an updated catalog is fundamental to attenuate the calibration offset uncertainty when employing a quasar as a flux calibrator, because of their intrinsic temporal variability. 
We verified that this is the case for our datasets, so that the flux calibration accuracy for the single-epoch observations can be taken indicatively as the nominal value, that is $\sim$5\% for Bands 3 to 6, and $\sim$10\% for Band 7.

\section{Free-free emission estimate}
\label{ap:ff}
\begin{figure}[h!]
 \centering
   \includegraphics[keepaspectratio=True,width=0.48\textwidth]{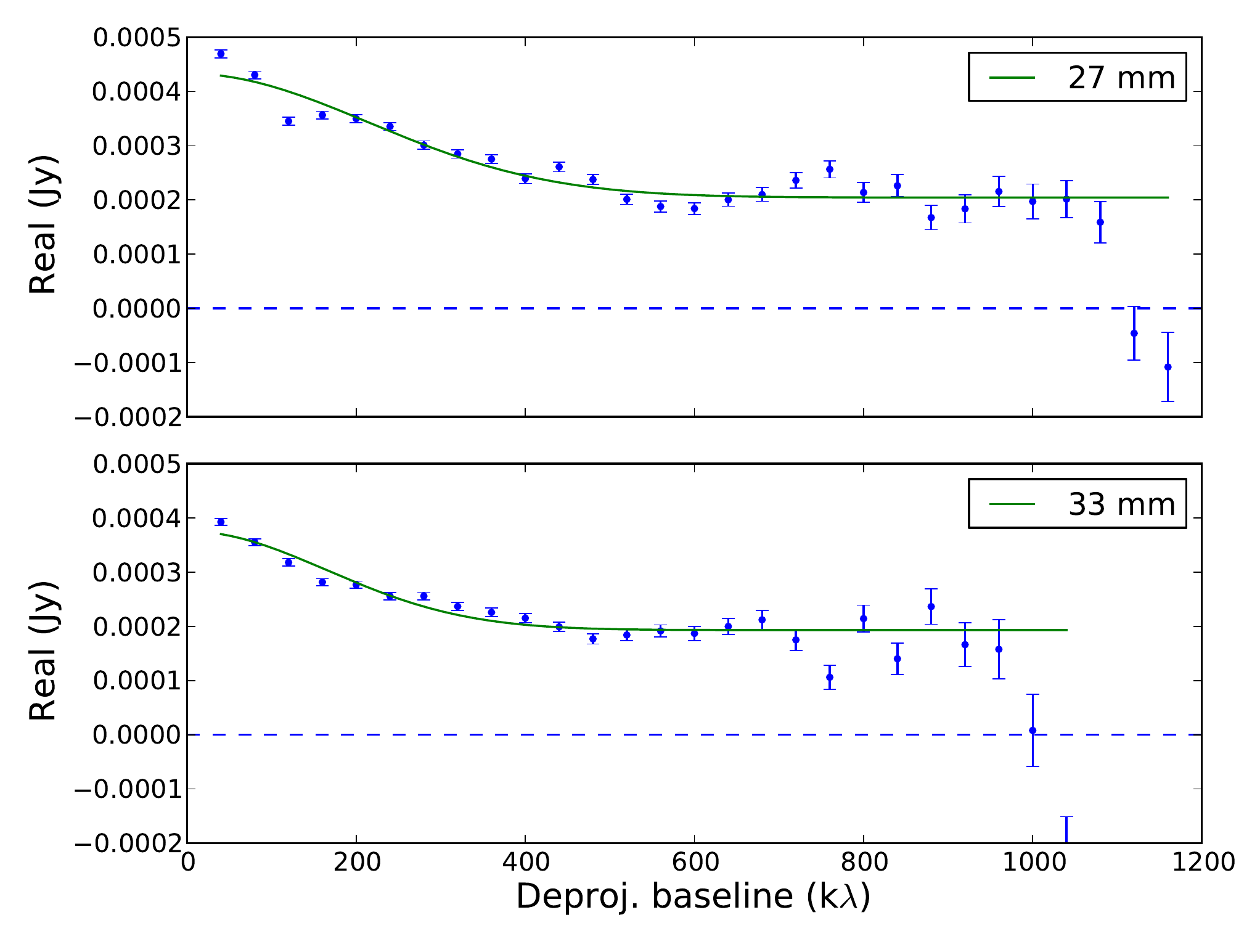}
 \caption{Visibilities in the VLA X Band: the green solid line represents the best fit for a gassian + a constant in the deprojected real visibilities.}
 \label{fig:ffX}
 \end{figure}
 
 \begin{figure*}[]
 \centering
 \includegraphics[keepaspectratio=True,width=\textwidth]{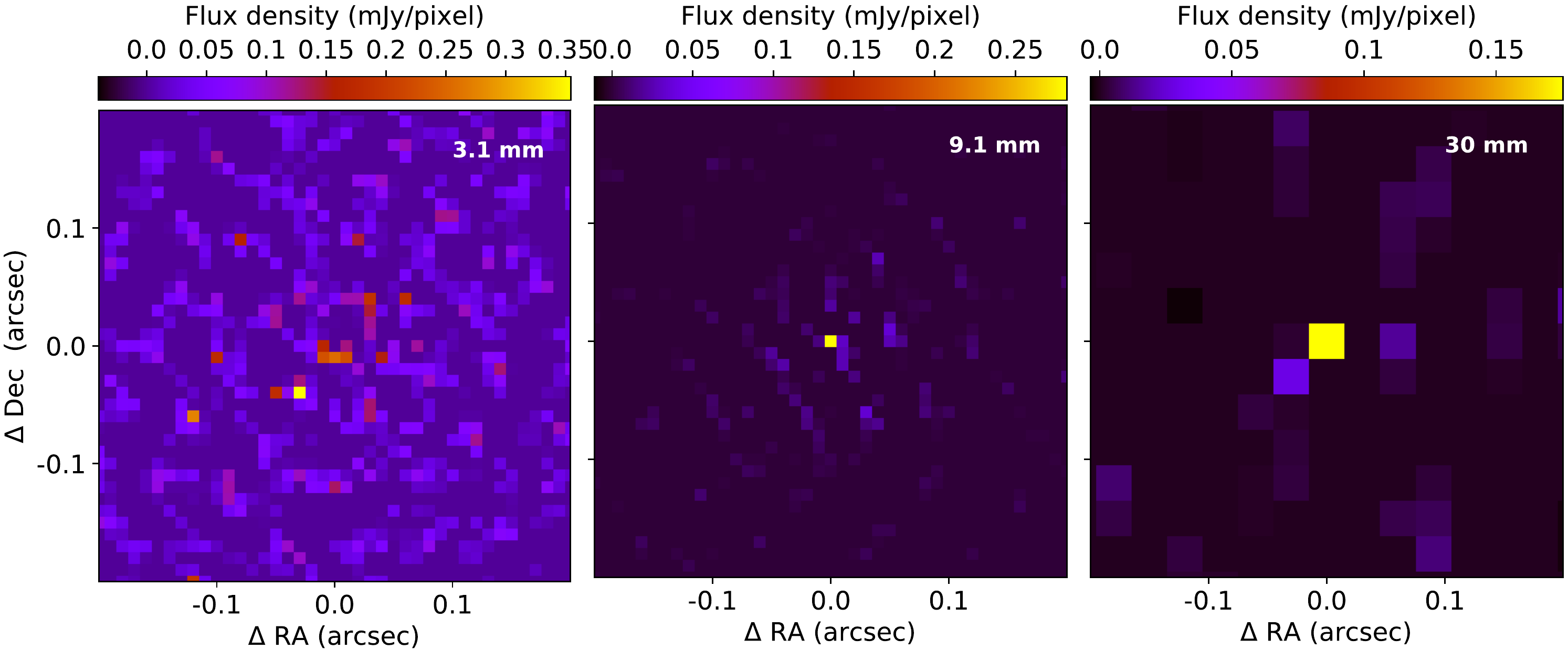}
 \caption{Model images obtained with uniform cleaning for the three long wavelength datasets. The field of view s 0\farcs2 for all images, while the pixel size depends on the dataset's resolution and is chosen according to the Nyquist sampling theorem, in order to have 2-3 pixel across one beam.}
 \label{fig:modff}
\end{figure*}
We inspected our ALMA and VLA datasets to detect and estimate a possible contamination from free-free electrons or other nonthermal processes. Despite the contribution below 7\,mm should be negligible \citep{natta04, Guidi2016}, we included in the analysis our ALMA dataset at the longest wavelengh ($\geq$3\,mm), taking advantage of the moderately high resolution to check for possible point-like sources in the center. 

If we assume that such emission comes from a very compact region close to the star (a Delta Dirac-like function in the image plane), we can describe it as a constant in the Fourier plane. i.e. we can fit an horizontal line in the deprojected visibility profile. 
We start analyzing the deprojected visibilities at Band~3: to choose the lower end of the spatial frequency range to include in the fit, we first fit a line of the form $y(\rho)_{\rho>\rho_0} = a \cdot \rho_{\rho >\rho_0}+b$ where $\rho$ is the uv-distance defined as $\sqrt{u^2+v^2}$. By varying the starting point $\rho_0$, we observe that at 3000 k$\lambda$ the slope starts to approach zero, i.e. the visibilities reach an asymptotic value. This latter is similar for both Band~3 sidebands and corresponds to 0.52 $\pm$ 0.20, where the error is calculated as the sum in quadrature of the covariance of the parameter and the systematic calibration error. 

We used the same method and starting point (3000\,k$\lambda$) for the VLA data at 8.6 and 9.7\,mm, and we find values of 0.17 $\pm$ 0.03 and 0.16 $\pm$ 0.20, respectively. 
 
The VLA data at 3\,cm cannot reach the high resolution of the observations at shorter wavelengths, and are sampled up to $\sim$1000~k$\lambda$ only. As the visibility profile appears as a combination of a compact source plus a gaussian-like distribution (Figure \ref{fig:ffX}), we fit a function composed by gaussian plus a constant over the whole range of visibilities. That would give us an estimate of the dust emission (gaussian) and a compact central contribution (constant). 
This results in a compact source of intensity (0.19 $\pm$ 0.02) for the low frequency side of the X band (33\,mm), and (0.20 $\pm$ 0.02) for the high frequencies (27\,mm).

As an additional step, we inspect the images to try and obtain further information on the spatial extent of a central contaminating emission. 
We produce images with uniform weighting (robust = -2 and nterms = 2 in the \texttt{tclean} task in CASA), to achieve the maximum resolution from aperture synthesis. The model images in Figure \ref{fig:modff} show the clean components resulting at the end of the cleaning iterative process, for the three wavelengths we considered: 3.2\,mm, 9.1\,mm and 30\,mm. In the VLA observations (central and right panels in Fig. \ref{fig:modff}, a bright central source is clearly dominating the emission, while this is not the case for ALMA Band~3 (left panel). 
In particular at 10~GHz (X band), the free-free is expected to dominate the total flux \citep[see. e.g.][]{2012ApJ...751L..42P}, and since the observations are marginally resolved we can obtain an estimate of the extent of this central bright emission. By deconvolving the image with the task \texttt{imfit} in CASA, we obtain a deconvolved size with FWHM $\simeq$ 0\farcs1, corresponding to 10~au. Therefore we can conclude that the free-free is dominating the 3~cm flux within a radius of about 5~au. Similarly, the deconvolved size at 33~GHz (Ka band) results $\sim$0\farcs09 or 9~au. 
The flux in the deconvolved model images inside this area (within a 5~au radius) can be translated in an upper limit for the free-free (as it will contain both dust emission and free-free). This results 2.97 $\pm$ 0.30 mJy at 3\,mm, 0.55 $\pm$ 0.06 mJy at 9\,mm and 0.21 $\pm$ 0.01\,mJy at 3\,cm, where the uncertainties are dominated by the calibration error. In the case of 3\,mm, we expect a putative free-free emission to be very small compared to the total flux, and the dust is likely to dominate the emission in the central regions. 
 
Finally, we subtract the fourier components corresponding to this central model emission from the original VLA visibilities, to obtain ``corrected'' visibilities that are used to extract the dust brightness profiles (see Sect.~\ref{sec:vis}). Given the uncertainties in the free-free estimation, the inner regions will be in any case masked for our multiwavelength analysis, but the removal of the central point-source is necessary for a good convergence of the forward modeling with the \texttt{frank} python package.

\FloatBarrier

\section{Brightness profiles extraction}
\label{app:vis}

We use the python package \texttt{frank} \citep{2020MNRAS.495.3209J} to derive the brightness profiles of HD~163296 from the ALMA and VLA observations, in the approximation that the emission is azimutally symmetric. This method takes advantage of the radial-only dependence of the intensity $I (r)$, which allows one to simplify the relation between this latter and the visibility function $V(\rho)$ (where $\rho$ is the $uv$-distance or spatial frequency) with a Hankel transform instead of a Fourier transform \citep{pearson99}. This consistently speeds up the inversion procedure from the visibility to the real plane, without the need of a discrete Fourier transform, and each fit is normally completed within minutes. 

The \texttt{frank} method follows a statistical approach and taking as an input the visibility data points, it retrieves a final brightness profiles with an associated confidence interval, that represent the statistical error resulting from the initial uncertainties on the visibility data. These latters are provided in the input tables as their square inverse value or $weight$ = $1/\sigma^2$, as extracted from the ALMA MS tables: such weights are associated with the RMS noise of each visibility and are first initialized with a value of $2\Delta \nu \Delta t$ (for the most recent CASA versions), and subsequently modified within the CASA software during the calibration process, when they are multiplied by the antenna-based gain factors calculated at each calibration step, and scaled for the system Temperature. 
Additional parameters entering the fit are related to the geometry of the system: for consistency we fix the inclination and position angle of the disk to the same values for each wavelength (46\degree and 133\degree, respectively), while we use the internal routine in the \texttt{frank} package to determine the offset in Right Ascension and Declination of each dataset, providing an initial guess obtained by comparing the positions of the phase center and the brightness peak in the synthesized images (see Figure \ref{fig:contmaps}) derived with a gaussian fit to the central regions of the maps with the CASA tool \texttt{imfit}. Finally, the outer radius is set to 2.2~arcseconds for the ALMA datasets and lower values for the VLA datasets (1\farcs5 for 9\,mm and 1\farcs2 for 3\,cm), and the number of radial bins is set to 300. We note that these last two parameters have very little influence on the final fit, as described in \citet{2020MNRAS.495.3209J}. 
\begin{figure*}
 \centering
 \includegraphics[keepaspectratio=True,width=0.48\textwidth]{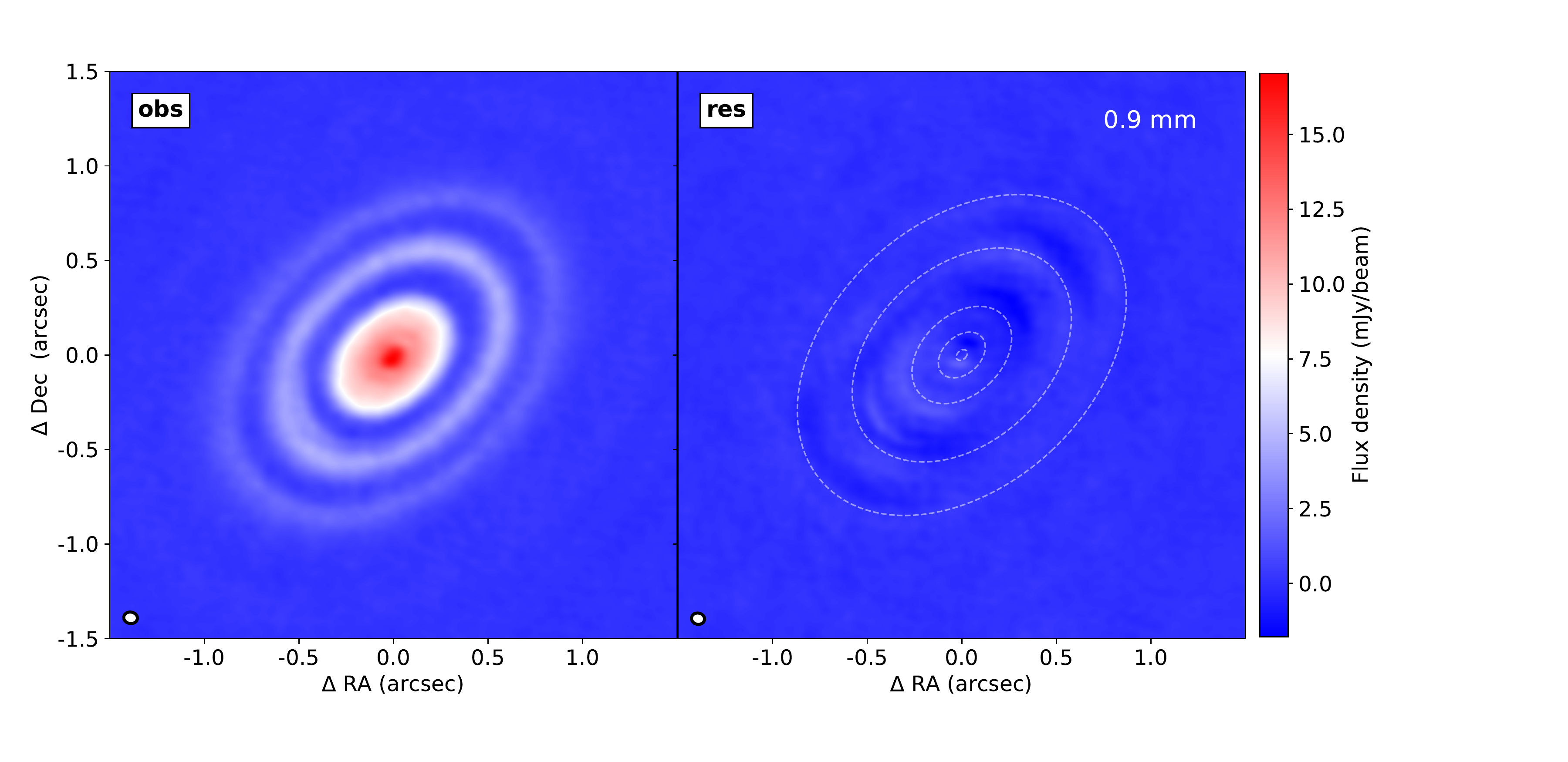}
 \includegraphics[keepaspectratio=True,width=0.48\textwidth]{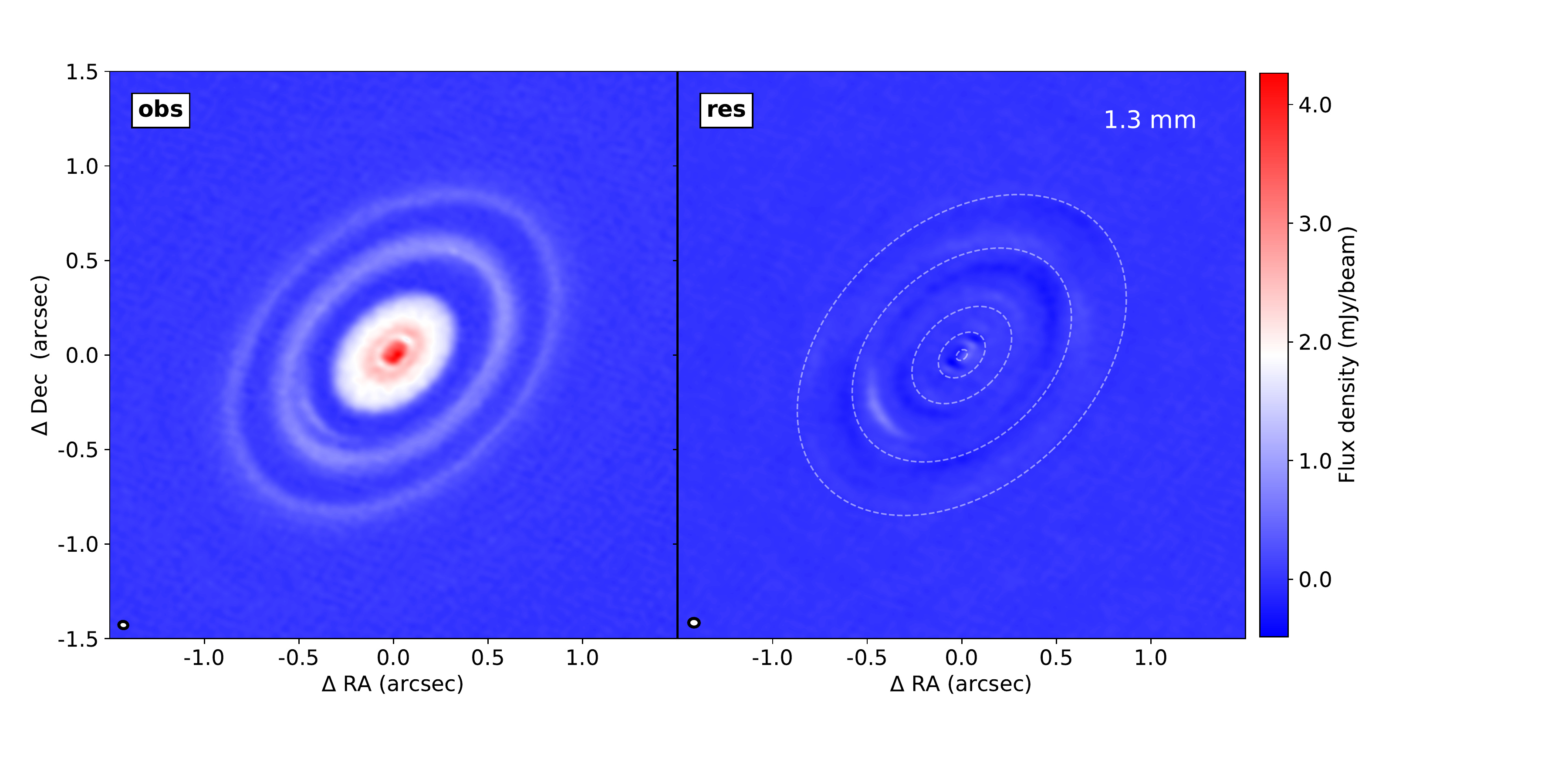}
  \includegraphics[keepaspectratio=True,width=0.48\textwidth]{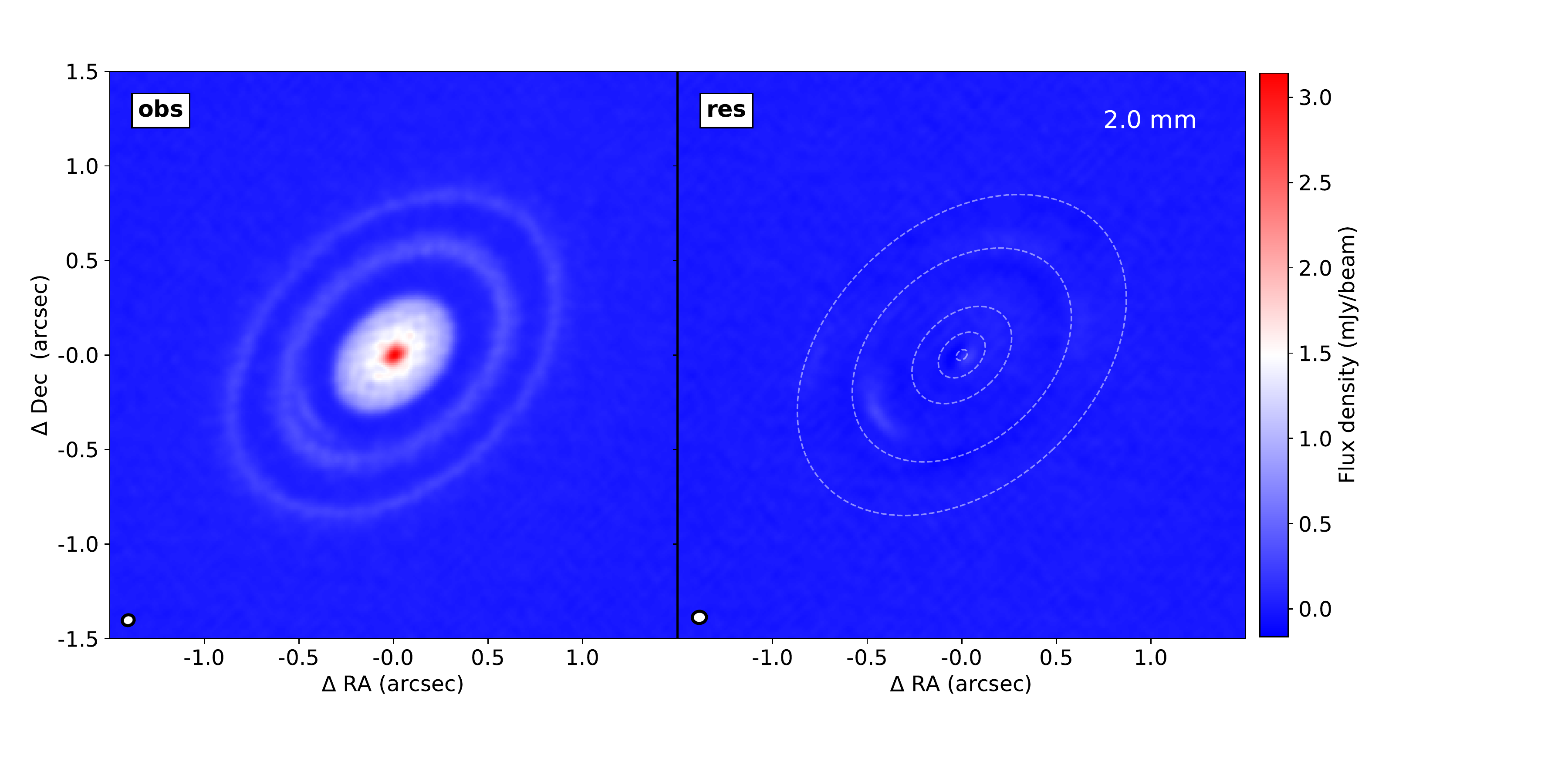}
   \includegraphics[keepaspectratio=True,width=0.48\textwidth]{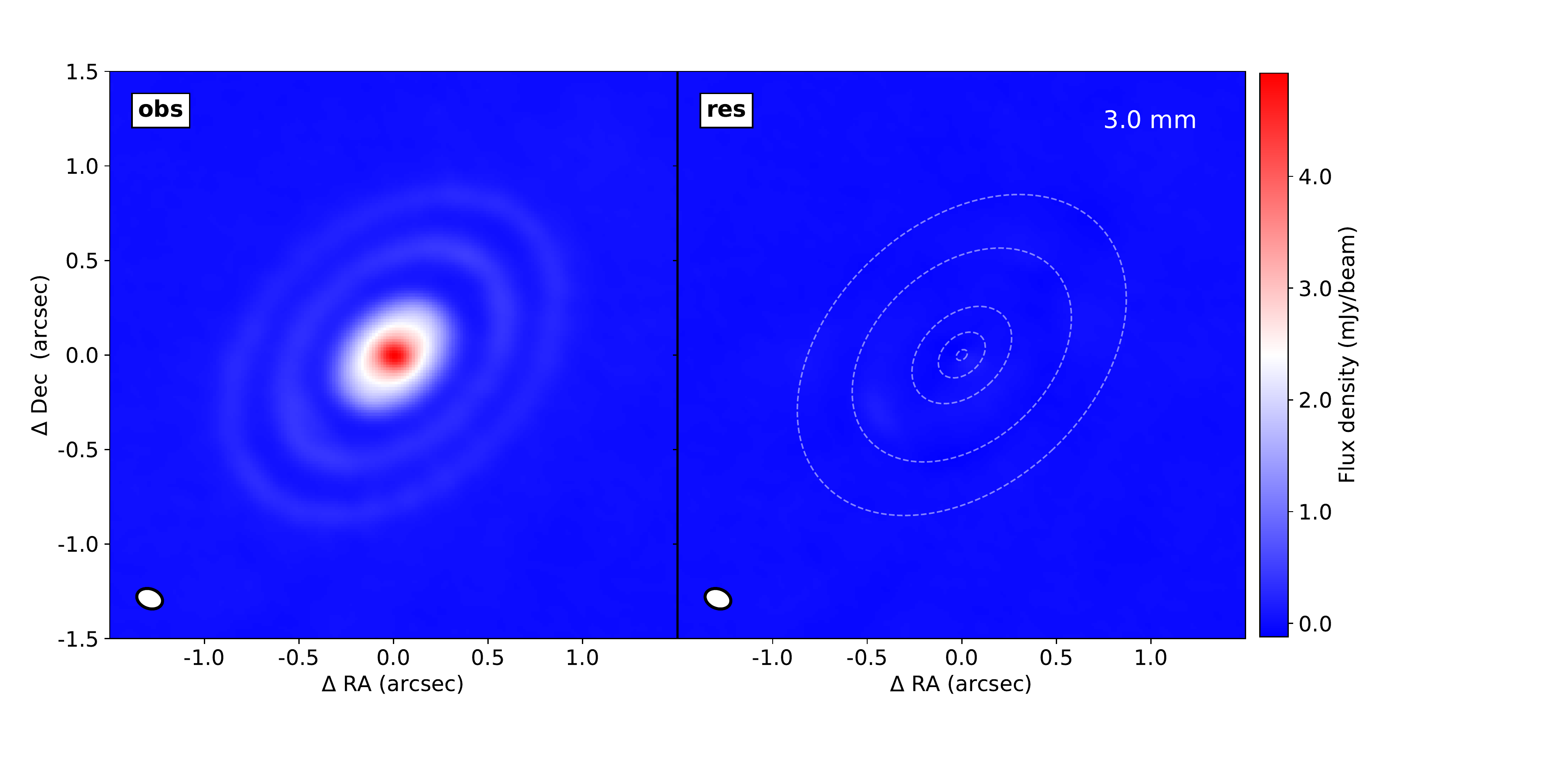}
     \includegraphics[keepaspectratio=True,width=0.48\textwidth]{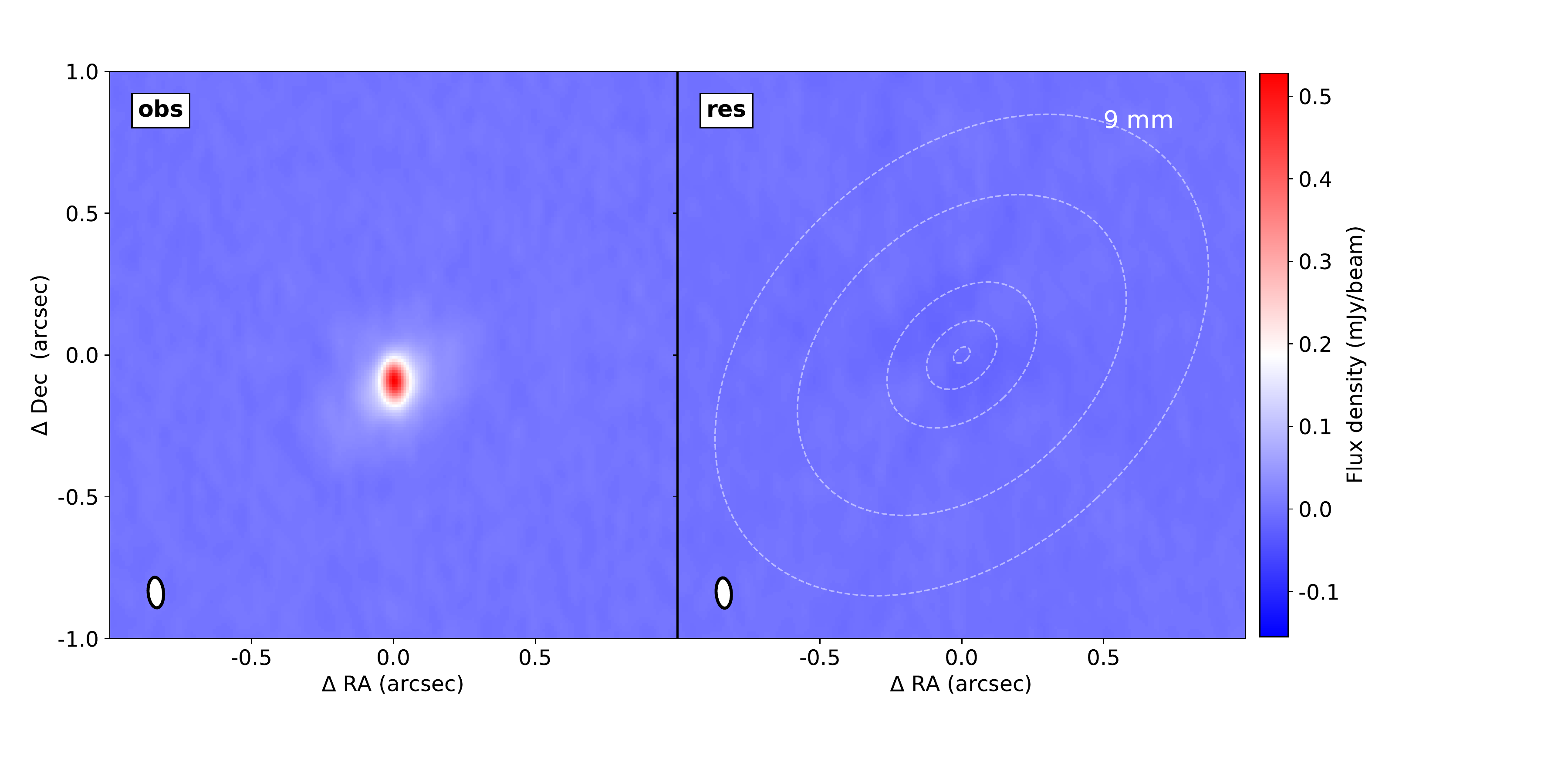}
    \includegraphics[keepaspectratio=True,width=0.48\textwidth]{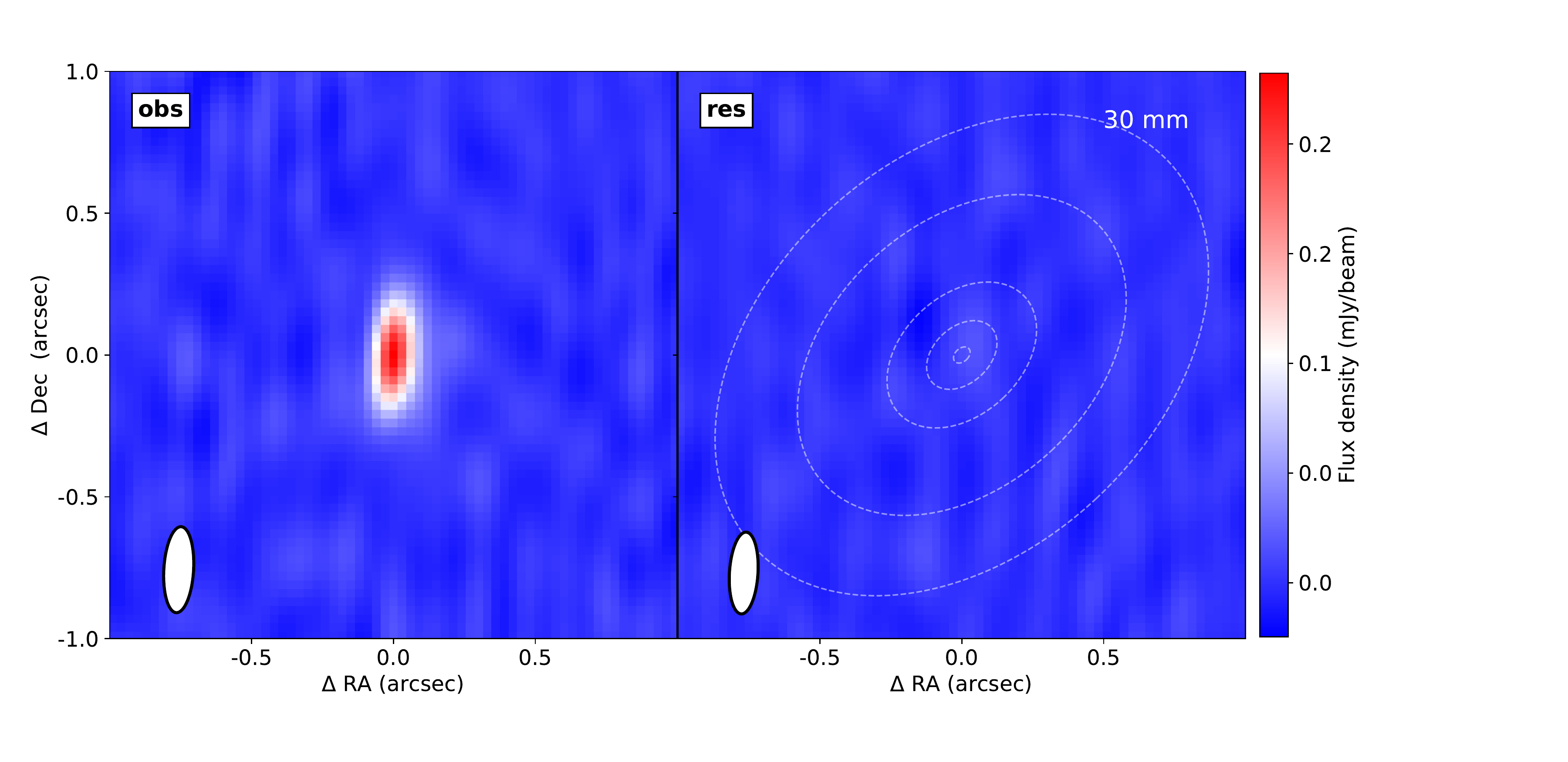}
 \caption{Observations and residual images for all wavelengths fit with \texttt{frank}, displayed with the same color scale.}
 \label{fig:res_frank}
\end{figure*}
For each dataset, we generate a grid of models varying the hyperparameters $\alpha$ and $w_{smooth}$, which corresponds to a change in the prior in this statistical framework. Using values of $\alpha$ = [1.05, 1.1, 1.2, 1.3] and $w_{smooth}$ = [0.0001, 0.001, 0.01, 0.1] we obtained 16 models for each dataset. The best-fits are chosen as the best compromise between reaching high spatial resolution (i.e. fitting to the long uv-distances) and limit the inclusion of noisy data points in fit, that generally correspond to the longest baselines and produce artificial oscillations in the extracted brightness profiles. 

A particular care is needed with the VLA datasets: as described in Sect.~\ref{sec:ff} these observations present a strong compact emission in the center, that in the Fourier space translates into a nonzero asymptotic value at the highest spatial frequencies. 

\begin{table}
	\caption{Parameters relative to the \texttt{frank} fits at each wavelength.}
	\label{tab:frankfits}
	\centering
	\begin{tabular*}{\columnwidth}{@{\extracolsep{\stretch{1}}}*{5}{c}@{}}
		\hline \hline
		$\lambda$ & $\alpha$ & $w_{smooth}$ & B$_{80}$ & $\theta$ \\
		{[mm]} &  &  & [k$\lambda$] & [arcsec] \\
		\hline
		0.89 & 1.1 & 0.001 & 2600 & 0.08 \\
		1.3 & 1.3 & 0.001 & 2940 & 0.07 \\
        2.0 & 1.1 & 0.01 & 2580 & 0.08 \\
        3.0 & 1.2 & 0.1 & 2240 & 0.09 \\
        9.1$^{\star}$ & 1.5 & 1.0 & 2180 & 0.09 \\
        30$^{\star}$ & 1.05 & 0.001 & 620 & 0.33 \\
		\hline
	\end{tabular*}
	\tablefoot{$\alpha$ and $w_{smooth}$ are the hyperparameters used for the retrieval of the brightness profiles with \texttt{frank}. The last two columns report the baseline up to which the model reproduces the visibilities within 20\% (B$_{80}$), and the corresponding spatial scale. The rows marked with a $\star$ symbol highlight the datasets that have been first corrected for a central compact emission. }
\end{table}
This impacts the \texttt{frank} solutions introducing strong oscillations in the intensity profiles, at scales corresponding to the uv-distance where the power spectrum (or the \texttt{frank} prior) drops to zero. The reason of this behavior is explained in \citep{2021arXiv210302392J}, and is related to the \texttt{frank} power spectrum that needs to converge to zero by construction. For this reason, we correct the input visibilities for this compact emission, which we already discussed it is not associated with dust (see Sect.~\ref{sec:ff}).  
We therefore subtract from the VLA 9\,mm dataset the constant value previously obtained by fitting the real  visibilities at the high uv-distances (see Appendix \ref{ap:ff}). This approach is suggested by the authors of the package in \citet{2021arXiv210302392J}, and seems to be working fairly well for this dataset, allowing us to unveil the 9\,mm emission in the 100~au ring, already hinted by the images produces with CASA \texttt{tclean}, but too noisy to be appreciated in the azimuthally averaged profiles of the cleaned images. We note that the central emission ($\lesssim$10~au) remains affected by the uncertainties on the free-free estimate and subtraction, so that we do not consider it a reliable representation of the dust emission. 
At 3\,cm, even after such a point-source correction, we still get significant oscillations and negative values in the final solutions. We find the best result by first subtracting from the visibilities the central emission obtained by the clean model with robust = -2 described in Appendix \ref{ap:ff}. This is done by transposing the clean model of the central source into the Fourier place with the task \texttt{ft} in CASA, and then subtracting it from the data. 
We note that this means subtracting the dust central emission as well, but since the uncertainty on the free-free contribution is anyway limiting our capability to give a precise estimate of the dust central flux, this does not represent an issue, as our goal is trying to unveil the emission at larger radii. 

The final models are displayed in Figure \ref{fig:bff}, the corresponding values of the hyperparameters are listed in Table \ref{tab:frankfits}, along with a metric proposed in \citet{2021arXiv210302392J} to characterize the spatial resolution accuracy of the fits: we measure B$_{80}$ as the shortest baseline beyond which the difference between the model and the observations has values $\geq$20\% for at least 200k$\lambda$ consecutively, and it is meant to represent the baseline where the fits starts to depart appreciably from the data. 
We note that we choose to fit the whole frequency band even at long wavelengths (larger than 3\,mm), as the fit obtained separating the datasets in higher and lower frequencies (as for the synthesized images in Figure \ref{fig:contmaps}) results in a much lower quality solutions (higher oscillations at large radii and larger RMS error). 
In Figure \ref{fig:res_frank} we show the residual maps from the modeling: these are produced from the \texttt{frank} residual visibilities and using the same parameters in CASA \texttt{tclean} as for the observations images (displayed next to the residual maps).

As explained in Sect.~\ref{sec:vis}, for the purpose of the spectral modeling a second round of fits is performed so that all the datasets are sampled at the same spatial scales. This way we obtained five spectral profiles with same accuracy B$_{80}$ = (2100 $\pm$ 100)~k$\lambda$ from 0.9\,mm up to 9\,mm. Such a resolution could not be achieved for he 30\,mm dataset, that is sampled up to lower spatial scales (as reported in Table \ref{tab:frankfits}), and was therefore not included in the SED modeling. However, we show the prediction of our models at this wavelength in Figure \ref{fig:fluxes_X}.

\section{SED fitting}
\label{app:sedfit}
\begin{figure*}
 \centering
 \includegraphics[keepaspectratio=True,width=0.45\textwidth]{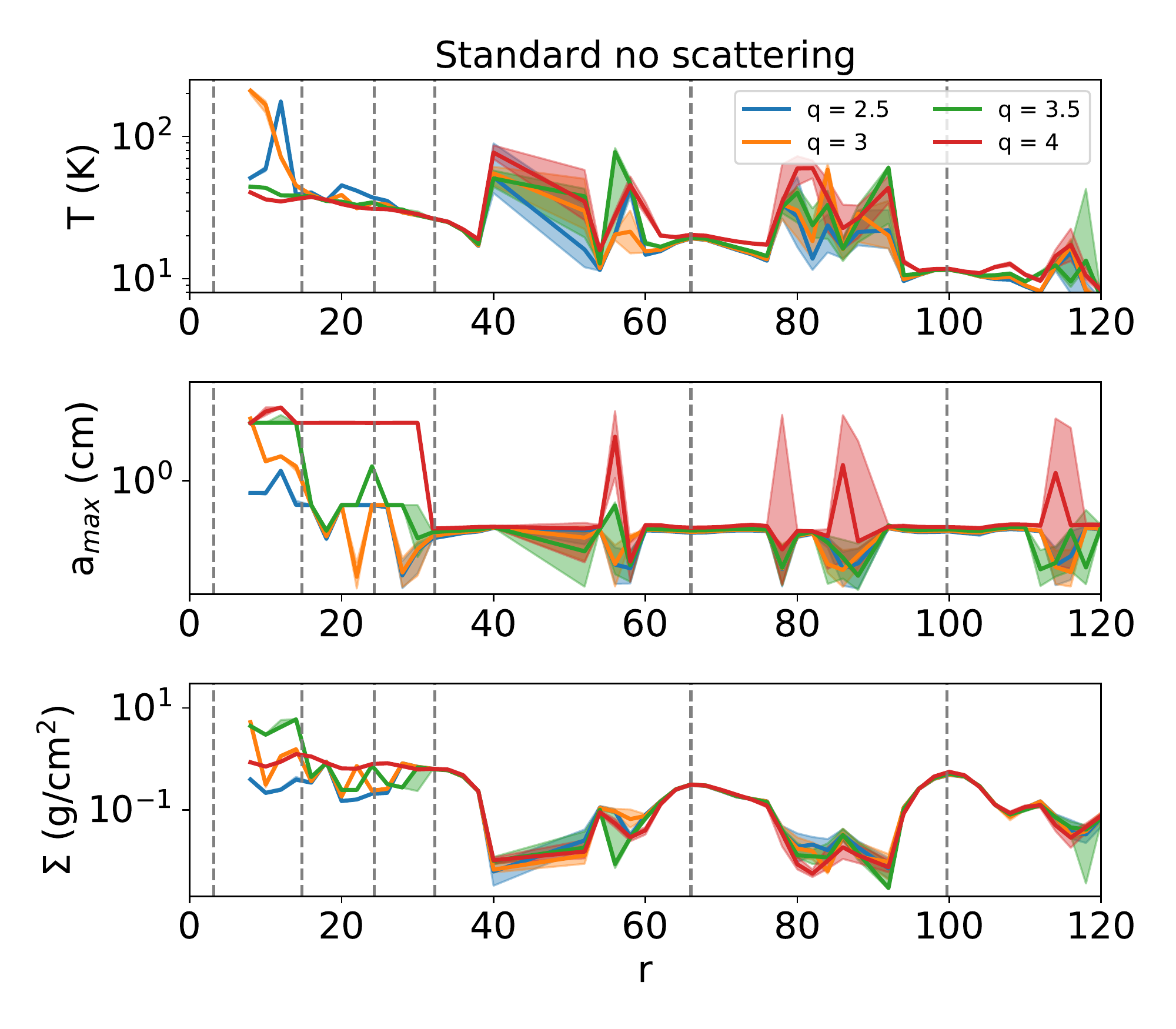}
  \includegraphics[keepaspectratio=True,width=0.45\textwidth]{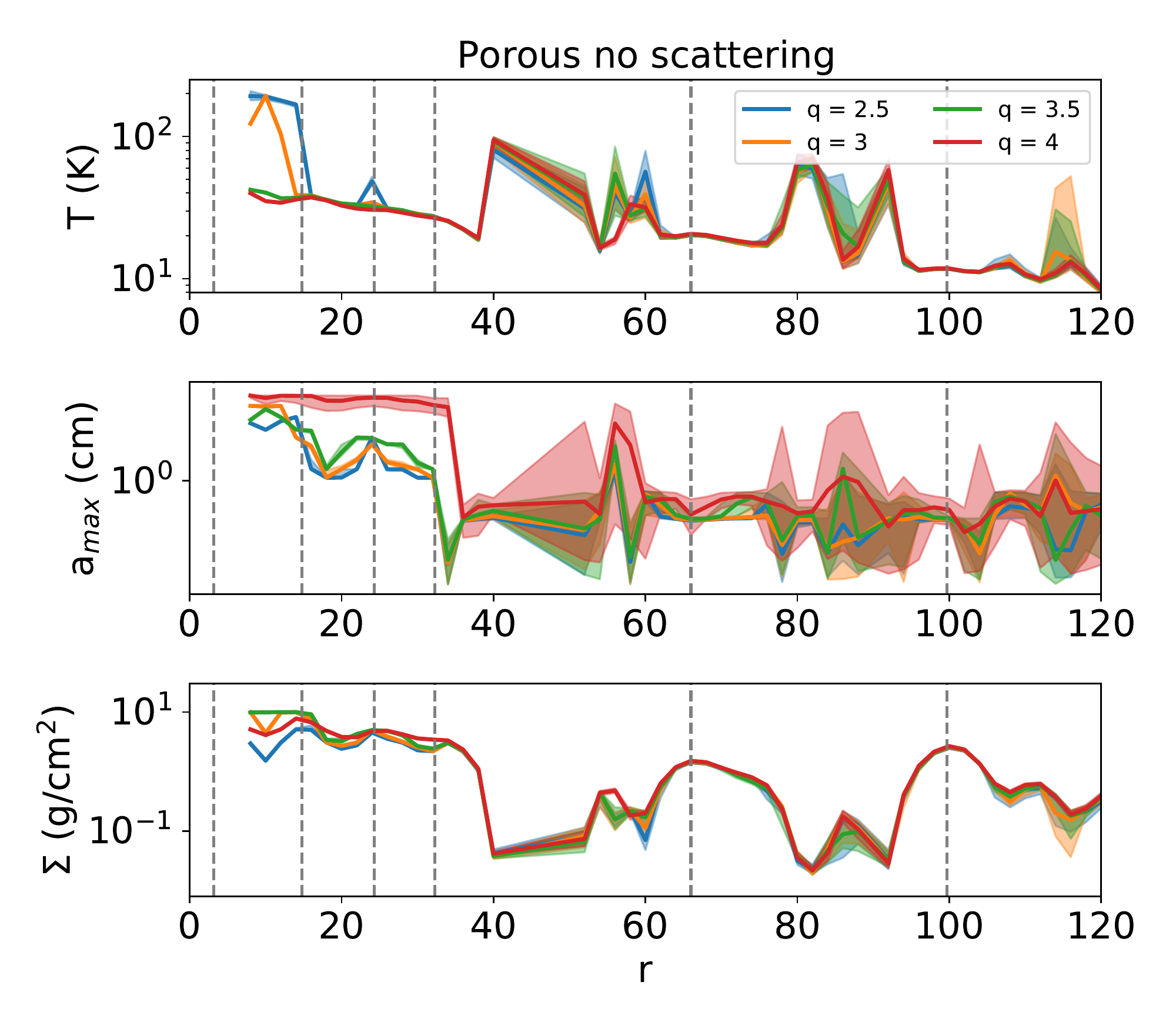}
 \caption{Results of the Monte Carlo fits for the nonscattering models. \textit{Left panel}: best-fit parameters for the standard composition and the four different size distributions for the nonscattering model. The shaded regions represent the 16th and 84th percentile of the posterior distributions at each radius, from the fits perfomed including only the statistical error in the likelihood evaluation. \textit{Right panel}: same as the left panel but for porous grains.}
 \label{fig:allbf_noscat}
\end{figure*}
\begin{figure*}
 \centering
 \includegraphics[keepaspectratio=True,width=0.45\textwidth]{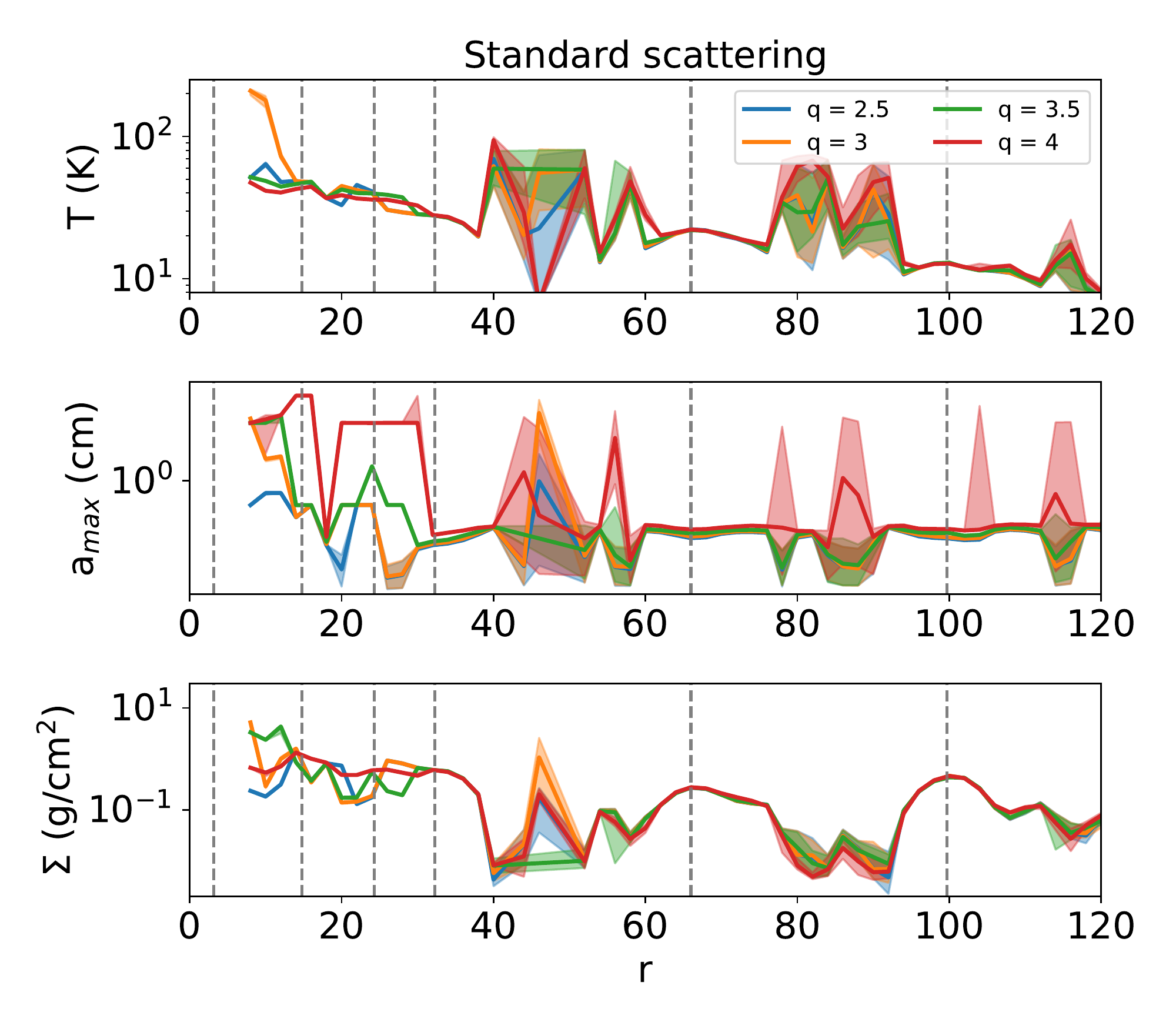}
  \includegraphics[keepaspectratio=True,width=0.45\textwidth]{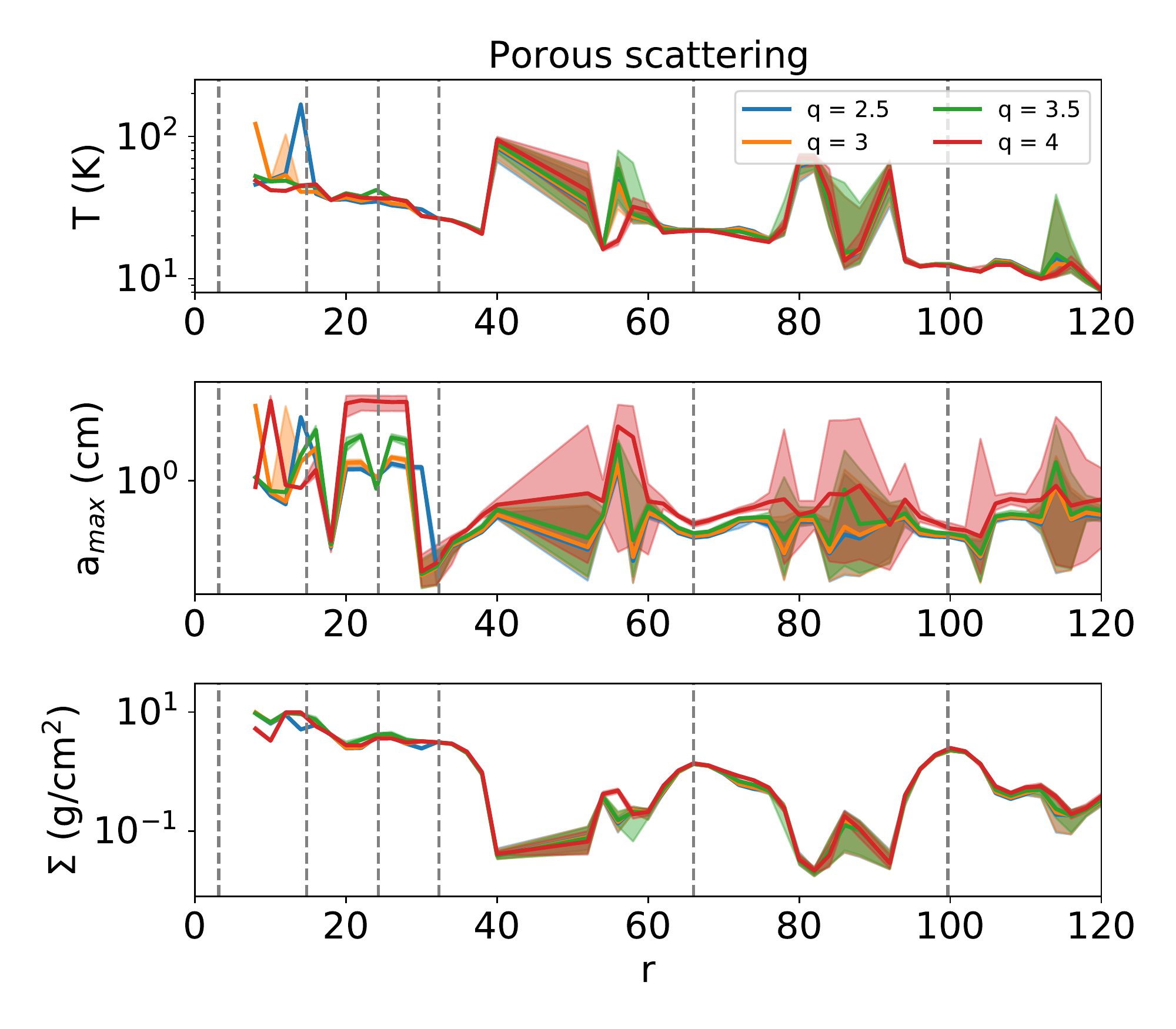}
 \caption{Same as in Figure \ref{fig:allbf_noscat} for the scattering model.}
 \label{fig:allbf_scat}
\end{figure*}
In Sect.~\ref{sec:parmodel} we show the results of fitting a power-law opacity (eq. \ref{eq:pl}) to the brightness profiles from 0.8\,mm to 9\,mm. 
In all the fits the uniform prior probabilities are set for all the three parameters, in the intervals $log_{10}\tau$ = [-4, 3], $\beta$ = [0 , 5], and T (K) = [3, T$_{up}$]. 
Similarly, the best-fit parameters of the physical model T, $a_\mathrm{max}$ and $\Sigma_d$ (Sect.~\ref{sec:pm}) are obtained using a uniform prior probabilities for T (K) = [3, T$_{up}$],  $log_{10}a_\mathrm{max}$  (cm) = [-4, 3] and $\Sigma_d$ (g/cm$^2$) = [0, 10]. 
The temperature prior upper limit varies with the radius, and corresponds to the temperature of the surface layer in the optically thin approximation, computed as in \citet{dullemond2001}, Equation 7. For the  effective temperature and stellar radius we use the values computed by \citet{2018ApJ...869..164S} (R$_*$ = 1.6 R$_{\odot}$ and T$_{*,\mathrm{eff}}$ = 9250\,K). The Planck mean opacities for protoplanetary disks are taken from \citet{2003A&A...410..611S} and computed with the fortran script \texttt{opacity.f} made available by MPIA Heidelberg\footnote{https://www2.mpia-hd.mpg.de/homes/semenov/Opacities/opacities.html}. These variable upper limits are useful especially where the Temperature is poorly constrained such as in the dust gaps, as they allow us to obtain physically meaningful values for the temperature that could otherwise undertake indefinitely high values. 
At each radius, we include the fluxes above 3 times the rms value at the single wavelengths, calculated as the standard deviation in the signal-free region of the disk. If this criteria is satisfied for less than 3 wavelengths we do not perform the fit at that location. 

At radii smaller than 8~au, some of the fitted parameters result highly unconstrained: in particular the posterior distributions of the surface density (in the physical model) and the $\beta$ index (in the parametric model) tend to accumulate on the upper and lower edge of the prior limits, respectively. This can be expected in the case of a highly optically thick emission, when the surface density or the opacity index are very loosely constrained because of the dominance of the Planck term in equation \ref{eq:pl} and \ref{eq:noscat}. In fact, even enhancing the upper limit of the prior to very high values (10$^5$ g/cm$^2$), we still observe the same behavior of the posterior distribution. Another explanation could be the failure of the simplified radiative transfer formulation in describing the emission in the very inner regions. In Figure \ref{fig:allbf_noscat} and \ref{fig:allbf_scat}  we show the best fits for all 8 dust populations introduced in Sect.~\ref{sec:mwle} for the nonscattering and scattering model, respectively. 

\begin{figure}
 \centering
 \includegraphics[keepaspectratio=True,width=0.5\textwidth]{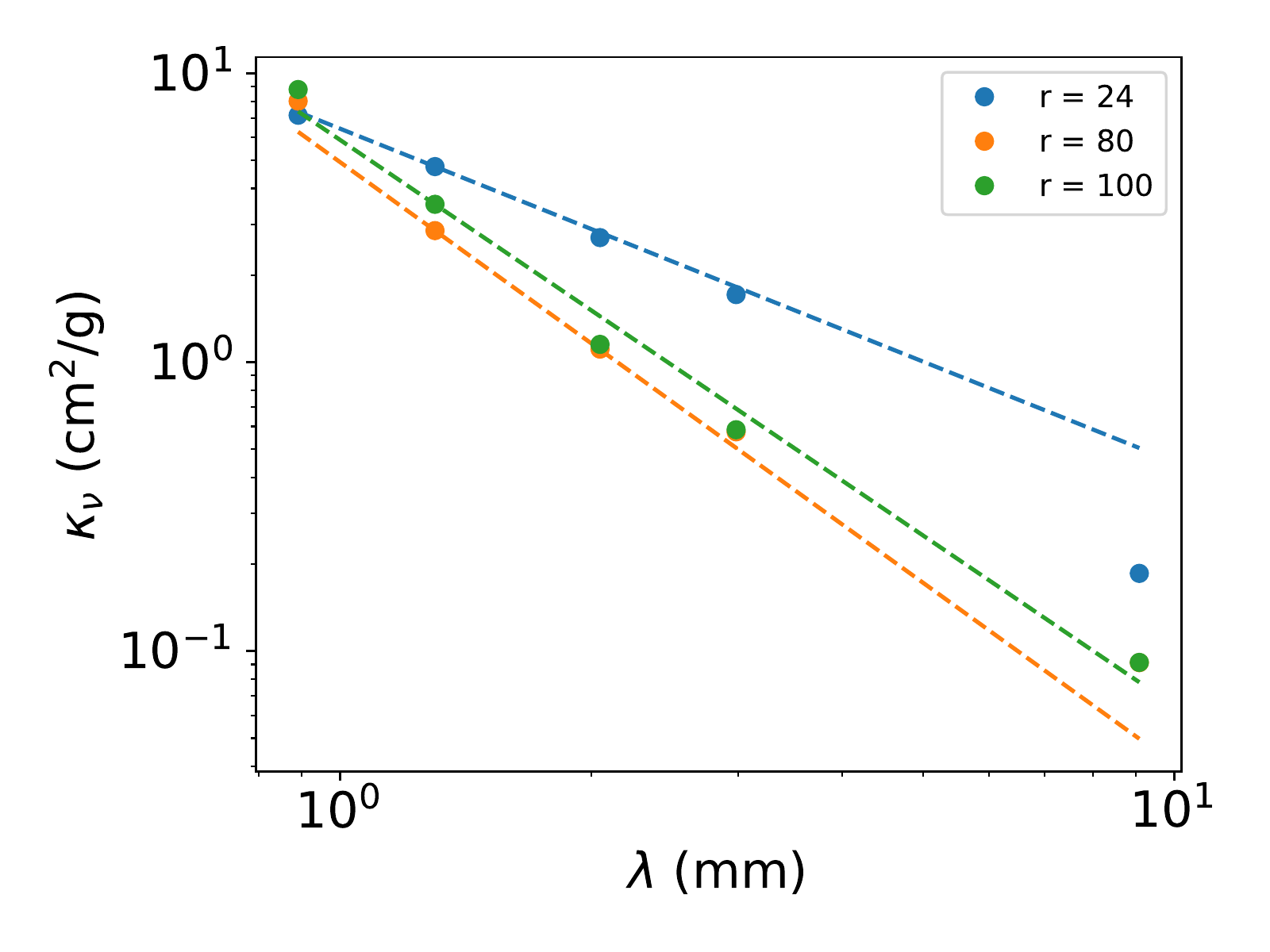}
 \caption{Absorption opacity at three different location (inner disk at 20\,au, gap at 80\,au and ring at 100\,au) from the nonscattering model plotted as circles. The dashed lines represent the power-law from the parametric fit, with $\kappa_{abs}$ = $\kappa_0$ ($\nu/\nu_0)^\beta$ where we used $\kappa_0$ = $\kappa_{1.3\mathrm{mm}}$ from the physical model. }
 \label{fig:kappa_plaw}
\end{figure}

In Figure \ref{fig:kappa_plaw} we show the absorption opacities derived from the physical model compared to a power-law prediction. Given the lower evidence of the parametric model shown in Sect.~\ref{sec:mcomp}, this indicates that the dust opacities are not well approximated with a power-law. 

\begin{figure}
 \centering
 \includegraphics[keepaspectratio=True,width=0.5\textwidth]{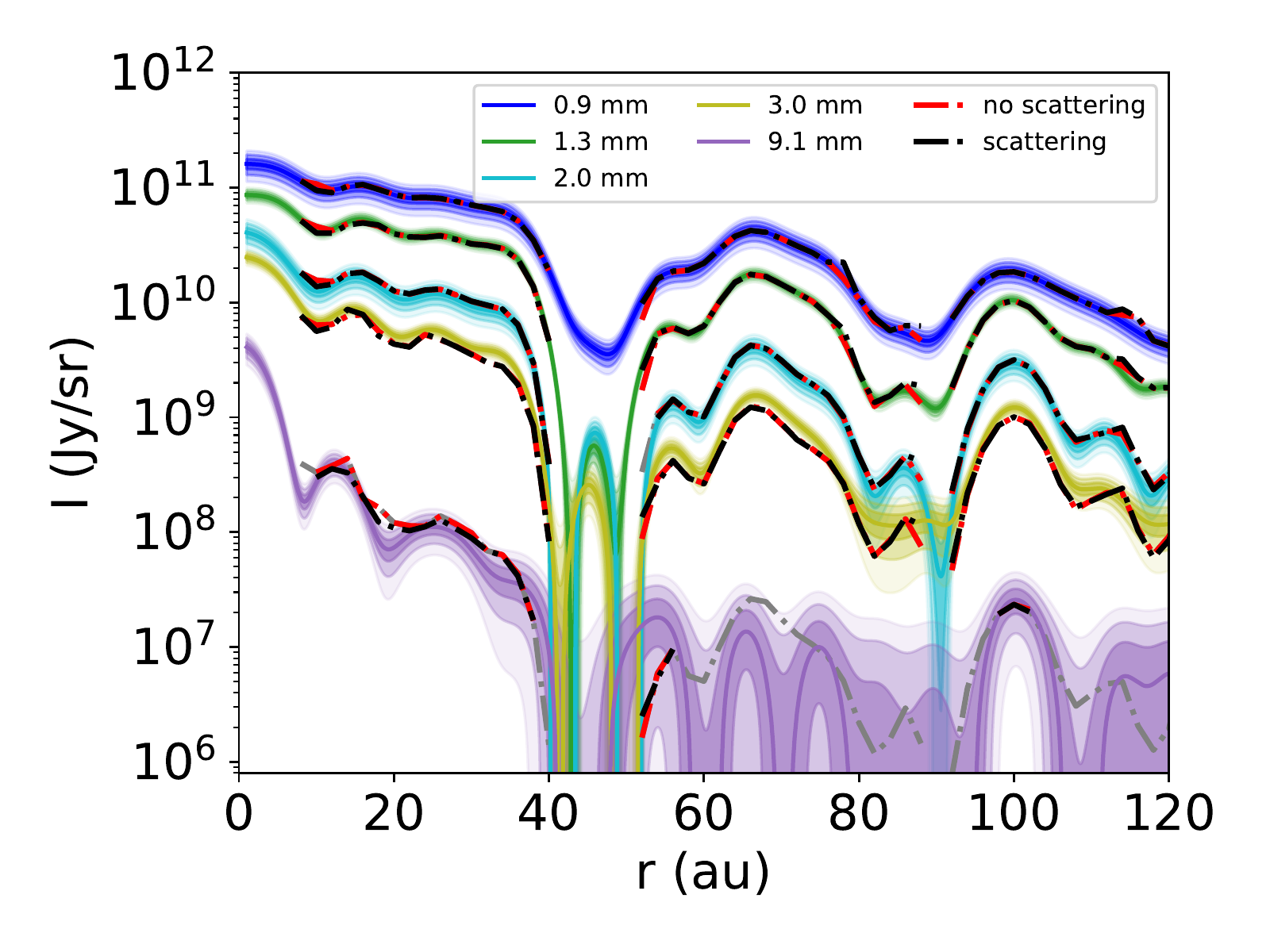}
 \caption{Flux densities at the five different wavelengths predicted by the best-fit physical models using standard grains in the nonscattering and scattering case. The red and black dashed-dotted curve shows the models predictions at the radii where the flux at each wavelength was included in the fitting procedure (only fluxes larger than 3 times the rms error were considered at each radius). The grey curve shows the scattering model prediction at all radii. }
 \label{fig:fluxes_pm}
\end{figure}

\begin{figure}
 \centering
 \includegraphics[keepaspectratio=True,width=0.5\textwidth]{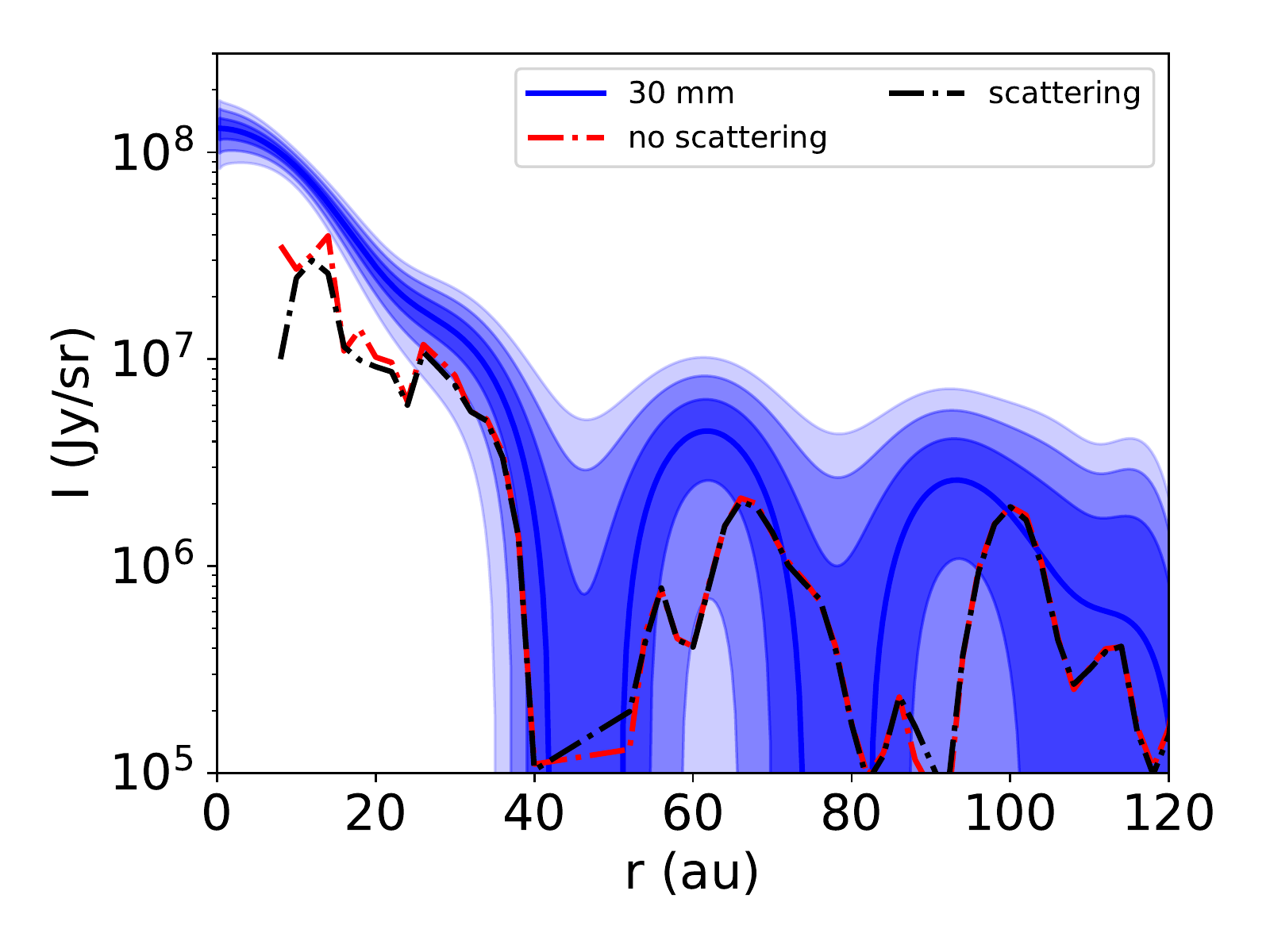}
 \caption{Flux densities at 30 mm predicted by the best-fit models in the nonscattering and scattering case, shown as red and black dashed-dotted lines, respectively. The observed brightness profiles at VLA Band X are plotted with the corresponding 1, 2 and 3 $\sigma$ errors (computed as the sum in quadrature of the statistical error and the flux calibration error) as shaded regions. }
 \label{fig:fluxes_X}
\end{figure}

\begin{figure*}
 \centering
 \includegraphics[keepaspectratio=True,width=\textwidth]{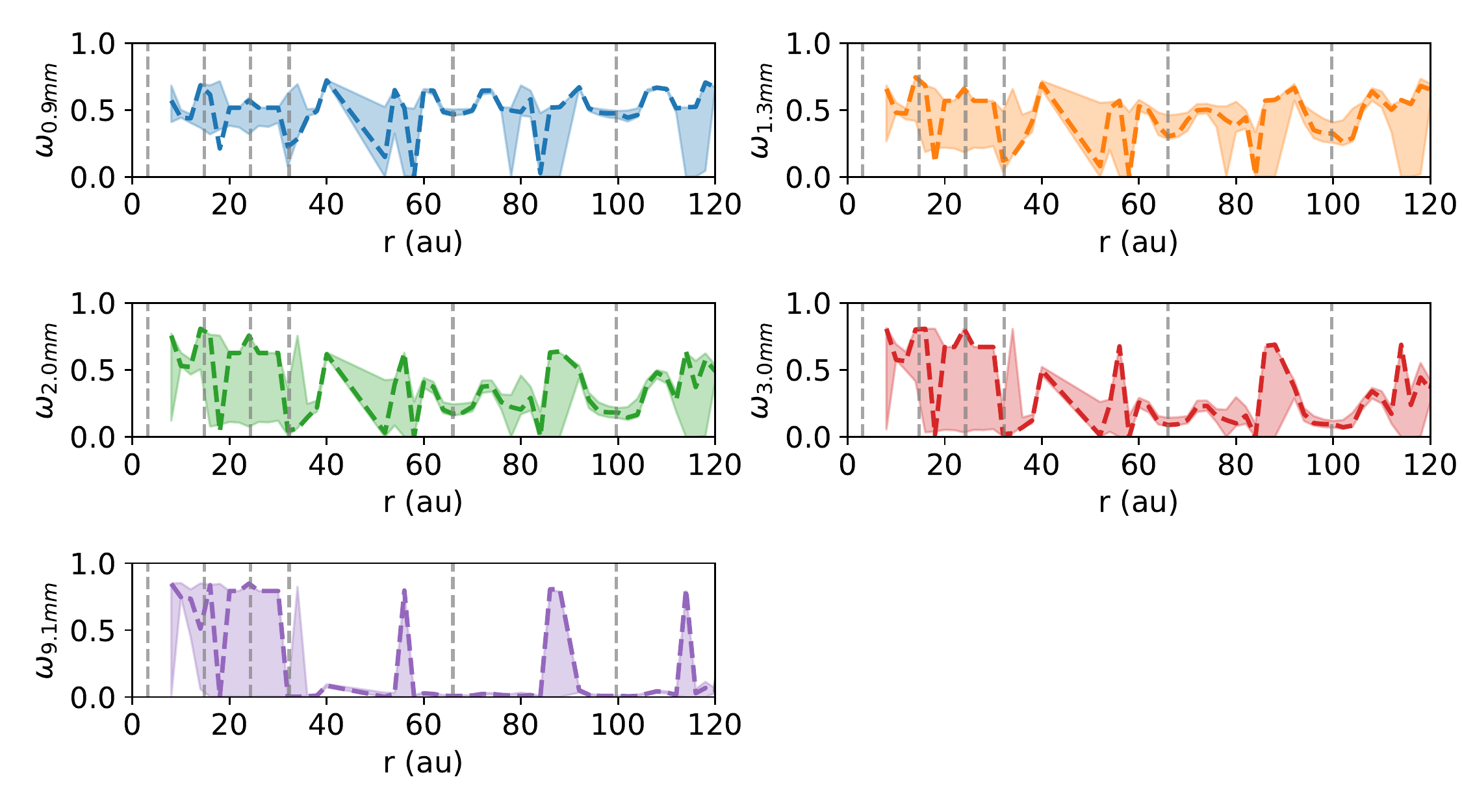}
 \caption{Scattering albedo as function of the radius for the five considered wavelengths. The shaded areas are drawn between the 16th and 84th percentile of the model distribution, as described in Sect.~\ref{sec:mwle}.}
 \label{fig:omegas_wle}
\end{figure*}

In Sect.~\ref{sec:pm} we show how the standard grains result to be a better model for describing our observations, and therefore we choose these latter as our reference model. 
The comparison between the predicted and observed brightness profiles is displayed in Figure \ref{fig:fluxes_pm}, while we show in Figure \ref{fig:fluxes_X} the predicted flux at 3\,cm. 
The scattering albedo at each wavelength for the scattering model are shown in Figure \ref{fig:omegas_wle}.

Finally, we show in Table \ref{tab:mrings_por} the masses we derived with the porous-grains model, as a comparison with the reference values displayed in Table \ref{tab:mrings}. While the total mass is higher by a factor of four, the mass in the 100\,au ring results 40\% of the total, similarly to what derived for the standard grains in Sect.~\ref{sec:dmass}.   

\begin{table}
	\caption{Dust mass and median dust temperature derived from the physical model with porous grains in the outer rings, and across the portion of the disk where the spectral analysis was carried out. The confidence intervals are calculated using the upper and lower estimates of the surface density and temperature from the best-fit model. The numbering of the rings is taken from Sect.~\ref{sec:vis}.}
	\label{tab:mrings_por}
	\centering
			Nonscattering model - porous grains\\
		\vspace{2pt}
		\begin{tabular*}{\columnwidth}{@{\extracolsep{\stretch{1}}}*{4}{c}@{}}
		\hline \hline
		 & R &  M$_d$ & T$_d$ \\
	     & [au] & [M$_{\oplus}$] & [K] \\
		\hline
		R5 & 58--84 & 271$^{+52}_{-100}$ & 20.1 \\
		\vspace{3pt}
		R6 & 92--108 & 507$^{+48}_{-145}$ & 11.7 \\
		\vspace{2pt}\\
		disk & 8--120 & 1535$^{+212}_{-551}$ & 20.8 \\
		\vspace{0.2pt}\\
		\hline
	\end{tabular*}
	\vspace{4pt}\\
	Scattering model - porous grains\\
	\vspace{2pt}
		\begin{tabular*}{\columnwidth}{@{\extracolsep{\stretch{1}}}*{4}{c}@{}}
		\hline \hline
		 & R &  M$_d$ & T$_d$ \\
	     & [au] & [M$_{\oplus}$] & [K] \\
		\hline
		R5 & 58--84 & 232$^{+65}_{-112}$ & 22.8 \\
		\vspace{3pt}
		R6 & 92--108 & 463$^{+78}_{-138}$ & 12.7 \\
		\vspace{2pt}\\
		disk & 8--120 & 1417$^{+304}_{-524}$ & 22.9 \\
		\vspace{0.2pt}\\
		\hline
	\end{tabular*}
\end{table} 
\FloatBarrier
\section{Comparison with \citet{2021ApJS..257...14S}}
\label{app:dsharp}
In a recent study \citet{2021ApJS..257...14S} found the presence of millimeter-grains in the disk of HD~163296 outside 40\,au and a local increase in grain size at $\sim$100\,au. Since in this work we do not find millimeter grains or signature of dust trapping in the outer rings, we run a series of test to determine what are the main factor responsible for this discrepancy.
First, we note that there are some difference between these two studies: in terms of datasets, this work relies on a larger wavelength coverage (4 ALMA and one VLA band, compared to 2 ALMA bands (1 and 3\,mm) in \citet{2021ApJS..257...14S}), and a higher spatial resolution of a factor of about 2.5. 
They also assume a midplane temperature profile from \citet{2021ApJS..257....5Z}, obtained by modeling the CO emission lines and continuum emission at 1\,mm. 
Finally, the assumptions on the dust composition are different: our absorption opacities are on average a factor 2--3 higher than the DSHARP opacities used in \citet{2021ApJS..257...14S}, while the scattering opacities are comparable (this is related to the presence of amorphous carbon from \citet{1996MNRAS.282.1321Z} in the DIANA opacities, that produce higher absorption coefficient partially suppressing the role of dust self-scattering). 

\begin{figure}
 \centering
 \includegraphics[keepaspectratio=True,width=0.45\textwidth]{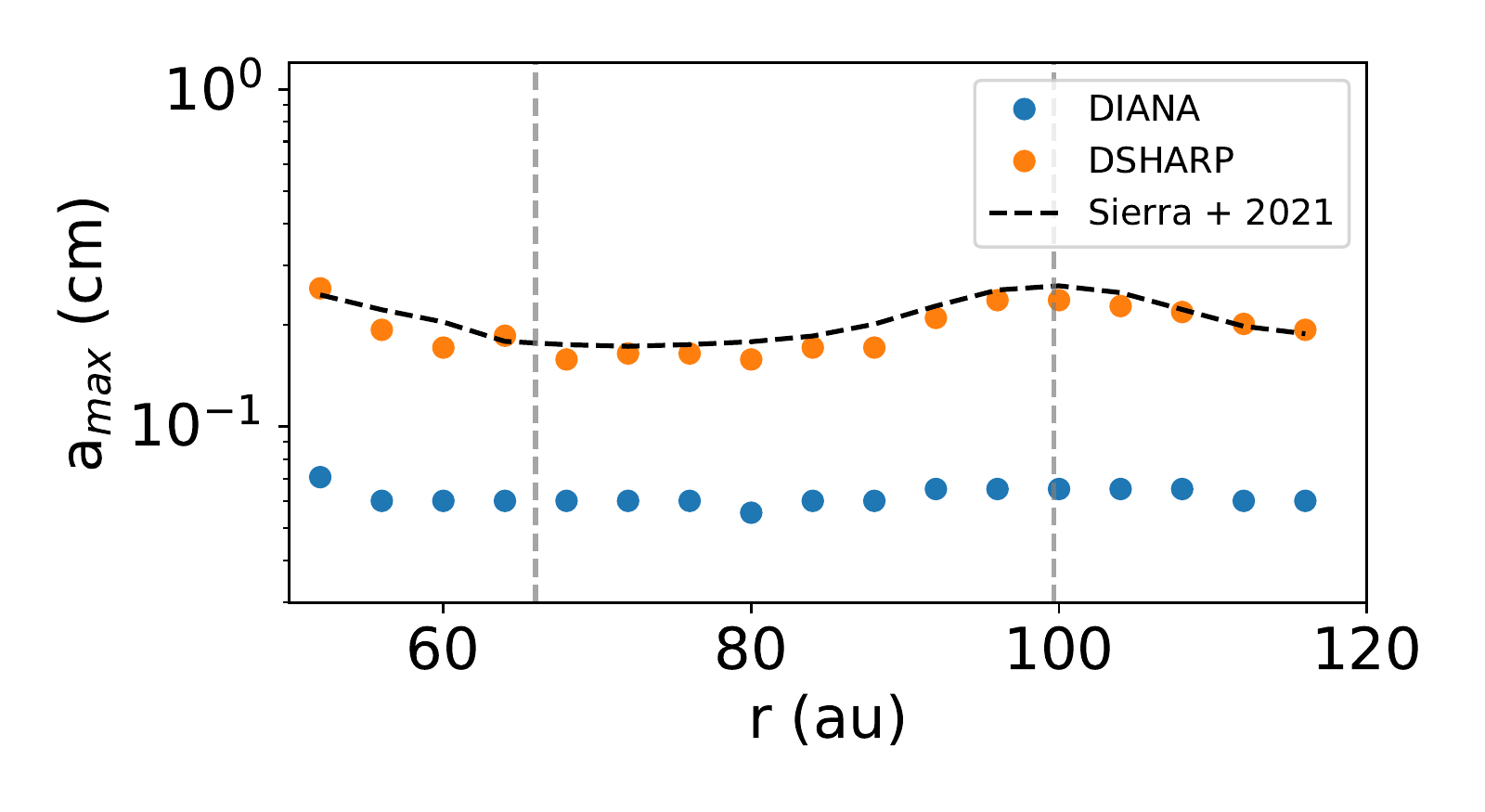}
 \includegraphics[keepaspectratio=True,width=0.45\textwidth]{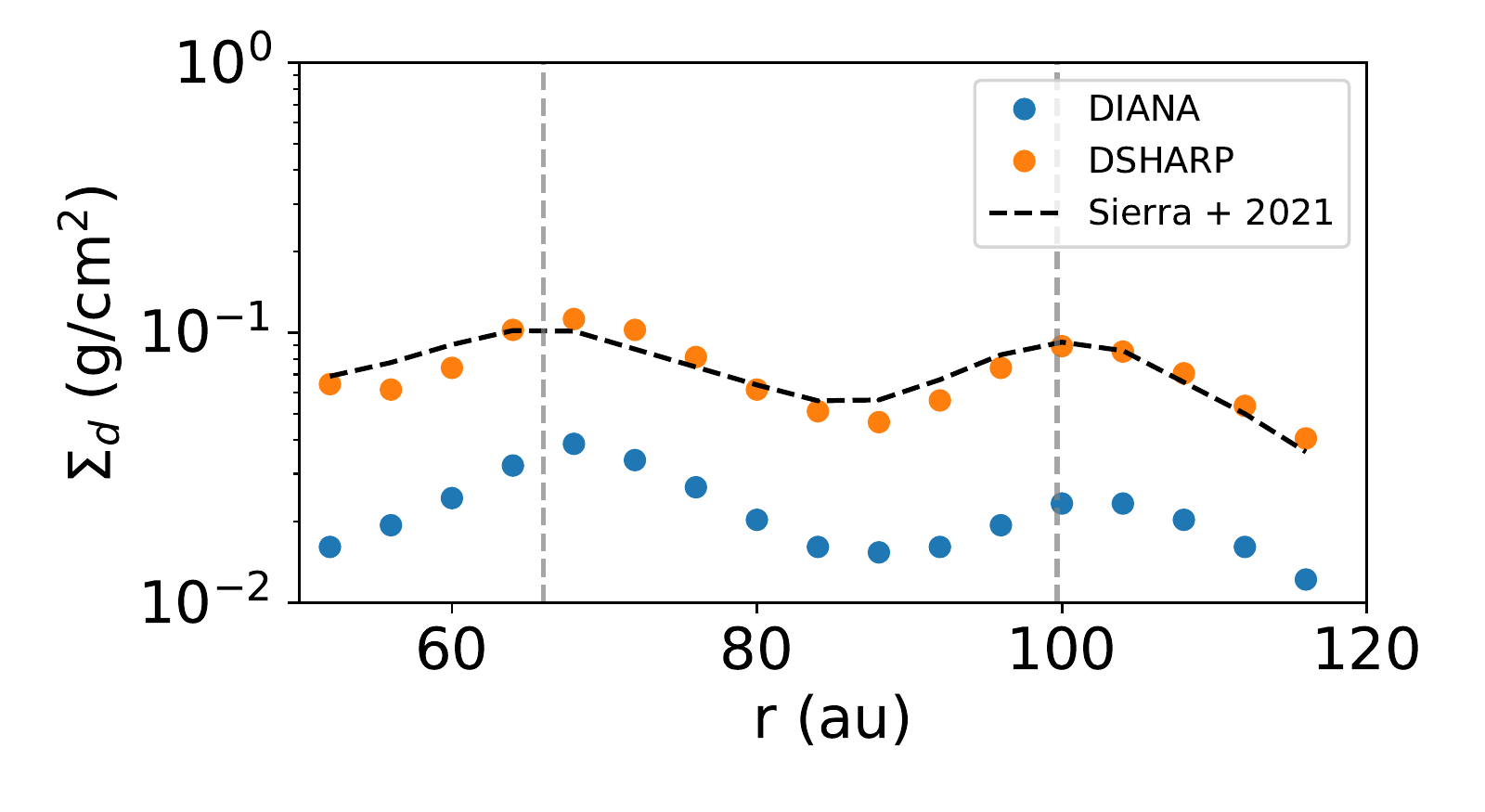}
 \caption{Parameters estimates using two different dust opacities. \textit{Top panel}: maximum grain size in the outer disk of HD~163296 computed from lower resolution data (see main text), using the DIANA and DSHARP opacities, with a grain size distribution with q = 2.5. We overplot with empty black circles the values from \citet{2021ApJS..257...14S}. \textit{Bottom panel}: same as for the top panel, but showing the surface density estimates.}
 \label{fig:diana_dsharp}
\end{figure}

To test whether the choice of dust opacity plays the major role in the final estimates of $a_\mathrm{max}$, we performed a quick check by taking the flux profiles at  1, 2.3 and 2.6\,mm published in \citet{2021ApJS..257...14S} and performing the same procedure that the author described but using the standard DIANA opacities described in this work - instead of the DSHARP - and with a size distribution q = 2.5. We fix the temperature profile as in \citet{2021ApJS..257...14S}, and approximate the statistical error to 5\% at each radius, summing in quadrature an additional 5\% for the calibration error (and giving a double weight to the Band~3 data). Using a two-dimensional grid with 200 bins for both log $a_\mathrm{max}$ [-4, 3] and log $\Sigma_d$ [-3,1], we calculate the likelihood as $L = exp(-0.5 \cdot \chi^2)$, with $\chi^2_r = \sum_{\nu_i} w_i (F_{obs,r,\nu_i}-F_{mod, r, \nu_i})^2/ \sigma_{r, \nu_i}^2$ and the model fluxes computed with eq. \ref{eq:scattering}. We derive the estimates for the two parameters as the values at the peak of the marginalized posterior distribution. We focus on the outer disk (r $\gtrsim$ 40\,au) and find a well defined posterior distribution for a$_\mathrm{max}$ and $\Sigma_d$ at all radii, as described in \citet{2021ApJS..257...14S}. 
The maximum grain size found using DIANA opacities are systematically lower than the one found using DSHARP opacities, by a factor of 3 on average (see Figure \ref{fig:diana_dsharp}, top panel). 
The DIANA surface densities are lower by a similar factor ( Figure \ref{fig:diana_dsharp}, bottom panel), which is expected because of the higher values of absorption opacities mentioned above. As a check, we plot the final values from \citet{2021ApJS..257...14S} in the scattering case, and confirm that we get the same results (the small difference with our best-fit values in orange is likely due to our approximation of the flux statistical error). 
We note that the maximum grain size calculated with the DIANA opacities does not seem to have a local increase at the 100\,au ring, opposite to what found with the DSHARP grains and in \citet{2021ApJS..257...14S}. 

\begin{figure}
 \centering
 \includegraphics[keepaspectratio=True,width=0.5\textwidth]{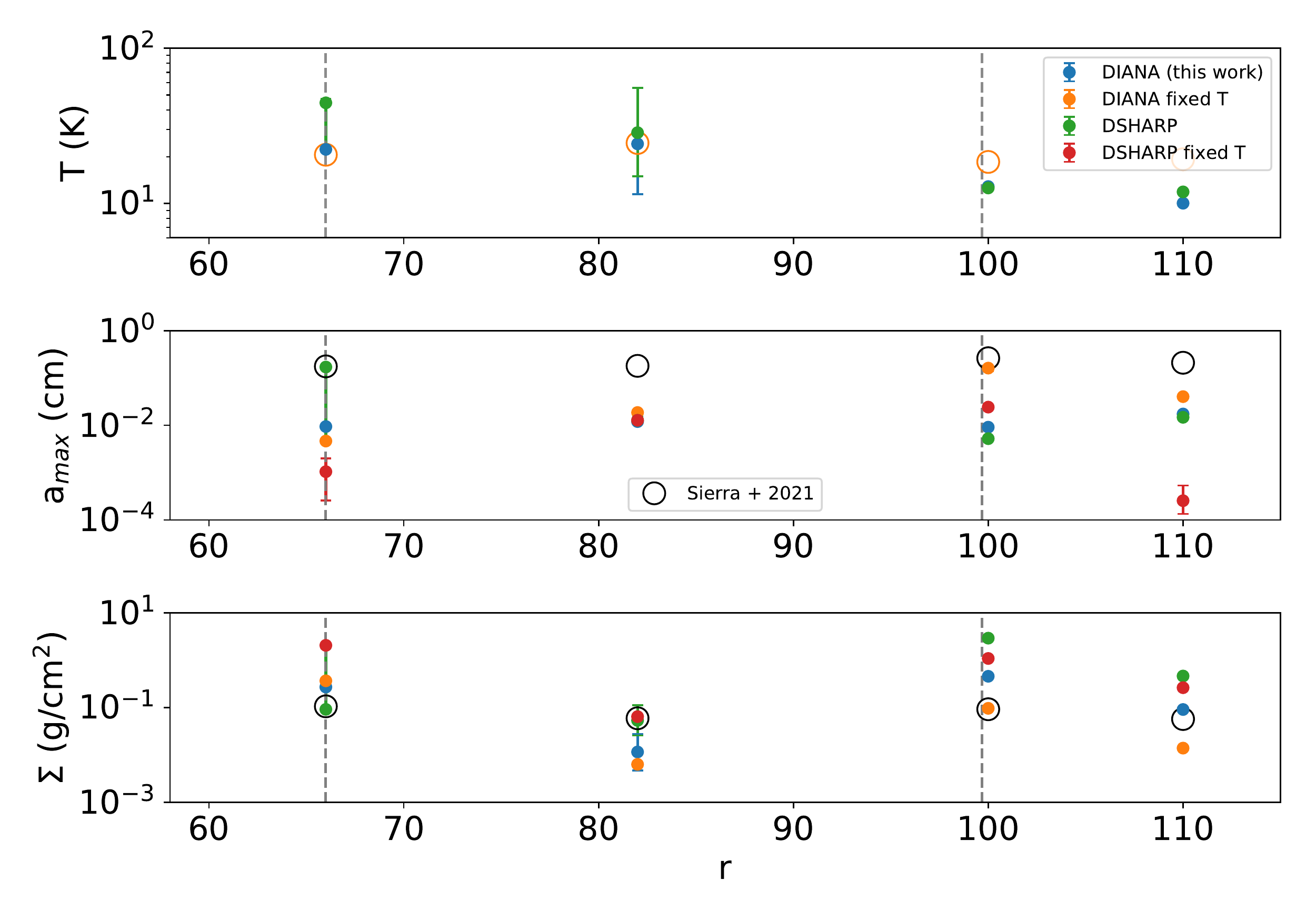}
 \caption{Comparison of the dust parameters in the outer disk estimated with a Monte Carlo nested sampling tool in the scattering case, using statistical error only and for a size distribution with q = 2.5. Empty markers represent the fixed Temperature n the upper panel, and the values of the best-fot parameters from \citet{2021ApJS..257...14S} in the middle and bottom panel.}
 \label{fig:fit_dsharp}
\end{figure}

Estabilished that the grain composition and dust opacities can significantly affect the grain size and surface density values, we need to consider that signal dilution - where the disk structures are not resolved - and the assumption of a fixed dust temperature can play a additional role. To get an approximate idea of these effects, we run our nested sampling routine fitting the datasets used in this work, first using the DSHARP opacities, and then adding the fixed temperature profile as in \citet{2021ApJS..257...14S}. We note that this is just a simple test that relies on the data statistical error only (we explained how we additionally computed the effect of calibration error with 30 additional fits in Sect.~\ref{sec:mwle}), so that it does not give a clear indication of the absolute uncertainties, but it can be useful for a relative comparison. 
We show the results in Figure \ref{fig:fit_dsharp}: we find that the spatial resolution and the assumed temperature profile have both an impact on the final parameters. We note that fixing the temperature profile (that at 100\,au corresponds to a $\sim$50\% increase compared to out bestfit values) can induce a significant variation in $a_\mathrm{max}$ resulting in a larger grain size in this ring.

\end{appendix}

\end{document}